\documentclass[aps,rmp,author-year,twocolumn]{revtex4}
\usepackage{dcolumn,graphicx}

\def\cf{{\it cf.}}
\def\eg{{\it e.g.}}
\def\etal{{\it et al.}}
\def\etc{{\it etc.}}
\def\ie{{\it i.e.}}

\def\spose#1{\hbox to 0pt{#1\hss}} 
\def\ltsim{\mathrel{\spose{\lower.5ex\hbox{$\mathchar"218$}}
     \raise.4ex\hbox{$\mathchar"13C$}}}
\def\gtsim{\mathrel{\spose{\lower.5ex \hbox{$\mathchar"218$}}
     \raise.4ex\hbox{$\mathchar"13E$}}}
\def\gtlt{\mathrel{\spose{\lower.5ex\hbox{$\mathchar"13E$}}
     \raise.5ex\hbox{$\mathchar"13C$}}}
\def\pmb#1{\setbox0=\hbox{$#1$}%
  \kern-0.25em\copy0\kern-\wd0
  \kern.05em\copy0\kern-\wd0
  \kern-0.025em\raise.0433em\box0}
\def\spmb#1{\setbox1=\hbox{${\scriptstyle #1}$}%
  \kern-0.25em\copy1\kern-\wd1
  \kern.05em\copy1\kern-\wd1
  \kern-0.025em\raise.0433em\box1}
\def\bJ{\;\pmb{\mit J}}
\def\bm{\;\pmb{\mit m}}
\def\bv{\;\pmb{\mit v}}
\def\bOmega{\;\pmb{\Omega}}
\def\sbm{\;\spmb{\mit m}}
\def\lcrit{\lambda_{\rm crit}}
\def\RCR{R_{\rm CR}}
\def\Rcore{R_{\rm core}}
\def\BTii{BT08}

\long\def\Ignore#1{\relax}

\begin{document}

\title{\hbox to 18cm{Secular Evolution in Disk Galaxies \hfil {\rm To appear in RMP}}}
\author{J A Sellwood}
\affiliation{Department of Physics \& Astronomy, Rutgers University,
136 Frelinghuysen Road, Piscataway, NJ 08854, USA}
\email{sellwood@physics.rutgers.edu}

\begin{abstract}
Disk galaxies evolve over time through processes that may rearrange
both the radial mass profile and the metallicity distribution within
the disk.  This review of such slow changes is largely, though not
entirely, restricted to internally driven processes that can be
distinguished from evolution driven by galaxy interactions.  It both
describes our current understanding of disk evolution and identifies
areas where more work is needed.  Stellar disks are heated through
spiral scattering, which increases random motion components in the
plane, while molecular clouds redirect some fraction of the random
energy into vertical motion.  The recently discovered process of
radial migration at the corotation resonance of a transient spiral
mode does not alter the underlying structure of the disk, since it
neither heats the disk nor causes it to spread, but it does have a
profound effect on the expected distribution of metallicities among
the disk stars.  Bars in disks are believed to be major drivers of
secular evolution through interactions with the outer disk and with
the halo.  Once the material that makes up galaxy disks is converted
into stars, their overall angular momentum distribution cannot change
by much, but that of the gas is generally far more liable to
rearrangement, allowing rings and pseudobulges to form.  While
simulations are powerful tools from which we have learned a great
deal, those of disks may suffer from collisional relaxation that
requires some results to be interpreted with caution.
\end{abstract}

\maketitle

\section{Introduction}
Galaxies are distributed throughout the Universe in a clustering
hierarchy.  A large majority of bright galaxies are disk-shaped,
with a significant minority being ellipsoidal.  The question of how
these objects came into existence is the subject of intense current
research.  However, it has become increasingly clear that the
present-day properties of galaxies were not exclusively laid down at
the time of their formation, and that internally-driven processes have
contributed significantly to their present properties.  This review
describes, from a theoretical perspective, the dynamical behavior that
is believed to be important in restructuring galaxy disks from their
initially endowed properties.  \citet{KK04}, updated in
\citet{Korm12}, give a comprehensive review of the same topic from an
observer's perspective.  The present-day properties of galaxy disks
were recently reviewed by \citet{vdKF11}.

It has to be said at the outset that galaxies do not first form, and
then evolve, in temporally distinct phases.  In fact, even today
formation is incomplete for many galaxies, such as the Milky Way.
However, the balance has clearly shifted from the rapid collapse and
merging picture that characterized the roughly first one third of the
life of the Universe to more quiescent evolution over the remaining
two thirds.  The vibrant topic of galaxy formation is too large to be
included in any detail in this review, yet it cannot be omitted
entirely as it provides the context for galaxy evolution.

After the initial collapse, and every subsequent major merger event,
gas begins to settle into a disk whose orientation is determined by
the angular momentum that it acquired from the tidal fields of other
nearby mass concentrations.  The thinness of galaxy disks requires
there to have been a protracted period of quiescent evolution, during
which a number of internally-driven processes gradually rearrange the
structure of galaxies.  These include disk growth through slow
accretion of gas, the formation and evolution of bars, recurring
spiral instabilities, the response of the stellar system to the radial
rearrangement of matter, especially the gas, \etc \ These, together
with the influence of the environment, drive what has become known as
``secular evolution,'' by which is meant the gradual restructuring of
a galaxy over time scales much longer than a crossing time.  Evolution
is mostly driven by the outward redistribution of angular momentum in
a galaxy, which enables it to reach a state of lower energy, and such
changes are prolonged by gas accretion.

Note that the word ``secular'' has a narrow meaning in classical
studies of rotating fluid spheroids by Maclaurin and Jacobi
\citep[summarized by][]{Chan87}.  They revealed that viscosity, a
dissipative process, can destabilize some rotating Maclaurin
spheroids, which become {\it secularly\/} unstable and evolve to
Jacobi ellipsoids.  However, the same Maclaurin spheroid could be {\it
  dynamically\/} stable in the absence of viscosity.  In this review,
I adopt the deliberately broader and vaguer meaning of secular
explained in the previous paragraph.

Of the many processes discussed in this review, I here highlight two
of particular significance.  Spiral patterns are probably the most
important agent of secular evolution.  They have long been known to
redistribute angular momentum and to cause the random motions of stars
to increase over time, but we now know that they cause extensive
radial mixing of both the stars and the gas, and they smooth
small-scale irregularities in the mass distribution.  Bars also cause
substantial secular changes.  Once formed, stellar bars in isolated,
gas-free disks simply rotate steadily with no tendency to evolve
\citep[\eg][]{MS79}, but interaction with gas and other mass
components of the galaxy can gradually alter the properties of the
bar, with evolutionary consequences for its host galaxy.  It is
noteworthy that the rate of secular evolution by both spirals and bars
is substantially accelerated by a moderate fraction of gas.

In order to keep to a manageable length, I focus here on secular
evolution in galaxy disks largely driven through internal processes.
In appropriate places I mention environmental factors, such as the
infall of debris, tidal interactions, \etc, which can also alter the
structure of a galaxy.  But describing the full extent to which
environmental factors may drive evolution would stray into the domain
of galaxy assembly and would add too much to the length of this
review.

\section{Background}
The purpose of this long section is to introduce the concepts and
mechanisms that are invoked in the main part of the review, which
begins in \S III.  Note that I refer back to the various parts of this
section where appropriate, so that those who start with the later
sections can quickly locate a summary of the background that is
assumed.

\subsection{Galaxy formation}
\label{sec.galform}
The current paradigm for galaxy formation is the $\Lambda$CDM (Lambda
Cold Dark Matter) model \citep{Spri06}, which holds that galaxies form
in dark matter halos whose distribution and properties were seeded by
a Gaussian random field of tiny density fluctuations in the early
Universe \citep{Bard86}.  Because the mean matter density was so close
to the closure density in the early Universe, even very mild initial
overdensities grew through self-gravity, and subclumps on all but the
very largest scales were gravitationally bound together.  In this
epoch, few overdense regions were isolated from their neighbors, and
the growth of structure was characterized by the development of a
``cosmic web'' of dense sheets, filaments and voids, that has been
simulated with ever increasing precision \citep[see][for a recent
  review]{FW12}.  During this phase, initially distinct collapse
centers underwent a considerable degree of merging, giving each major
overdensity a treelike origin as different leaves, branches and more
major boughs join to the main density peak.  Later, a little after
redshift $z \sim 1$, some mysterious ``dark energy'', which has many
of the properties of Einstein's cosmological constant, has caused the
originally slowing universal expansion to reaccelerate \citep{Ries98};
\citep{Perl99}.  Reacceleration increases the isolation of different
pieces of the clustering hierarchy, reducing the later merging rate of
halos and allowing galaxies to settle in a more dynamically quiescent
period.

As the first collapses began, the mutual tidal stresses between the
extended overdense regions impart some angular momentum about each
collapse center.  A dimensionless measure of the total angular
momentum $L$ is $\lambda_L = L|E|^{1/2}/GM^{5/2}$, where $E$ and $M$
are the system's total energy and mass.  Halos formed in hierarchical
simulations are found to have a log-normal distribution of this
parameter with a most probable value of $\lambda_{L,0} = 0.037$
\citep{Bull01}.  Stochastic hierarchical growth leads to a net angular
momentum of a halo that varies in magnitude and direction both with
distance from the point of highest density and over time.

The dynamical evolution just described is driven by the dark matter,
popularly supposed to be some relic, weakly interacting particle that
became nonrelativistic in the early Universe and is therefore
described as ``cold.''  It is believed \citep{Hins12} to have a cosmic
density more than 6 times that of the ``baryonic'' matter, comprised
of the familiar protons, neutrons and electrons.  The small mass
fraction of hydrogen and helium, which combined from the primordial
plasma to become neutral gas at $z \sim 1100$, is known to have
subsequently reionized sometime around $z \sim 10$ \citep{Lars11} as
the first objects dense enough to support thermonuclear reactions
began to radiate.

Gas collected in overdense regions, either by cooling of shock heated
material or through flows of cold gas accreted along filaments of the
cosmic web, and spun up as it settled into rotationally supported
disks at the centers of the dark matter halos.  The on-going formation
of stars in these gaseous disks gave rise to the galaxies we observe
today.  Numerical simulations of the galaxy formation process lack the
dynamic range \citep{Spri06} to resolve the complicated gas physics of
fragmentation, star formation, and the subsequent release of energy
through supernovae (see \S\ref{sec.ISM}).  Thus the rate, efficiency,
and precise location of star formation, described as ``subgrid
physics,'' is included in the simulations through rules that are both
motivated by observational evidence and tuned to achieve the desired
outcome.  The difficulty that simulations have in making galaxylike
objects with the properties we observe today is widely attributed to
inadequacies in the implementation of star formation and feedback.

As halos continue to merge, the disks of stars that had begun to form
in them are transformed into amorphous ellipsoidal components in the
violently changing gravitational potential \citep{BH92}.  Where some
combination of shocks, supernovae, and active galactic nuclei has
heated most remaining gas to very high temperatures, the merged
remnants are believed to become the ``red sequence'' galaxies that
have little gas that can cool and reconstitute an active star-forming
disk.\footnote{The terms ``red sequence'' and ``blue cloud'' refer to
  distinct groupings in the color-magnitude diagram for galaxies
  \citep[\eg][]{Bald04} and reflect, more objectively, the distinction
  between early- and late-type galaxies already known to Hubble
  \citep[\eg][]{Sand61}.}  Where gas can cool and resettle, the
ellipsoidal ball of stars becomes a spheroidal bulge at the center of
a newly assembling disk that generally continues to grow.  Disk
galaxies that are actively forming stars are the ``blue cloud''
galaxies.

This current picture is widely accepted as broadly correct because it
accounts for the distribution of galaxies throughout the Universe and
some of their properties \citep{Spri06}.  Yet there are quite a number
of important predictions of the model that seem to be inconsistent
with known galaxy properties.  Perhaps the foremost is that (1) many
present-day galaxies lack the types of bulges produced by mergers
\citep{Shen10}; \citep{Korm10}, whereas the merging hierarchy of the
model predicts substantial, dense bulges in most large galaxies.
Other problems are (2) the central density of dark matter in the halos
of galaxies today seem less than the models predict \citep{Sell09};
\citep{KS11}, (3) the number and properties of dwarf satellite
galaxies surrounding each major galaxy seems inconsistent with what we
observe \citep{Boyl12}, (4) the old stars in galactic disks reside in
a thinner layer that cannot have been stirred by a minor merger for a
very long period \citep{Wyse09}, \etc \ See \citet{SM12} for a more
detailed critique.

\subsection{Relaxation time in the disks of galaxies}
\label{sec.relax}
Except in the immediate neighborhood of a star, the gravitational
attraction of nearby stars is generally negligible compared with that
from the large-scale distribution of matter in a galaxy.  While the
argument that establishes this point can be found in many text books
\citep[\eg][hereafter \BTii]{BT08}, the usual derivation requires some
modification for disk systems that is generally omitted even though it
involves several nontrivial issues that are important both to the
scattering of stars by mass clumps in the disk and to the proper
interpretation of simulations, as noted elsewhere in this review.

\subsubsection{Standard theory for spherical systems}
A test particle moving at velocity $\bv$ along a trajectory that
passes a stationary field star of mass $\mu$ with impact parameter $b$
is deflected by the attraction of the field star.  For a distant
passage, it acquires a transverse velocity component $|\bv_\perp|
\simeq 2G\mu/(bv)$ to first order (\BTii\ eq.\ 1.30).  Encounters at
impact parameters small enough to produce deflections where this
approximation fails badly are negligibly rare and relaxation is driven
by the cumulative effect of many small deflections.

If the density of field stars is $n$ per unit volume and uniform in
3D, the test particle encounters $\delta n = 2\pi b\delta b\,nv$
stars per unit time with impact parameters between $b$ and $b+\delta
b$.  Assuming stars to have equal masses, each encounter at this
impact parameter produces a randomly directed $\bv_\perp$ that will
cause a mean square net deflection per unit time of
\begin{equation}
\langle \delta v_\perp^2 \rangle \simeq \left( {2G\mu \over bv
}\right)^2 \times 2\pi \, b\delta b\,nv = {8\pi G^2\mu^2n \over
  v}{\delta b \over b}.
\label{eq.elerate}
\end{equation}
The total rate of deflection from all encounters is the integral over
impact parameters, yielding
\begin{equation}
\langle v_\perp^2 \rangle = {8\pi G^2\mu^2n \over v}\int_{b_{\rm min}}^{b_{\rm max}}{db \over b}
= {8\pi G^2\mu^2n \over v} \ln\Lambda,
\label{eq.trate}
\end{equation}
where $\ln\Lambda \equiv \ln(b_{\rm max}/b_{\rm min})$ is the Coulomb
logarithm.  Typically one chooses the lower limit to be the impact
parameter of a close encounter, $b_{\rm min} \simeq 2G\mu/v^2$, for
which $|\bv_\perp|$ is overestimated by the linear formula, while the
upper limit is, say, the half mass or {\bf effective radius}, $R_e$,
of the stellar distribution beyond which the density decreases
rapidly.  The vagueness of these definitions is not of great
significance to an estimate of the overall rate because we need only
the logarithm of their ratio.  The Coulomb logarithm implies equal
contributions to the integrated deflection rate from every decade in
impact parameter simply because the diminishing gravitational
influence of more distant stars is exactly balanced by their
increasing numbers.

Note that the first order deflections that give rise to this steadily
increasing random energy come at the expense of second order
reductions in the forward motion of the same particles that we have
neglected \citep{Heno73}.  Thus the system does indeed conserve
energy, as it must.

We define the {\bf relaxation time} to be the time needed for $\langle
v_\perp^2 \rangle \simeq v^2$, where $v$ is the typical velocity of a
star.  Thus
\begin{equation}
\tau_{\rm relax} = {v^3 \over 8\pi G^2\mu^2n \ln\Lambda}.
\label{eq.relax}
\end{equation}
To order of magnitude, a typical velocity $v^2 \approx GN\mu/R_e$, where
$N$ is the number of stars each of mean mass $\mu$, yielding $\Lambda
\approx N$.  Defining the dynamical time to be $\tau_{\rm dyn} = R_e/v$
and setting $n \sim 3N/(4\pi R_e^3)$, we have
\begin{equation}
\tau_{\rm relax} \approx {N \over 6 \ln N}\tau_{\rm dyn},
\label{eq.trelax}
\end{equation}
which shows that the collisionless approximation is well satisfied in
galaxies, which have $10^8 \ltsim N \ltsim 10^{11}$ stars.  Including the
effect of a smooth dark matter component in this estimate would
increase the typical velocity, $v$, thereby further lengthening the
relaxation time.

\subsubsection{Applications to disk systems}
This standard argument, however, assumed a pressure-supported
quasispherical system in several places.  \citet{Rybi72} pointed out
that the flattened geometry and organized streaming motion within
disks affects the relaxation time in a number of important ways. 

First, the assumption that the typical encounter velocity is
comparable to the orbital speed $v=(GN\mu/R_e)^{1/2}$ is clearly
wrong; stars move past each other at the typical random speeds in the
disk, say $\beta v$ with $\beta \sim 0.1$, causing larger deflections
and decreasing the time for $\langle v_\perp^2 \rangle \simeq
\beta^2v^2$ by a factor $\beta^3$.

Second, the distribution of scatterers is not uniform in 3D, as was
implicitly assumed in eq.~(\ref{eq.elerate}).  Assuming a razor-thin
disk changes the volume element from $2\pi v\,b\delta b$ for 3D to
$2v\,\delta b$ in 2D, which changes the integrand in Eq.~(\ref{eq.trate})
to $b^{-2}$ and replaces the Coulomb logarithm by the factor $(b_{\rm
  min}^{-1} - b_{\rm max}^{-1})$.  In 2D therefore, relaxation is
dominated by close encounters when the forces are Newtonian, and the
system could {\it never\/} be collisionless.

Real galaxy disks are neither razor thin, nor spherical, so the
spherical dependence applies at ranges up to the typical disk
thickness $z_0$, beyond which the density of stars drops too quickly
to make a significant further contribution to the relaxation rate.
Thus we should use $\Lambda \simeq z_0/b_{\rm min}$ for disks.

Third, the local mass density is also higher, so that $N \sim \pi
R_e^2z_0n$.  These three considerations shorten the relaxation time
(Eq.~\ref{eq.trelax}) by the factor
\begin{equation}
\beta^3\left({z_0 \over R_e} \right){\ln\left(R_e / b_{\rm min}\right)
  \over \ln\left(z_0 / b_{\rm min}\right)},
\label{eq.2D}
\end{equation}
or $\sim 10^{-4}$ for $\beta \simeq 0.1$ and reasonable $z_0/R_e$.
Note that star-star encounters in galaxy disks remain unimportant, even
with this large reduction in the relaxation time scale.  But
significant relaxation can occur in simulations of stellar disks, and
the issues originally raised by Rybicki are of importance for
scattering by clouds, as developed below.

A fourth consideration for disks is that the mass distribution is much
less smooth than is the case in the bulk of pressure-supported
galaxies.  A galaxy disk generally contains massive star clusters and
giant molecular clouds whose influence on the relaxation rate turns
out to be non-negligible and also determines the shape of the
equilibrium velocity ellipsoid (see \S\ref{sec.clouds}).

\begin{figure}[t]
\includegraphics[width=.7\hsize,angle=180]{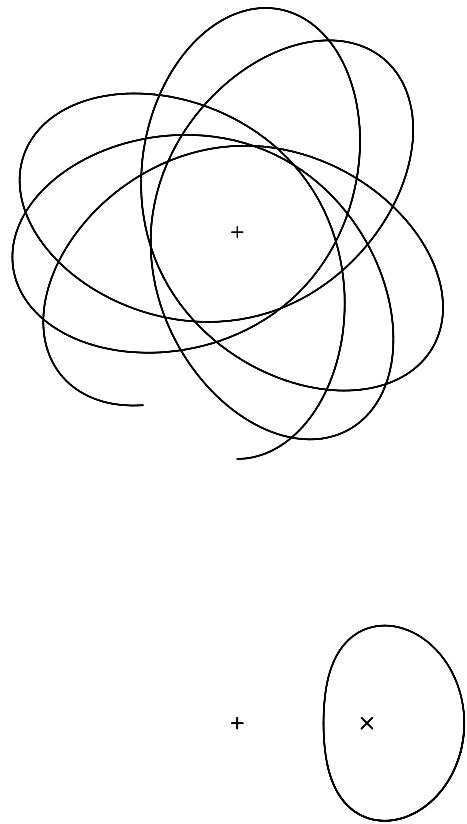}
\caption{An eccentric orbit in the midplane of an axisymmetric
  potential.  The center of the potential is marked with a plus.  The
  orbit is drawn in the inertial frame above and below in a frame that
  rotates with the guiding center, marked with a cross.  Since the
  star conserves $L_z$, the motion around the epicycle is in the
  retrograde sense.}
\label{fig.epi}
\end{figure}

\subsection{Stellar orbits in disks}
\label{sec.orbs}
While the above considerations should be borne in mind, it is
nevertheless useful to regard the gravitational potential of a galaxy
as a smooth function of position to a first approximation.  If this
assumption holds, the stellar component of a galaxy behaves as a
collisionless fluid (\BTii).  I extend the discussion to include
relaxation processes present in galaxy disks in \S\ref{sec.scatt}.

A star, or test particle, moving in the symmetry plane ($z=0$) of a
steady axisymmetric potential $\Phi(R, z)$ must conserve its specific
energy $E$ and specific angular momentum $L_z$ about the symmetry axis
($R=0$); these are the only two nontrivial integrals of motion when
$L_x=L_y=0$.  In general, the orbit of a star is a rosette, as shown
in Fig.~\ref{fig.epi}, but when viewed from a frame rotating about the
center of the galaxy at an angular frequency $\Omega_\phi$ (lower
panel), we see that the motion can be decomposed into a radial
oscillation about a {\bf guiding center}, which is marked with a
cross, plus the uniform angular motion of the guiding center about the
center with a period $\tau_\phi = 2\pi /\Omega_\phi$.  The guiding
center radius $R_g$ is where the radial acceleration of the particle
passes through zero, \ie\ where the central attraction matches the
centripetal acceleration:
\begin{equation}
\left({\partial\Phi \over \partial R}\right)_{(R_g, 0)} = {L_z^2 \over R_g^3}.
\label{eq.rguide}
\end{equation}
For eccentric orbits, such as that shown, $\Omega_\phi \ltsim \Omega_c
\equiv L_z/R_g^2$, the angular frequency of circular motion for the
same $L_z$.  The radial oscillation is anharmonic and we simply define
the radial frequency $\Omega_R \equiv 2\pi/\tau_R$, where $\tau_R$ is
the period of a full radial oscillation.  In all realistic galactic
potentials $\Omega_\phi < \Omega_R < 2\Omega_\phi$, and stars
therefore make more than one, but less than two radial oscillations
per orbit period.  For near-circular orbits, $\Omega_\phi \rightarrow
\Omega_c$, and the radial oscillation becomes harmonic about the
guiding center, with $\Omega_R \rightarrow \kappa$, the Lindblad
epicyclic frequency defined through
\begin{equation}
\label{eq.kappa}
\kappa^2(R_g) = \left({\partial^2\Phi\over\partial R^2}\right)_{(R_g, 0)} + {3
\over R_g} \left({\partial\Phi \over \partial R}\right)_{(R_g, 0)}.
\end{equation}

Motion in the third dimension can also be oscillatory, with a
well-defined period $\tau_z$.  When both the radial and vertical
excursions are small, the vertical oscillation is decoupled from the
in-plane motion, and has an angular frequency $\nu$ given by
\begin{equation}
\nu^2(R_g) = \left({\partial^2\Phi \over \partial z^2}\right)_{(R_g, 0)}.
\label{eq.nudef}
\end{equation}
Naturally, $\Omega_z \equiv 2\pi/\tau_z \rightarrow \nu$ in this limit.

In a general static, axisymmetric potential, the motion of most stars
can be decomposed into three separate oscillations at the three
different frequencies, $\Omega_R$, $\Omega_\phi$, and $\Omega_z$.
Such orbits are described as regular orbits.  However, there is a
generally small fraction of irregular orbits, whose motion is more
complicated and cannot be decomposed into three oscillations and
another fraction that are truly chaotic.

In addition to the two classical integrals $E$ and $L_z$, regular
orbits respect a third, nonclassical integral.  It is described as a
nonclassical integral because it cannot be expressed as a simple
function of the phase-space variables.

\subsection{Action-angle variables}
\label{sec.actions}
The formal clutter that usually accompanies any introduction to
action-angle variables makes it hard to grasp what they really are and
why they are useful.  In an attempt to clarify these points, I here
give a brief qualitative discussion, and refer the interested reader
to \BTii\ for a more mathematical, but still not fully rigorous,
treatment.

The 2D axisymmetric case, for which there are just two actions, is
easiest to visualize and was illustrated already in
Fig.~\ref{fig.epi}.  The azimuthal action $J_\phi$ is simply the
angular momentum, which is a measure of the size of the orbit or
equivalently the radius of the guiding center (Eq.~\ref{eq.rguide}).
The radial action $J_R$ is a second conserved quantity that is a
measure of the radial extent of the oscillation; thus $J_R=0$ for a
circular orbit and nonzero values can be calculated using
Eq.~(\ref{eq.JRdef}).  The orbit is uniquely determined by the values
of the actions $(J_R,J_\phi)$, which are an alternative pair of
integrals to $(E,L_z)$.

The doubly periodic motion is described by two angles $w_R$ and
$w_\phi$, which specify respectively the phase of the orbit around its
epicycle and the phase of the guiding center around the galaxy center.
One major advantage of this apparatus is that each angle variable
increases at the constant rate $w_i(t) = w_i(0) + t\Omega_i(E,L_z)$,
even though the $(R,\phi)$ coordinates of a star vary nonuniformly
with time.\footnote{\citet{LB62} pointed out that while $w_i(0)$ is a
  constant of the motion, it is a nonisolating integral, and
  therefore is of no importance to the overall structure of phase
  space.}

This approach becomes far more valuable when used to describe the 3D
motion of a regular orbit, which respects three integrals.  Even
though one or more integrals cannot be expressed as simple functions
of the phase-space variables in either Cartesian or polar coordinate
systems, we {\it can\/} fully describe regular 3D motion in an
arbitrary smooth potential using three actions that are conserved
quantities, \ie\ they are a set of integrals. The triply-periodic
motion is described by three angles that again increase at uniform
rates.  The actions in an axisymmetric potential are $J_R$ and
$J_\phi$, the radial and azimuthal actions as for 2D, and $J_z$, which
quantifies the up-and-down motion about the midplane.  For each
regular orbit, the three oscillation frequencies are $\Omega_i =
dw_i/dt = 2\pi/\tau_i$ (\S\ref{sec.orbs}), with $i$ denoting
any of the three cylindrical polar coordinates.

Not only do we now have a set of integrals and can describe the motion
as the three angles increase in time at uniform rates, but these
variables have two further advantages.  The actions are the adiabatic
invariants when the system is subject to slow changes, and the orbit
is not close to a resonance.  Finally, a more mathematical advantage
is that perturbation theory is greatly simplified when using these
variables (\eg\ \S\ref{sec.Lchange}).

\subsection{Orbital tori}
\label{sec.tori}
Were motion confined to a plane, as for the 2D orbit shown in
Fig.~\ref{fig.epi}, the star would move in the 4D phase space
$(R,\phi,v_r,v_\phi)$.  However, because both $E$ and $L_z$ are
conserved, the star's motion is confined to a 2D surface within the 4D
phase space, since both velocity components $v_\phi = L_z/R$ and $v_R
= \{2[E-\Phi(R)] - (L_z/R)^2\}^{1/2}$ are determined for every value
of $R$, except for the sign ambiguity of $v_R$.

To see that the motion is confined to the surface a torus, imagine
that we add to the rosette orbit shown in the upper panel, a third
coordinate that is the star's radial velocity $v_R$, which is positive
above the page and negative below.  As the star moves forward in time
from its pericentric distance, say, $v_R$ first rises to a maximum
height above the page as it crosses $R_g$, and then decreases to zero
as it reaches its apocentric distance.  Then $v_R$ changes sign and
the inward motion is below the page.  As the star moves out and in, it
also advances around the galactic center.  Thus the motion in the 3D
space of $(R,\phi,v_R)$ is confined to the surface of a torus in that
space.  It is impossible to visualize a fourth dimension, but while the
speed $v_\phi$ around the torus varies with $R$, it does so within a
restricted range that does not alter the topology.

In 3D, stars move in a 6D phase space, and every conserved quantity,
or isolating integral, confines its motion to a hyper-surface of one
lower dimension.  The regular orbit of a star that possesses three
integrals is confined to the surface of a 3D hyper-torus in the 6D
phase space, and again the motion within each dimension of the
hyper-torus is an independent oscillation.

Fewer quantities are conserved for chaotic orbits, whose motion cannot
be decomposed into three independent oscillations.  For example, a
star that moves chaotically is not confined to a 3D surface, but
explores a 5D space if only $E$ is conserved, which is typical in a
nonaxisymmetric potential.

All three actions are quantities having the dimensions of angular
momentum, and each is defined as the cross-sectional area of the
appropriate slice through the star's orbit torus in 6D phase space,
\ie
\begin{equation}
J_i \equiv {1 \over 2\pi} \oint \dot x_idx_i,
\label{eq.actdef}
\end{equation}
where $i$ labels a particular coordinate and the integral is around
one complete loop in this slice through the torus.  In an axisymmetric
potential, $J_\phi \equiv (2\pi)^{-1} \int_0^{2\pi} v_\phi Rd\phi$
and, since the integrand $Rv_\phi = L_z$ is a constant, we have
$J_\phi \equiv L_z$, but Eq.~(\ref{eq.actdef}) generally must be
evaluated numerically for the other actions.

Since stars oscillate at differing frequencies in each of the three
coordinate directions, one way to estimate the $i$-th action is to
integrate the orbit and measure the area inside the closed curve
delineated by the locus of points, or consequents, as the star
crosses the 2D surface $(x_i,\dot x_i)$, known as the {\bf surface of
section}.  \citet{MB08} described a superior method of ``torus
fitting'' that yields all three actions simultaneously in an arbitrary
potential.

For those who find torus fitting intimidating, useful approximations
can be obtained more easily for disk star orbits.  We first assume
axial symmetry, so that $J_\phi = L_z$, and write the energy of a star
as
\begin{equation}
E = \textstyle{1\over2}\left(\dot R^2 + \dot z^2 \right) + \Phi_{\rm eff}(R,z),
\label{eq.effect}
\end{equation}
with the effective potential being $\Phi_{\rm eff} \equiv \Phi(R,z) +
L_z^2/(2R^2)$.  The approximation is to assume that the radial and
vertical oscillations are decoupled, and that the radial action can
be estimated from motion in the midplane as
\begin{equation}
J_R \simeq  \left. {1 \over \pi} \int_{R_p}^{R_a}\left\{2\left[E - \Phi_{\rm eff}(R,0)\right]\right\}^{1/2} \, dR \; \right|_{z=0}.
\label{eq.JRdef}
\end{equation}
The integration limits are the pericentric and apocentric values of
$R$ in the midplane, where the argument of the square root is zero.
This integral is for half the radial period, and we drop the factor 2
in the denominator because the return half makes an equal
contribution.  Similarly,
\begin{equation}
J_z \simeq \left. {1 \over \pi} \int_{z_{\rm min}}^{z_{\rm max}} \left\{2\left[E - \Phi_{\rm eff}(R_g,z)\right]\right\}^{1/2}\,dz \; \right|_{R = R_g}.
\label{eq.Jzdef}
\end{equation}
These equations are exact for stars lacking vertical or radial
oscillation, respectively, but in general they are slight overestimates
since they assume that a star moving in 3D explores the full extent of
the region that is energetically accessible, which is not quite the
case when the orbit is integrated.

The epicyclic approximation for small-amplitude excursions assumes
that both the radial and vertical oscillations are harmonic.  If this
is valid, $J_{R,\rm epi} = \kappa a^2/2$, with $a$ being the radial
excursion of the star \citep{LBK72} and $J_{z,\rm epi} = \nu z_{\rm
  max}^2/2$.  Since most stars in a disk have vertical excursions that
take them outside the harmonic region of the disk potential well,
$J_{z,\rm epi}$ should be avoided in favor of Eq.~(\ref{eq.Jzdef}),
which is readily evaluated.

\subsection{Distribution function}
\label{sec.DF}
The density of stars in the 6D phase space of position and velocity is
given by a {\bf distribution function}, $f$, hereafter DF.  The DF for
an equilibrium stellar system must be a function of the integrals only
(Jeans theorem).  The set of actions is a possible set of integrals,
and the density of regular orbits could be written as
$f(J_1,J_2,J_3)$, but formally only if every possible orbit respects
three integrals and there are no irregular or chaotic orbits.

For axisymmetric systems, if there were no third integral, the DF
would be a function of the two classical integrals, $E$ and $L_z$
only.  If this were true, the ratio of the second moments of the
radial and vertical velocity components,
\begin{equation}
{\langle v_R^2 \rangle \over \langle v_z^2 \rangle} =
{\int f(E,L_z) \dot R^2 dv^3 \over \int f(E,L_z) \dot z^2 dv^3} = 1,
\end{equation}
as both $\dot R$ and $\dot z$ enter equally in $E$
(Eq.~\ref{eq.effect}).  Since we observe that $\sigma_R \neq
\sigma_z$, we conclude that large parts of phase space of disks are
regular and the effective DF must depend upon three integrals.

A few analytic expressions for DFs are known for 2D disks, but because
the possible third integral is not a simple function of the
phase-space coordinates, the development of analytic three-integral
DFs for realistic flattened disks \citep[\eg][]{Binn10} is much more
difficult.

\subsection{Nonaxisymmetric disturbances}
\label{sec.nonax}
Consider the potential of a small-amplitude disturbance in the $z=0$
midplane that is the real part of
\begin{equation}
\Phi_1(R, \phi, t) = \Psi(R) e^{i(m\phi - \omega t)}.
\label{eq.pbn}
\end{equation}
This potential has the following properties: it varies sinusoidally
with the azimuthal coordinate $\phi$, it is $m$-fold rotationally
symmetric (\eg, $m=2$ for a bi-symmetric spiral or bar), it rotates
about the $z$-axis at the angular rate $\Omega_p = \Re(\omega)/m$
which is described as the {\bf pattern speed}, and grows exponentially
at the rate $\gamma = \Im(\omega)$.  The complex function $\Psi(R)$
describes the radial variation of the amplitude and phase of the
perturbation.

The density distribution that gives rise to this disturbance potential
is not easily computed.  Generally, Poisson's equation requires the
potential spiral to be less tightly wound than the density spiral, and
the phase relation between the density and potential therefore varies
systematically with radius.  The density and potential are in phase
when the tight-winding (or WKB) approximation is employed, but spirals
in galaxies are sufficiently open that this approximation gives only a
rough guide to the dynamics of real spirals.

\subsection{Resonances}
\label{sec.resnc}
Stars orbiting in the midplane of an almost axisymmetric galaxy are
in resonance with a weak nonaxisymmetric disturbance of the form
(\ref{eq.pbn}) when
\begin{equation}
\Omega_p = \Omega_\phi + {l \over m}\Omega_R.
\label{eq.res}
\end{equation}
The unperturbed orbit frequencies of the stars are defined in
\S\ref{sec.orbs} and $l$ is a signed integer.  Equation~(\ref{eq.res})
is satisfied for $l=0$ when the guiding center of a star rotates
synchronously with the disturbance, which is described as the {\bf
  corotation resonance}.  When $l = \pm 1$, Eq.~(\ref{eq.res}) defines
the locations of the {\bf Lindblad resonance}s, which arise because
the Doppler shifted frequency at which the star encounters the wave
$m|\Omega_\phi - \Omega_p|$ is equal to its unforced frequency of
radial oscillation $\Omega_R$, or $\kappa$ for nearly circular orbits.
Interior to corotation, the stars overtake the wave, and $l=-1$ at the
inner Lindblad resonance (ILR).  Outside corotation, the stars are
overtaken by the wave, and the outer Lindblad resonance (OLR) occurs
where $l=+1$.  Resonances for larger values of $|l|$, if they occur at
all, are generally of little dynamical interest, since spiral patterns
can exist only between the Lindblad resonances; a steady perturbing
potential does not elicit a supporting response from the stars outside
this radial range.

{\bf Ultraharmonic resonances} arise where Eq.~(\ref{eq.res}) is
satisfied for $l=\pm1$ and $m$ replaced by $2m$.  At these resonances,
which are closer to corotation than are the Lindblad resonances, the
star completes two radial oscillations as it moves between
wave-crests.  Yet higher-order resonances exist for larger integral
numbers of radial oscillations; they are located still closer to
corotation as stars drift ever more slowly relative to the pattern.
Ultraharmonic resonances are dynamically unimportant in linear
perturbation theory, but their nonlinear generalizations do play a
role in finite-amplitude perturbations, especially bars
(\S\ref{sec.bars}).

{\bf Vertical resonances} will occur where the Doppler shifted
frequency
\begin{equation}
m(\Omega_p - \Omega_\phi) = n\Omega_z,
\end{equation}
with $n$ being a signed integer; the $n=0$ case (corotation) is of no
special significance for vertical motion.  In the epicycle
approximation, $\Omega_z \rightarrow \nu$, which is a higher frequency
in the massive part of the disk than is $\kappa$.  Therefore, the
$n=\pm1$ vertical resonances are farther from corotation than
are the Lindblad resonances.  In linear perturbation theory, spiral
perturbations do not extend beyond the Lindblad resonances, making
these vertical resonances uninteresting because the perturbation
potential is tiny there.  However, it should be noted that the
effective vertical frequency $\Omega_z \equiv 2\pi / \tau_z$ can be
much smaller than $\nu$ for stars with vertical excursions extending
well outside the region where the potential is approximately harmonic,
and such stars could, in principle, experience a vertical resonance.

Linear perturbation theory holds even at resonances for
small-amplitude disturbances that grow exponentially, for then the
resonances have a Lorentzian width determined by the growth rate.
However, it breaks down for stars in resonance with a steady, or
slowly growing, perturbation.  Stars can be trapped by the resonance,
and the size of the trapped region in phase space increases with the
amplitude of the perturbing potential.

\subsection{Local stability}
\label{sec.local}
The problem of computing the gravitational potential of an arbitrary
spiral disturbance is one reason that the global normal modes of a
stellar disk cannot be computed in a straightforward manner
\citep{Kaln77}; \citep[][\BTii]{Jala07}.  While a WKB (local)
approach, in which the local spatial variation of the disturbance can
be approximated as part of a plane wave, is generally a poor
approximation, results obtained using it do give some indication of
the global behavior.

\citet{Toom64} used this approximation to show that {\it axisymmetric\/}
oscillations in a razor-thin disk of surface density $\Sigma$ are
stabilized by rotation on scales
\begin{equation}
\label{eq.lcrit}
\lambda > \lcrit = {4 \pi^2 G \Sigma \over \kappa^2}, 
\end{equation}
where $\kappa$ is the epicyclic frequency defined in
Eq.~(\ref{eq.kappa}).  In the complete absence of random motion, a
disk is unstable to gravitational clumping into rings on all scales
smaller than $\lcrit$.  Physically, $\lcrit$ decreases with increasing
$\kappa$ because stars are held more tightly to their guiding center
radii.  The value of $\lcrit \approx 6\;$kpc in the solar
neighborhood, or three fourths of the Sun's distance from the Galactic
center, indicating that the WKB approximation is indeed stretched.

Random motions of the stars prevent gravitational instabilities when
the disturbance disperses more rapidly than it grows.  The tendency of
random motions to provide stability on small scales while rotation
suppresses instability on large scales, led Toomre to the following
celebrated {\bf local stability criterion}: if the stellar radial
velocities have a Gaussian distribution with spread $\sigma_R$, then
the system is axisymmetrically stable on all scales if
\begin{equation}
\label{eq.Qdef}
Q \equiv {\sigma_R \over \sigma_{R, \rm crit}} > 1, \quad{\rm where}\quad
\sigma_{R, \rm crit} = {3.36 G\Sigma \over \kappa}.
\end{equation}
Adopting solar neighborhood values, we find $\sigma_{R, \rm crit}
\approx 20\;$km~s$^{-1}$.  In a razor-thin, isothermal gas disk
$\sigma_R$ is replaced by the sound speed $c_s$ and the constant
$3.36$ is replaced by $\pi$.  The constant is also slightly reduced in
finitely thick disks since the destabilizing gravitational forces are
diluted by the vertical spread of matter \citep[\eg][]{Rome92}.  The
local stability of combined gas and stellar disks was calculated by
\citet{Rafi01}, while \citet{RW11} offer an approximate formula for
$Q$ in thickened two-component disks.  Global axisymmetric stability
may be guaranteed if the disk is locally stable everywhere
\citep{Kaln76}.

It cannot be emphasized too strongly that criterion (\ref{eq.Qdef}) is
for local axisymmetric stability only, and that disks that meet this
criterion can still be unstable to nonaxisymmetric modes.  In fact, no
general criterion for nonaxisymmetric stability is known.

\begin{figure}[t]
\includegraphics[width=\hsize]{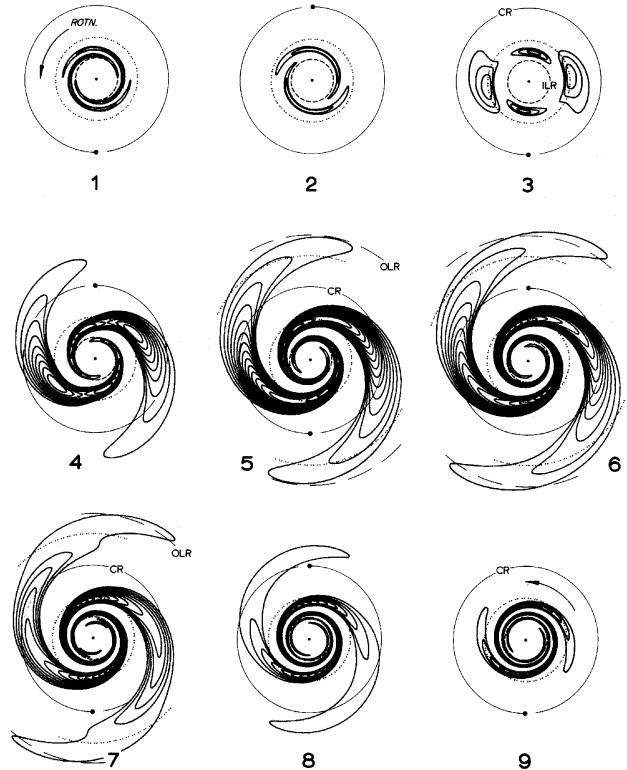}
\caption{The time evolution of an input leading wave packet in the
  half-mass Mestel disk.  The unit of time is half a circular orbit
  period at the radius marked corotation.  From Toomre 1981.}
\label{fig.dust}
\end{figure}

Local nonaxisymmetric stability was investigated by \citet{GLB65} and
by \citet{JT66}, who independently discovered the process of
swing-amplification.  Figure~\ref{fig.dust}, from a global calculation
due to \citet{Toom81}, illustrates the fate of an arbitrary input
leading spiral inserted by hand and given a pattern speed so that it
is localized near the ILR.  In this linearly stable, $Q=1.5$ disk, the
disturbance initially propagates away from the ILR and unwinds due to
the differential rotation until it ``swings'' from leading to
trailing.  The disturbance is amplified by a large factor during this
period when it is least wound because rotational support, which is a
critical part of axisymmetric stability, is compromised briefly.  The
disturbance propagates radially at the group velocity \citep{Toom69},
which is away from corotation for trailing waves, and the inner part
returns toward the ILR.  The part of the disturbance outside
corotation fades quickly as it spreads over a wider area, while the
opposite behavior affects the inner part until it is gradually
absorbed by wave-particle interactions as it approaches the resonance.
Thus the whole episode is a transient response that, to first order,
causes no lasting change to the disk, although there are second order
changes.

The amplification of a wave packet at corotation can be calculated in
a variety of local approximations \citep{Toom81}, while \citet{Drur80}
computed the relationship between a continuous wave train incident on
corotation and the super-reflected and transmitted waves.  In the
notation of \citet{JT66}, the most important parameter is
\begin{equation}
\label{eq.Xdef}
X \equiv {\lambda_y \over \lcrit} 
  = {2\pi \RCR \over m}{\kappa^2 \over 4\pi^2G\Sigma}, 
\end{equation}
where $\lambda_y$ is the wavelength of the disturbance with angular
periodicity $m$ around the corotation circle of radius $\RCR$.  For a
flat rotation curve, amplification is significant for $1 \ltsim X
\ltsim 3$ and is strongest for an unwrapped wavelength that is about
twice $\lcrit$.  If the rotation curve declines, amplification extends
to larger values of $X$ and, conversely, it is confined to smaller $X$
values in rising cases.  Of course, the range of $X$ for strong
amplification shrinks to zero in the absence of shear (uniform
rotation).

The amplification factor also decreases rapidly with increasing $Q$
(Eq.~\ref{eq.Qdef}).  The reflected wave can be 100 times as strong
as the incident wave for $X \simeq 2$ and $Q \simeq 1.2$, but only
a few times greater when $Q \simeq 2$.

Notice that $X \propto (m\Sigma)^{-1}$, implying that for a fixed
radius and rotation curve, amplification will be strong for higher $m$
values when the disk surface density $\Sigma$ is low -- \ie\ we expect
bisymmetric spirals in heavy disks and multi-arm spirals in strongly
submaximal disks \citep{SC84}.  Thus the number of spiral arms in a
galaxy could be an indicator of the relative contribution of the disk
to the total central attraction \citep{ABP87}.  This argument should
not be applied to flocculent galaxies \citep{EE82}, which have many
small arm fragments, where the small spatial scale of the arms
probably indicates that the responsive part of the disk is a low-mass
component that has become dynamically decoupled from a hotter,
underlying stellar disk.

\subsection{Angular momentum changes}
\label{sec.lzchange}
The response of the stars to a weak potential perturbation is most
easily calculated in action-angle variables \citep{LBK72};
\citep{Dekk76}; \citep{CS85}; \citep{BL88}.  \citet{LBK72} showed that
the first order changes in the angular momenta of stars average to
zero everywhere.  However, to second order, the rate of change in
angular momentum of a group of stars is
\begin{equation}
{d L_{\rm tot, 2} \over dt} = 4\pi^2\gamma e^{2\gamma t}
\sum_{\sbm}m \int \bm{\partial f \over \partial \bJ}
{|\Phi_{1,\sbm}^2| \over |\bm \cdot \bOmega - \omega|^2} \; d\bJ, 
\label{eq.LBK}
\end{equation}
plus a boundary term.  Here the ranges of integration are unperturbed
angular momenta $L_{z, 1} \leq J_\phi \leq L_{z, 2}$ and radial action
$0 \leq J_R \leq \infty$.  The $\Phi_{1,\sbm}$ are Fourier
coefficients of the perturbing potential $\Phi_1$ (Eq.~\ref{eq.pbn}),
$\bm = (l, m)$, $\bJ = (J_R, J_\phi)$, \etc\ \ Resonant denominators
arise in Eq.~(\ref{eq.LBK}) where $\bm \cdot \bOmega = \Re(\omega)$
(the same condition as Eq.~\ref{eq.res}), which pick out important
locations in phase space where substantial changes take place.  Note
that the magnitude of each $\bm$-term depends on the gradient of the
DF with rrespect to the actions at that resonance; thus the net change
depends on the imbalance between those stars that lose on one side of
the resonance compared with those that gain on the other side.

Generally, we expect $f$ to be a decreasing function of all the actions
in any reasonable galaxy; \ie\ for a given $L_z$, there are more stars
with small $J_R$ and $J_z$ and $f$ falls off steeply with increasing
values of either of these actions.  Also the density of disk stars
generally rises toward the center, and therefore $f$ rises with
decreasing $J_\phi \equiv L_z$, which is usually the shallowest of the
three gradients.

A self-excited spiral involves no external torque, and this expression
must therefore integrate to zero over the whole disk.  However,
\citet{LBK72} showed that the mean angular momenta of stars inward of
corotation is lowered, and those outward are raised, by the growth of
the disturbance.  This feature allows a mode to extract energy from
the gravitational potential well of the galaxy, enabling it to grow.
Unfortunately, these angular momentum changes cannot be equated to the
gravity torque between the misaligned density and potential because
angular momentum can also be transported by a Reynolds-like advective
stress \citep[dubbed ``lorry transport'' by][]{LBK72}.  The Reynolds
stress is probably of minor importance in the vigorously growing modes
that galaxies seem to support, but would be significant were
quasi-steady spiral modes important.

\subsection{Gas}
\label{sec.ISM}
The stars of a galaxy move on ballistic orbits that are affected only
by gravitational forces.  The fraction of the total baryonic mass
contained in gas is generally less than 10\% in large disk galaxies
today.  Over time, gas is converted into stars, but is replenished
partly by returned material as massive stars end their lives, and also
by on-going infall in spiral galaxies.  The interstellar gas is
collected into clouds, the diffuse ones being composed largely of
neutral atomic hydrogen and helium with a sound speed $c_s \sim
1.3\;$km~s$^{-1}$, while dense molecular gas clouds are colder with
$c_s \ltsim 0.5\;$km~s$^{-1}$.

Typical orbital speeds in galaxies are 100 -- 200 km/s, while typical
velocity spreads of clouds about the mean orbital motion appear to
have a lower bound of some 6 -- 8 km~s$^{-1}$, rising to twice this
value in the bright, star-forming parts of disk galaxies.  This
supersonic turbulence \citep[see][for a review]{SE04} is maintained by
a variety mechanisms, the most important of which is mechanical energy
input through supernovae and, to a lesser extent, stellar winds
(streams of particles accelerated from the surfaces of massive stars).
When many massive stars are born at similar times in an exceptional
burst of star formation, the ensuing rapid succession of supernovae
can create a galactic wind that drives some of the gas out of the disk
plane and perhaps, in the cases of small galaxies with shallow
potential wells or young galaxies with high rates of star formation,
right out of the galaxy.  Large-scale dynamical phenomena, such as
spiral activity, tidal interactions, and gas infall are other sources
of turbulence.

The medium is also stirred by the magneto-rotational instability of
weakly-magnetized differentially rotating fluid disks \citep{BH98};
\citep{SB99}, which maintains a lower level of trans-Alfv\'enic
turbulence in parts of disks that have few young stars, and
correspondingly few supernovae, where the dispersion remains about
6~km~s$^{-1}$ \citep[\eg][]{Dick90}; \citep{Tamb09}.

High spatial resolution simulations of this medium in small volumes
\citep{Ston98}; \citep{MacL99} suggest that the magnetic field plays
at most a secondary role in the dynamics of the gas clouds, which have
a small filling factor.  Collisions between clouds are highly
supersonic and therefore strongly dissipative, with the thermal energy
being radiated efficiently.

This complex medium is radiatively heated by stars to an extent that
varies strongly with the proximity to clusters of hot young stars.  It
is also cooled radiatively through processes such as thermal
bremsstrahlung, recombination lines from excited electronic states at
rates that depend strongly on the fraction of heavy elements, various
molecular rotational and vibrational transitions, and thermal emission
from dust.

All these processes are intensely localized on spatial scales that are
tiny compared with the overall size of a galaxy, and therefore well
below the resolution limits of most simulations that attempt to model
the formation and evolution of galaxies (see \S\S\ref{sec.galform} and
\ref{sec.flows}).

However, despite the complicated microphysics of this heated, cooled,
magnetized, and stirred multiphase medium, the crucial point is that
turbulence cascades down to small scales where it is dissipated.
Dissipation of random energy is the most important role of gas in the
overall dynamics of the star plus gas disk.  Galaxies lacking even a
small fraction of mass in gas barely evolve.  I emphasize the role of
gas in appropriate places in this review.

\subsection{Gravity softening in simulations}
\label{sec.soft}
Computer simulations are powerful tools that have proved indispensible
for unraveling the sometimes mystifying behavior of disk galaxies.
Yet even with present-day computational power, simulations cannot
routinely employ as many particles as there are stars in a galaxy.
Thus some degree of smoothing of the mass distribution is needed,
which also prevents strong accelerations during close encounters
between the particles that would otherwise require adaptive time
steps.  Smoothing can be introduced directly through ``softening'' the
interparticle force law at short range, or indirectly through the use
of a grid, which similarly weakens the forces between particles on
scales of the cell size \citep[see the appendix of][]{SM94}.  A third,
but less general, method of smoothing is to determine forces from an
expansion in some set of basis functions that is truncated at low
order.

Note that shot noise from the particle distribution remains the most
important limitation of simulations.  The contribution of distant
particles to the relaxation rate is unaffected by softening, which
smooths fluctuations on only the smallest scales, and changes nothing
in the formulae for relaxation but the value of the denominator of
$\Lambda$ in the Coulomb logarithm (\S\ref{sec.relax}).  Noise-driven
density variations on larger scales can also excite non-negligible
collective responses \citep{Sell83}; \citep{TK91}; \citep{Wein98}.

\subsubsection{2D simulations}
Simulating galaxy disks with the motion of particles confined to a
plane has the obvious advantage of reduced computational cost over
fully 3D simulations.  The most appropriate gravity softening rule for
2D simulations is the Plummer law, for which the potential at distance
$d$ from a point mass is:
\begin{equation}
\Phi_{\rm P} = -G\mu (d^2 + \epsilon^2)^{-1/2},
\label{eq.Plum}
\end{equation}
where $\epsilon$ is the gravitational softening length.

An advantage of the Plummer softening rule for this application is
that it provides an approximate allowance for disk thickness, as
follows.  Convolution of Eq.~(\ref{eq.Plum}) with the mass
distribution of a razor-thin disk yields the exact Newtonian field in
a plane offset by a vertical distance $\epsilon$ from that containing
the mass.  In real finitely thick galaxy disks, the field everywhere
is the sum of the Newtonian fields of the various mass elements spread
in layers parallel to the midplane.  The Newtonian forces experienced
by the stars are therefore weaker than if the mass distribution were
razor thin.  Thus the value chosen for $\epsilon$ in a 2D simulation
should be closely related to the finite thickness of the disk
\citep{Rome98}.

Note that gravity softening weakens nonaxisymmetric instabilities
\citep{Sell83}.  Since the Newtonian potential of an arbitrary
razor-thin mass distribution can be determined by expansion in Bessel
functions (\BTii\ \S2.6.2), the potential of each radial wavenumber,
$k$, of the expansion decays away from the plane as $\exp(-|kz|)$.
Further, since softened gravity is equivalent to sampling the field of
a 2D mass sheet in a plane offset vertically by a distance $\epsilon$,
the disturbance potential of each term is weaker by the factor
$\exp(-|k|\epsilon)$.  Hence instabilities are less vigorous.
However, this weakening is physically realistic because softening
provides an approximate allowance for the real finite thickness of
galaxy disks as explained above.

To estimate the time for peculiar velocities to be randomized by
encounters in 2D simulations, we replace Eq.~(\ref{eq.relax}) with
\begin{equation}
\tau_{\rm relax} = {\beta^3v^3 \epsilon \over 8 G^2\mu^2n},
\end{equation}
where $n$ is now the number of particles per unit area, $\epsilon =
b_{\rm min}$ and we assumed $b_{\rm max}^{-1} \ll b_{\rm min}^{-1}$.
This formula, without the $\beta^3$ factor, was already given by
\citet{Hohl73}.  Setting $N = \pi R_e^2 n$, with $R_e$ being the
half-mass radius of the disk, we find for 2D disks
\begin{equation}
\tau_{\rm relax} \approx {\beta^3\pi \over 8}{\epsilon \over R_e}N\tau_{\rm dyn}.
\end{equation}
This time is estimated for particles that interact with the forces
derived from the potential of Eq.~(\ref{eq.Plum}).

An advantage of computing forces through a cylindrical polar grid is
that one can further smooth the mass distribution by restricting the
sectoral harmonics $m$ that contribute to the forces acting on each
particle.  The effect of restricting force terms to include only the
range $0 \leq m \leq m_{\rm max}$ is to replace each point particle by
an azimuthally extended mass, providing some additional smoothing of
the density distribution.

\subsubsection{Simulations of thickened disks}
In 3D simulations, the Plummer softening law (Eq.~\ref{eq.Plum}) has
the undesirable property of weakening the interparticle force at all
distances from the source particle, and a softening kernel that
weakens forces only to a finite-range is greatly preferred.  All
that is needed for a serviceable softening kernel is that it should
join smoothly to the Newtonian law at some distance $\epsilon$ and
yield an inter-particle force for $d < \epsilon$ that smoothly
approaches zero as $d \rightarrow 0$.  The precise form of the force
at short-range should not matter because forces in a collisionless
fluid are dominated by the distant mass distribution.  Thus, if the
behavior of the $N$-body system is to mimic that of a galaxy, its
evolution should be insensitive to the adopted force law at
short-range.  Put another way, if the choice of the softening kernel
affects the behavior, then the simulation is not collisionless.

Of its very nature, gravity softening limits the sharpness of forces
that arise from steep density gradients.  While the in-plane density
distribution of galaxy disks varies on spatial scales that greatly
exceed the values of $\epsilon$ generally adopted, the disk mass is
strongly confined to a plane.  Unless the value of $\epsilon \ll z_0$,
the restoring force to the midplane will be weakened significantly,
which has adverse consequences for the correct representation of
buckling instabilities (see \S\ref{sec.buckle}).

In quasispherical mass distributions, the relaxation rate is given by
Eq.~(\ref{eq.trelax}), with $b_{\rm min} = \epsilon$ in the Coulomb
logarithm.  Following the discussion in \S\ref{sec.relax}.2, we
replace Eq.~(\ref{eq.trelax}) for disks having a characteristic
thickness, $z_0$, with
\begin{equation}
\tau_{\rm relax} \approx {\beta^3\ln\left(R_e / b_{\rm min}\right)
  \over 8\ln\left(z_0 / \epsilon\right)} {z_0 \over R_e} N \tau_{\rm dyn},
\label{eq.trdisk}
\end{equation}
where the factor $\beta^3$ is appropriate for the peculiar velocities
to be randomized by encounters.  For typical disk values of $\beta
\sim 0.1$ and $z_0/R_e \sim 0.1$, this time is almost $\sim 10^4$
times shorter than for quasispherical, pressure-supported systems
with the same $N$ \citep[see also][]{Sell13b}.

\section{Transient Spiral Modes}
The large majority of disk galaxies manifest beautiful spiral patterns
of some form or other.  The patterns are sometimes quite coherent and
symmetric, which are described as ``grand design'' spirals, or the
overall pattern can have little clear symmetry with individual pieces
of spiral arm being hard to trace over long distances because they
bifurcate or fade.  The more coherent patterns are often seen in
galaxies that are barred or have recently suffered a tidal interaction
with a passing companion galaxy \citep{KN79}; \citep{KKC11}.  However,
the ubiquity of the spiral phenomenon, and the fact that similar
patterns develop in simulations of stellar disks even when the
influences of bars and companions are excluded \citep{SC84};
\citep{Rosk08a}; \citep{Fuji11}; \citep{Wada11}, suggest that spirals
in galaxies can also be self-excited.

Spirals are important to secular evolution because they transport
angular momentum to a limited extent (see \S\ref{sec.Lchange}),
scatter stars at Lindblad resonances, which increases random motion,
cause radial mixing, and smooth rotation curves.  I discuss each of
these processes in turn.

\subsection{Origin and Recurrence}
The precise mechanism that causes spiral patterns to develop is not
fully understood and a thorough survey of the various ideas would
require too long a digression here \citep[see][for a recent
  review]{Sell13a}.  There is general agreement among theorists that
spirals are gravitationally driven density waves in the stellar disk,
for which there is substantial body of supporting observational
evidence, both photometric \citep{Schw76}; \citep{GGF95};
\citep{GPP04}; \citep{ZCR09} and kinematic \citep{Viss78};
\citep{Chem06}; \citep{Shet07}.  While the idea that spiral patterns
could be long-lived, or quasisteady, features has been advocated for
some time \citep[\eg][]{BL96}, it seems increasingly certain that
an individual spiral pattern does not persist for more than a few disk
rotations \citep{Sell11}.  The supporting evidence has to be indirect,
since we cannot observe the time evolution of real galaxies, and is
based on the behavior in simulations, which has not changed as their
quality has improved, and is supported by the arguments developed
below that disk evolution makes more sense when spirals are
short-lived.

\begin{figure}[t]
\includegraphics[width=\hsize,angle=270]{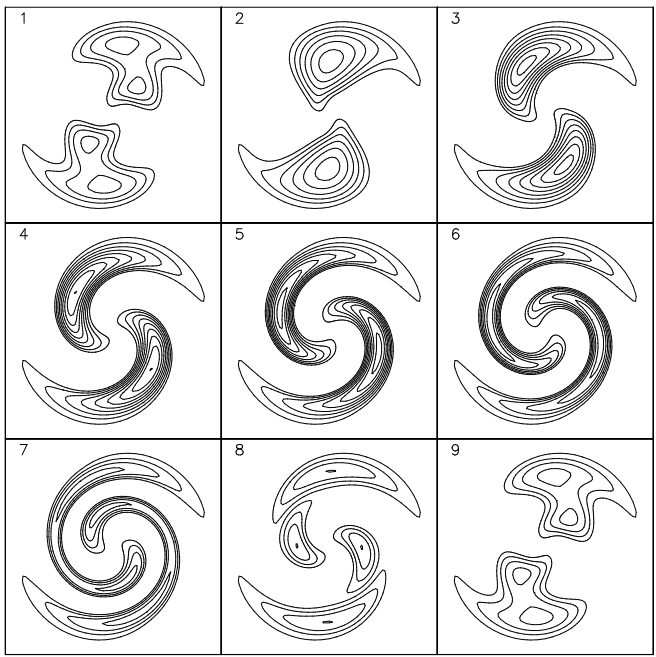}
\caption{Contours of the combined positive overdensity that results
  from the superposition of two open spiral patterns that each has
  constant amplitude and rotates at a steady, but different, rate.
  This is purely a set of drawings, with no underlying dynamics.  The
  numbered sequence illustrates one full beat period, and is shown in
  a frame that rotates such that the outer wave does not appear to
  move.  Notice that the largest net overdensities occur when the
  spiral is moderately wrapped.  An animation is available at
  \hfil\break {\tt
    http://www.physics.rutgers.edu/$\sim$sellwood/spirals.html}}
\label{fig.blobs}
\end{figure}

Most simulations manifest spiral patterns whose appearance changes on
time scales of less than one rotation of the disk.  However, power
spectrum analysis \citep{Sell89a} showed that the extreme variability
of the spirals reported by \citet{SC84} was caused by the
superposition of a few underlying longer-lived waves, as has
subsequently been found by others \citep[\eg][]{Rosk12};
\citep{Gran12}; \citep{Minc12}.

\citet{Wada11}, \citet{Gran12}, \citet{Baba13}, and \citet{Roca13}
reported that spirals in their simulations are almost material features
that wind up over time according to the local shear rate of the disk.
This behavior could also result from the superposition of multiple
waves, as illustrated in Fig.~\ref{fig.blobs} which shows the visual
appearance of the combined density of two separate fixed-amplitude
waves that rotate at different angular frequencies.  As long as the
inner wave has the higher angular speed, the combined density has most
of the properties reported by these authors.

The title of this section contains the word ``modes'' to distinguish
what is meant from the ``transient spiral'' illustrated in
Fig.~\ref{fig.dust}, which shows the time evolution of the vigorous
disk response to a particularly provoking peturbation.  That transient
response is neither an instability, because it does not grow
indefinitely, nor a mode, because it does not have a fixed shape.  A
mode, by contrast, is a standing wave oscillation of the system having
a constant shape and frequency.

A few authors have risen to the challenge of solving for the normal
modes of a smooth stellar disk with random motion.  They found
vigorously unstable bar-forming \citep{Kaln72, Kaln78}; \citep{Jala07}
and lop-sided \citep{Zang76}, \citep{ER98} modes.  However, when these
instabilities are avoided, perhaps by embedding the disk in a halo
(see \S\ref{sec.barorig}), studies of smooth disks generally do not
reveal milder spiral modes \citep{Toom81}.  An exception was the study
by \citet{Bert89}, who found slowly-growing bi-symmetric spiral modes
in low-mass, cool disks.  But \citet{Sell11} showed that their adopted
disk models would not survive, since they were subject to more
vigorous multiarm instabilities that caused the background state of
the disk to heat quickly.  Thus these normal mode analyses have been
useful to understand global disk stability, but have not yielded any
promising spiral-causing modes.

However, the waves that underlie the rapidly changing patterns in the
snapshots and movies from simulations {\it do} appear to be genuine
modes of the disk.  Each has a fixed shape and well-defined pattern
speed, and grows and decays on timescales of a few disk rotations.
As they are not truly long-lived, they are best described as
``transient spiral modes.''

Each spiral mode in the simulations is a vigorously growing
instability that saturates at an overdensity of some 20\% -- 30\%
relative to the local unperturbed disk density.  After it saturates it
fades just about as quickly as it grew, with all the wave action that
had been extracted from the particles during its growth being carried
radially away from corotation at the group velocity \citep{Toom69}.
Wave action is finally absorbed at the Lindblad resonances
(\S\ref{sec.resnc}), where wave-particle interactions occur
\citep{LBK72}; \citep{Mark74}.  The changes to the underlying disk
caused by the scattering of stars at these Lindblad resonances seed
conditions for a new instability to develop \citet{Sell12}, and thus
the cycle recurs.  These instabilities were missed by the authors of
the above-cited stability analyses because they studied the modes of
an assumed smooth, featureless disk.  \citet{Sell10} found some
evidence in the velocities of stars in the solar neighborhood to
support this picture.

\citet{DVH12} introduced a collection of heavy, co-orbiting particles
into their large-$N$ simulations of low-mass disks, which seeded
spiral responses.  In one experiment, they introduced a single heavy
perturber and removed it again after some evolution; they found that
spiral activity continued, which they attributed to additional
responses to the fluctuations caused by responses to the earlier
forcing particle, in a bootstrap fashion, that they described as
``nonlinear'' effects, although it was unclear that the behavior they
observed depended in any way on the amplitudes of the disturbances.
It is also possible to regard their result as the superposition of
multiple spiral modes of the underlying disk, which were triggered at
moderate amplitude by the original perturbing heavy particle.
Whatever the correct explanation, they concurred with \citet{Sell12}
that spiral activity at one instant is directly influenced by the
immediately preceding activity.

In the absence of dissipation, the recurring spirals drive up the
level of random motion in the disk (see \S\ref{sec.sh}).  As $Q$
rises, the disk becomes ever less able to support collective
oscillations, and activity weakens on a time scale of perhaps ten disk
rotations \citep{SC84}.\footnote{\citet{Fuji11} suggested the time-scale
  could be longer, but the spirals in their simulations are quite
  faint at late times.  Furthermore, the dominant halo they use
  results in multiarm spirals that heat the disk more slowly than
  would patterns of lower $m$ (\S\ref{sec.sh}).}  At this point the
minor gas component (\S\ref{sec.ISM}) takes on a dynamically important
role; while dense clouds of gas are accelerated by the spirals in the
same manner as are the stars, they are able to dissipate random motion
quickly through supersonic collisions that allow the excess energy to
be radiated away.  The gas clouds themselves, and the stars that form
within them, therefore constitute a low velocity dispersion component
that is able to maintain the dynamical responsiveness of the combined
star-gas disk. \citet{SC84} estimated that a birth rate of a few stars
per year over the entire disk of a galaxy would be sufficient to
sustain spiral activity indefinitely.  Subsequent work \citep{CF85};
\citep{Toom90}; \citep{Rosk08a} on isolated disks, and fully
cosmological simulations \citep[\eg][]{Ager11}, seems to confirm that
no matter how the dissipation is mimicked, the disk continues to
support transient spiral patterns.  This behavior provides an
attractive explanation for the long-noted \citep[\eg][]{Oort62}
contrast between the striking spirals manifested by galaxies having
abundant gas to the featureless appearance of S0 galaxies that have
very little gas.

\begin{figure}[t]
\begin{center}
\includegraphics[width=.9\hsize, clip=]{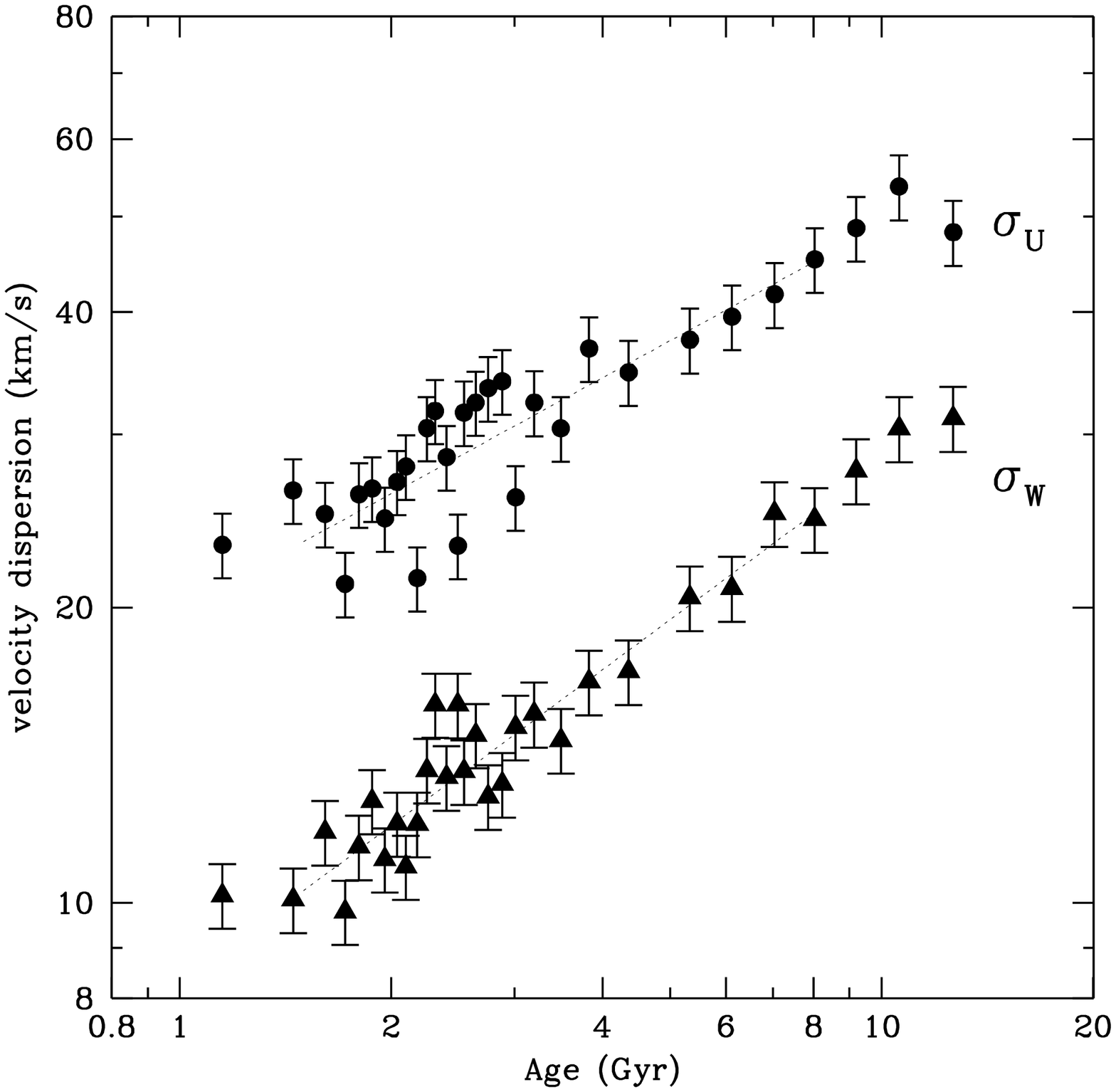}
\includegraphics[width=.9\hsize, clip=]{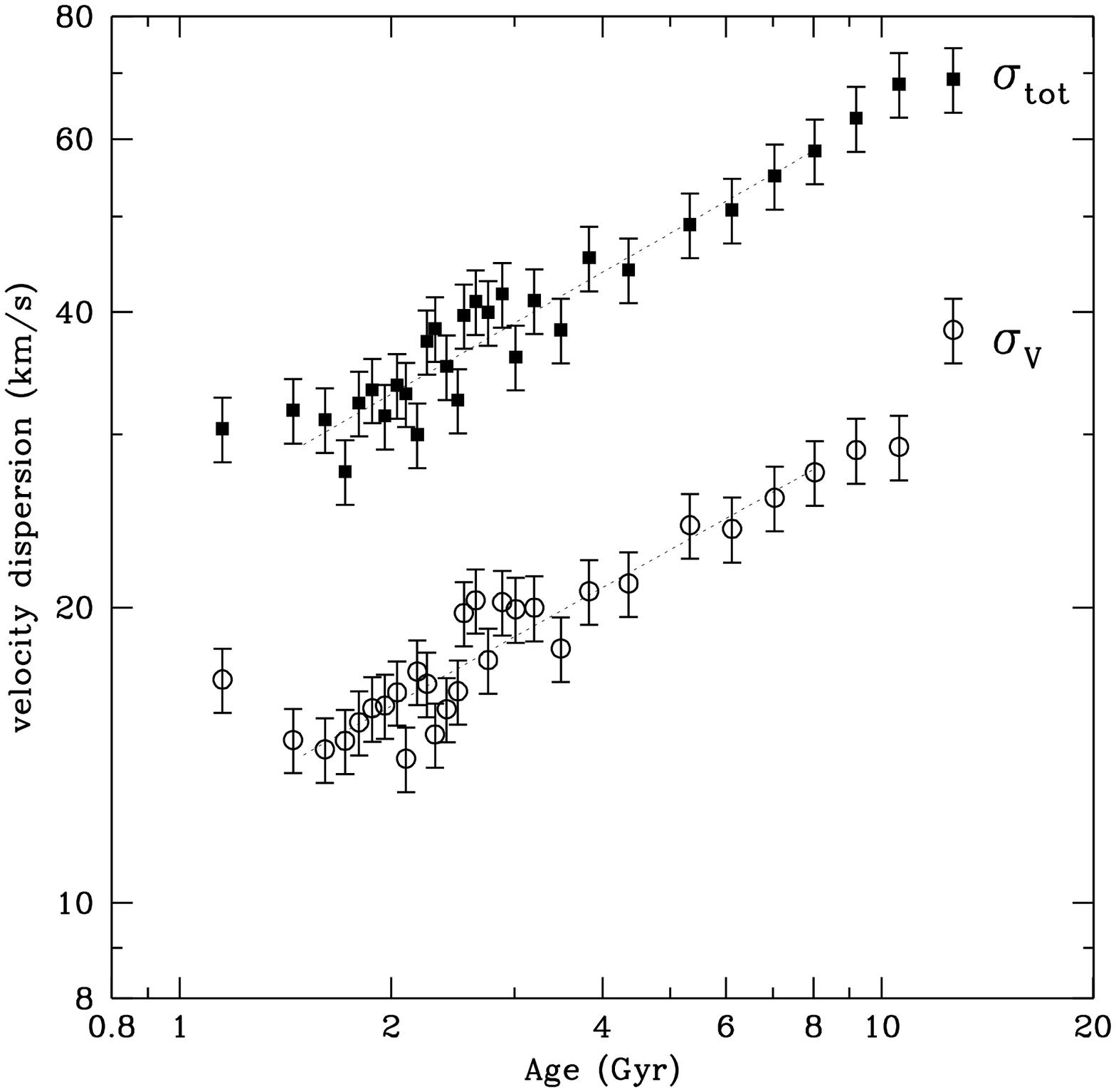}
\end{center}
\caption{Estimates of the second moments of stellar velocities from
  the Geneva-Copenhagen survey.  The symbols with error bars show the
  estimated spreads of the three velocity components in Galactic
  coordinates: in the upper panel, $U$ is in the radial direction and
  $W$ in the direction normal to the plane.  The spread of $V$ in the
  azimuthal direction and the total dispersion are shown in the lower
  panel.  The stars were divided by estimates of their ages and the
  fitted lines ignore the two youngest and the two oldest groups.
  From \citep{HNA09}.}
\label{fig.avr}
\end{figure}

\subsection{Scattering of stars}
\label{sec.scatt}
It has been clear for over 50 years that older stars in the
neighborhood of the Sun have larger velocity spreads than do younger
stars \citep{Wiel77}; \citep{Nord04}.  It seems unsatisfying to
suppose that older stars were born with larger random velocity
components, since it requires us to live at a special time when random
motions at birth have just become small, but this suggestion has been
advocated \citep[\eg][]{Krou02}.  Some initial random motion seems
likely in the disturbed conditions of disks in the early Universe when
the oldest stars formed, but the progressive increase of random
motions of disk stars with increasing intermediate ages is generally
attributed to scattering processes.  Both massive gas clumps
\citep{SS53} and spiral patterns \citep{BW67} are still considered
viable as scattering agents.

\subsubsection{Solar neighborhood data}
Fig.~\ref{fig.avr} shows the variation of stellar velocity dispersion
components with estimated age, as presented by \citet{HNA09}.  (The
small scatter about the trend among the points for the older stars is
somewhat odd.)  These results are synthesized from the heroic
Geneva-Copenhagen survey (herafter GCS) of $\sim 14 000$ F- and
G-dwarf stars by \citet{Nord04}, with repeated radial velocity
measurements of all the stars to eliminate binaries, as well as
improved stellar parameter and age calibrations \citep{HNA07}, and the
revised {\it Hipparcos\/} distances and proper motions \citep{vLee07}.
\citet{HNA09} use the usual notation $\sigma_U$, $\sigma_V$, and
$\sigma_W$ for the radial, azimuthal, and vertical velocity spreads as
seen from the Sun, where others may use $\sigma_R$, $\sigma_\phi$, and
$\sigma_z$ for the radial, azimuthal and vertical dispersions
anywhere.  \citet{Casa11} reanalyzed the same sample adding infrared
fluxes for about half the stars to obtain new estimates of stellar
parameters and ages, and again found the total velocity dispersion
rose steadily with age (their Fig.~17), even when they excluded
metal-poor stars.

Assigning ages to individual stars is highly controversial
\citep[see][for a review]{Sode10}, and the precise trend with age has
therefore been the subject of much debate.  \citet{Reid07} suggest
that the ages of individual stars assigned by \citet{Nord04} and
revised by \citet{HNA07} are compromised by large random errors.  Were
this the case, then the real trends with age would have to be even
steeper than shown in Fig.~\ref{fig.avr}, since large age errors will
flatten a trend, as found by \citet{Casa11} when they included
stars with more uncertain ages.

On the other hand, \citet{Edva93} and \citet{QG01} claimed a more nearly
constant velocity dispersion with age, except for very old stars that
have a larger dispersion.  \citet{Soub08} also found an almost flat
dispersion with age, although it is unclear whether their exclusion of
probable thick disk and halo stars, based at least in part on their
kinematics, is affecting the trend.  If a constant dispersion with
age were correct, and if the small samples of stars used in these
studies were drawn from the same population as those selected by
\citet{HNA09}, then the age errors that give rise to the steady trends
seen here in Fig.~\ref{fig.avr} and in Fig.~17 of \citet{Casa11} would
have to correlate with the kinematics.  It therefore seems more likely
that claims of a flat trend result from large age errors or selection
against stars with large peculiar velocities.  Furthermore
\citet{AB09} found that blue main-sequence stars, which must be young,
have much smaller random motions than do red main-sequence stars,
which can have a wide range of ages.  Using color as a proxy for age
assumes a well-behaved star-formation rate in the disk, which has
probably declined slowly over time \citep[\eg][]{FT12}.  \citet{AB09}
constructed a model to fit the data that also favored a steady
increase of velocity dispersion with time.

\begin{figure}[t]
\begin{center}
\includegraphics[width=.9\hsize, clip=]{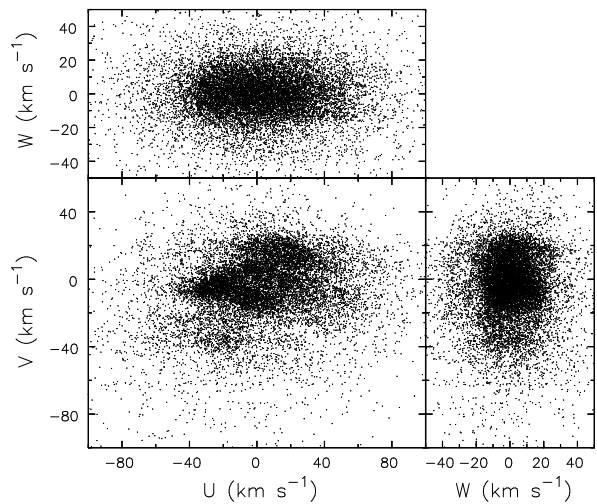}
\end{center}
\caption{The stellar velocities from the Geneva-Copenhagen survey
  \citep{HNA09}, corrected for solar motion $(U_\odot,V_\odot,W_\odot)
  = (11.1,12.24,7.25 )\;$km s$^{-1}$ \citet{SBD10}.  Note the
  substantial substructure in the $(U,V)$-plane that is not reflected
  in $W$.}
\label{fig.phspac}
\end{figure}

In fact, the in-plane components of the GCS stars do not have simple
Gaussian velocity distributions (Fig.~\ref{fig.phspac}), as first
deduced from the {\it Hipparcos} data by \citet{Dehn98}.  Not only is
the overall distribution of the $V$ components quite skew with an
``asymmetric drift'' \citep[$\bar V < 0$ as expected, \BTii;][]{SB12},
but the distribution is characterized by multiple ``streams'' that are
distinct at a high level of significance \citep{Bovy09}.  The streams
are both far too massive and have a spread in metallicities to be
dissolved star clusters \citep{Fama07}; \citep{Bens07};
\citep{Pomp11}.  \citet{Hahn11} examined the nearby stars of the Sloan
Digital Sky Survey \citep[SDSS][]{York00} and RAdial Velocity
Experiment \citep[RAVE][]{Stei06} finding similar, but less distinct,
substructure.  Blurring of the velocity structures is to be expected
for stars in these larger surveys, which do not have {\it
  Hipparcos}-quality astrometry and distances, with consequent loss of
precision in the sky-plane velocity components.  \citet{Anto12} traced
some of these features in somewhat more distant stars of the RAVE
survey.

The substructure in Fig.~\ref{fig.phspac} probably arises from the
dynamical influence of density perturbations in the disk and a number
of attempts have been made to model it.  \citet{dSWT} found that
multiple, imposed transient spiral perturbations were able to create
qualitatively similar substructure in the stellar velocity
distribution.  On the other hand, individual features have been
interpreted as responses to assumed models for the bar \citep[][see
  also Kalnajs 1991]{Dehn00, MBSB10}, or to specific spiral models
\citep{QM05}; \citep{Pomp11}; \citep{Anto11}, or both \citep{Quil03};
\citep{Chak07}; \citep{Anto09}.  Finally, \citet[][see also Hahn
  \etal\ 2011 and McMillan 2011, 2013]{Sell10} did not need to adopt a
perturbing potential, but instead used action-angle variable analysis
to identify the Hyades stream as resulting from scattering by a recent
Lindblad resonance.  It is likely that the different streams have
different origins and a combination of these ideas would be needed to
explain all the features.

Returning to Fig.~\ref{fig.avr}, the data show that the second moments
of the peculiar velocities differ in all three components -- \ie, the
velocity ellipsoid of nearby disk stars has a triaxial shape that
apparently grows roughly homologously with age.  \citet{SG07} argue
that the multiple streams in the $U-V$ plane (Fig.~\ref{fig.phspac})
make the second moment a poor measure of the velocity spread.
However, the distribution in the $U-W$ plane is much smoother and it
is worth noting that the rising trend of the $V$ component in
Fig.~\ref{fig.avr} maintains a constant fraction of the radial
component at the ratio expected from epicycle theory (\BTii,
Eq.~8.117) in a galaxy with an approximately flat rotation curve.

In Galactic components, the dispersion in the radial direction is the
largest, the azimuthal component is intermediate, while the smallest
is the component normal to the disk plane, being only about half as
large as the radial velocity dispersion.  A flattened shape appears to
be representative of that in other disk galaxies \citep{Bott93};
\citep{Gers00}; \citep{HC09}; \citep{Bers11}, although \citet{GS12}
claim evidence that the axis ratio varies along the Hubble sequence.

\citet{Smit12} present a study of local disk kinematics using the
``Stripe 82'' data from SDSS, although they make no attempt to assign
ages to stars.  Instead they divide stars by metallicity and
present-day $z$-height below the disk plane and devise a procedure to
estimate separate dispersions of the thin- and thick-disk stars that,
however, must become increasingly difficult for metal-poor stars and
those at large distances from the midplane.  While their results for
thin-disk stars with $-0.5 < \hbox{[Fe/H]} < 0.2$ are consistent with
those from other studies, they found the velocity ellipsoid is
distinctly rounder for more metal-poor stars that also have higher
velocity dispersions.

\citet{LF89} found that the velocity dispersion in the Milky Way disk
has a steep outward gradient over a wide radial range, as the
above-cited studies also found in other galaxies.  A gradient is, of
course, expected on local stability grounds (Eq.~\ref{eq.Qdef}), but
the radial gradient must somehow combine with the velocity ellipsoid
shape and disk surface density to create a vertical thickness scale
that appears to be independent of radius for many galaxies
\citep{vdKS81}; \citep{Kreg02}.  This conspiracy of disk properties
has yet to be fully explained.

\subsubsection{Scattering by spirals}
\label{sec.sh}
\citet{LBK72} showed that, away from resonances (\S\ref{sec.resnc}), a
spiral perturbation that grows and decays adiabatically (on a time
scale long compared with orbital and epicycle periods) leaves the
stellar motions unchanged.  Stars do work on, or receive energy from,
a potential perturbation as it grows, but these changes are undone as
the wave decays, leaving only oscillatory ripples in the phase-space
density that average to no change \citep{CS85}, except at resonances.

Wave-particle interactions at resonances do, however, cause lasting
changes to the orbits of stars, and \citet{LBK72} showed that stars at
the ILR lose angular momentum on average, while those at the OLR gain.
Changes at corotation could be of either sign, depending on the sign
of the gradient of the angular momentum density of stars around the
resonance (see \S\ref{sec.DF}).

\begin{figure}[t]
\includegraphics[width=\hsize]{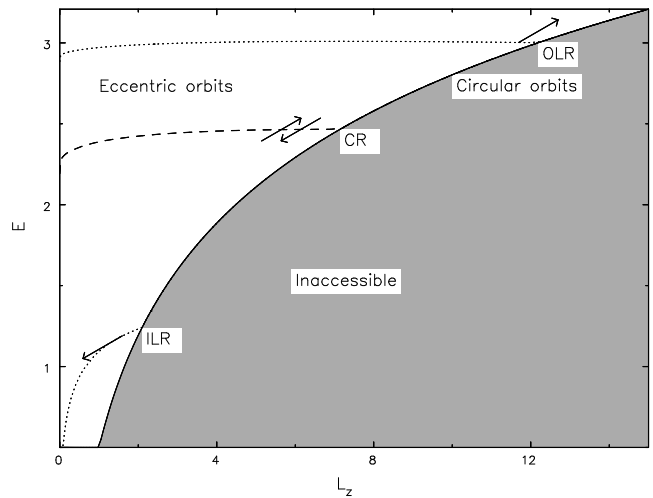}
\caption{The Lindblad diagram for a razor-thin disk galaxy model.
  Circular orbits lie along the full-drawn curve and eccentric orbits
  fill the region above it.  Angular momentum and energy exchanges
  between a steadily rotating disturbance and particles move them
  along lines of slope $\Omega_p$ as shown.  The dotted and dashed
  lines are the loci of resonances for an $m=2$ perturbation of
  arbitrary pattern speed.}
\label{fig.lind}
\end{figure}

The changes given by Eq.~(\ref{eq.LBK}) become $\delta$-functions at
resonances in the limit $\Im(\omega) = \gamma \rightarrow 0$, which
\citet{LBK72} take.\footnote{Note the important aspect of trapping at
  resonances by a steady or slowly growing disturbance is not captured
  by Eq.~(\ref{eq.LBK}).}  However, changes are smooth when broadened
by time dependence, and \citet{CS85} compute the lasting changes under
the assumptions that the wave both grows and then decays
exponentially.

In a rotating, but otherwise steady, nonaxisymmetric potential test
particles conserve neither their energy, nor their angular momentum, 
but Jacobi's integral,
\begin{equation}
I_J \equiv E - \Omega_pL_z, 
\label{eq.jacobi}
\end{equation}
is conserved.  Thus changes in energy and angular momentum are related
as $\Delta E = \Omega_p \Delta L_z$.  This is illustrated in
Fig.~\ref{fig.lind}, which is drawn for the midplane of an
axisymmetric potential \citep[see also][their Fig.~2]{LBK72}.
Circular orbits have the minimum energy $E_c$ for a given $L_z$, which
marks the boundary of the shaded region, while orbits with $E>E_c$ are
eccentric.  Equation~(\ref{eq.jacobi}) constrains particles that are
scattered by a nonaxisymmetric perturbation to move along lines of
fixed slope $\Omega_p$ as illustrated by the arrows, which are marked
at the principal resonances because those are the only places where
lasting changes occur.

There is an important difference between scattering at Lindblad
resonances and at corotation.  A star near corotation may suffer quite
a large change in its angular momentum, but because $dE_c/dL_z =
\Omega_p$ at this radius (Fig.~\ref{fig.lind}), it neither gains nor
loses random energy (to first order); all the energy change is
invested in changing the radius of the guiding center
(Eq.~\ref{eq.rguide}).  This is a characteristic feature of radial
migration (\S\ref{sec.rm}).  The situation is different away from
corotation, where there is an excess of energy available to increase
random motion, provided that stars inside corotation lose $L_z$ while
those outside gain.

\citet[][their Eq.~6]{SB02} showed that the first order change in
radial action of a star caused by a single spiral wave is related to
its change in angular momentum via
\begin{equation}
\Delta J_R = -{l \over m}\Delta L_z.
\label{eq.Lchange}
\end{equation}
Since $l=0$ at corotation, any $\Delta L_z$ causes no change to $J_R$,
as noted.  Furthermore, $\Delta L_z <0$ at the ILR where $l=-1$ and
$\Delta L_z > 0$ at the OLR where $l=+1$.  Thus outward transfer of
angular momentum from the inner to the outer Lindblad resonance causes
$\Delta J_R > 0$, or heating, at both, as is also clear from
Fig.~\ref{fig.lind}.  A succession of transient spiral modes with a
variety of pattern speeds will cause the in-plane components of
stellar random motion to increase generally over the disk.

\subsubsection{Vertical heating}
Note that the changes caused by transient spiral modes increase only
the in-plane random motions, not the component normal to the plane as
predicted by \citet{Carl87} and confirmed by \citet{Sell13b}.  The
reason (\S\ref{sec.resnc}) is that coupling between the spiral
perturbation and vertical motion is expected to be very weak because
the frequency $\nu$ of small vertical oscillations of a star near the
midplane (Eq.~\ref{eq.nudef}) is generally high compared with the
Doppler shifted frequency $m|\Omega_\phi - \Omega_p|$ at which it
encounters the perturbation, making its vertical motion adiabatically
invariant.  While a majority of stars rise out of the harmonic region,
the fraction that have a low enough vertical frequency to experience a
vertical resonance with a spiral perturbation is believed to be quite
small.

The discussion in the previous paragraph assumed Newtonian
gravitational forces, and softening in simulations (\S\ref{sec.soft})
can change the behavior.  Increasing the gravitational softening
length weakens the restoring forces to the midplane, decreasing the
vertical frequency and possibly allowing vertical resonances to become
dynamically important \citep[see][their Fig.~9]{SSS12}.  On the other
hand, simulations with small softening, but modest numbers of
particles, may thicken due to relaxation \citep{Sell13b}.  Thus the
modeling of disk thickening in simulations is somewhat delicate.

Coherent bending waves are another possible mechanism to increase the
vertical velocity dispersion.  The mechanics of bending waves is
complicated \citep[see][for a recent review]{Sell13a}.  However, we do
know that a bending wave may travel across a stable disk
\citep{Toom83}; \citep{Wein91} until it is damped as it approaches a
vertical resonance \citep{SNT98}, with the wave energy going into
localized vertical heating.  It is also known \citep{Toom66};
\citep{Arak85} that a disk in which the velocity ellipsoid is
flattened such that $\sigma_R \gtsim 3 \sigma_z$ will buckle and
thicken until the axis ratio is approximately this value
\citep{Sell96};\citep{RS13}.\footnote{This behavior is also affected
  by gravity softening.}  However, the velocity ellipsoid of local
stars in the Milky Way is not flattened enough to be near this
stability boundary.

Another possible heating mechanism is infall of cosmic substructure
\citep[\eg][]{Kaza09}.  While infall of massive clumps in their
simulations, and those of others, is quite disruptive and can probably
be excluded (\S\ref{sec.thin}), a gentler bombardment by smaller clumps
may cause more gradual heating.  However, a prediction of
\citet{Kaza09} is that satellite bombardment should create velocity
dispersions that are roughly constant with radius, whereas data on the
Milky Way \citep[\eg][]{LF89} indicate a strong decline with radius to
distances well beyond the solar circle.  Thus, while some heating by
infall cannot be excluded, it is clearly not the dominant process.

Since the data (Fig.~\ref{fig.avr}) show that the out-of-plane motions
rise with about the same slope as the in-plane part, it seems unlikely
that spiral waves, neutral bending waves, or buckling instabilities
are important in setting the shape of the local velocity ellipsoid in
the local Milky Way.  It therefore seems that scattering by collective
waves cannot be the whole story.

\subsubsection{Scattering by dense mass clumps}
\label{sec.clouds}
\citet{SS53} argued that massive clumps of gas were needed to account
for the increase of peculiar stellar velocities with their ages, and
therefore hypothesized the existence of giant molecular cloud
complexes (GMCs) long before their existence was established.  Their
original calculation of scattering by dense mass clumps was extended
to 3D by \citet{Lace84}.  In his analysis, as in the earlier work by
\citet{SS53}, the star-clump interaction was computed in the impulse
approximation, in which scattering is assumed to occur over a distance
that is short compared with both the size of the star's epicycle and
the scale on which the galactic gravitational potential changes.

\citet{Lace84} found that co-orbiting mass clumps are quite efficient
at redirecting peculiar motions out of the plane, but rather
inefficient at increasing them.  He also concluded that cloud
scattering should cause the vertical dispersion $\sigma_W$ to be
intermediate between the radial $\sigma_U$ and azimuthal $\sigma_V$
components.  This result seemed physically plausible on energy
equipartition grounds: scattering by massive clouds redirects the
peculiar motions of stars through random angles, and therefore
isotropizes the motions as far as the epicycle gyrations allow.

However, this again is inconsistent with the data
(Fig.~\ref{fig.avr}), where $\sigma_W$ is the smallest component.  In
order to account for the observed flattened shape, \citet{Carl87} and
\citet{JB90} developed the plausible argument that the in-plane
dispersion is driven-up more rapidly by spiral heating than scattering
is able to redirect those motions into the vertical direction.  Their
argument seemed to offer strong support for the transient spiral mode
picture \citep{Sell00}, but it now appears to be incorrect.

\citet{Ida93} found that cloud scattering alone would lead to the
vertical component being the smallest, with the precise axis ratio of
the velocity ellipsoid depending on the local slope of the rotation
curve.  Simulations by \citet{SI99}, and others \citep[see
  \eg][]{Vill85}; \citep{HF02}, confirmed their expectation.

\begin{figure}[t]
\begin{center}
\includegraphics[width=.9\hsize, clip=]{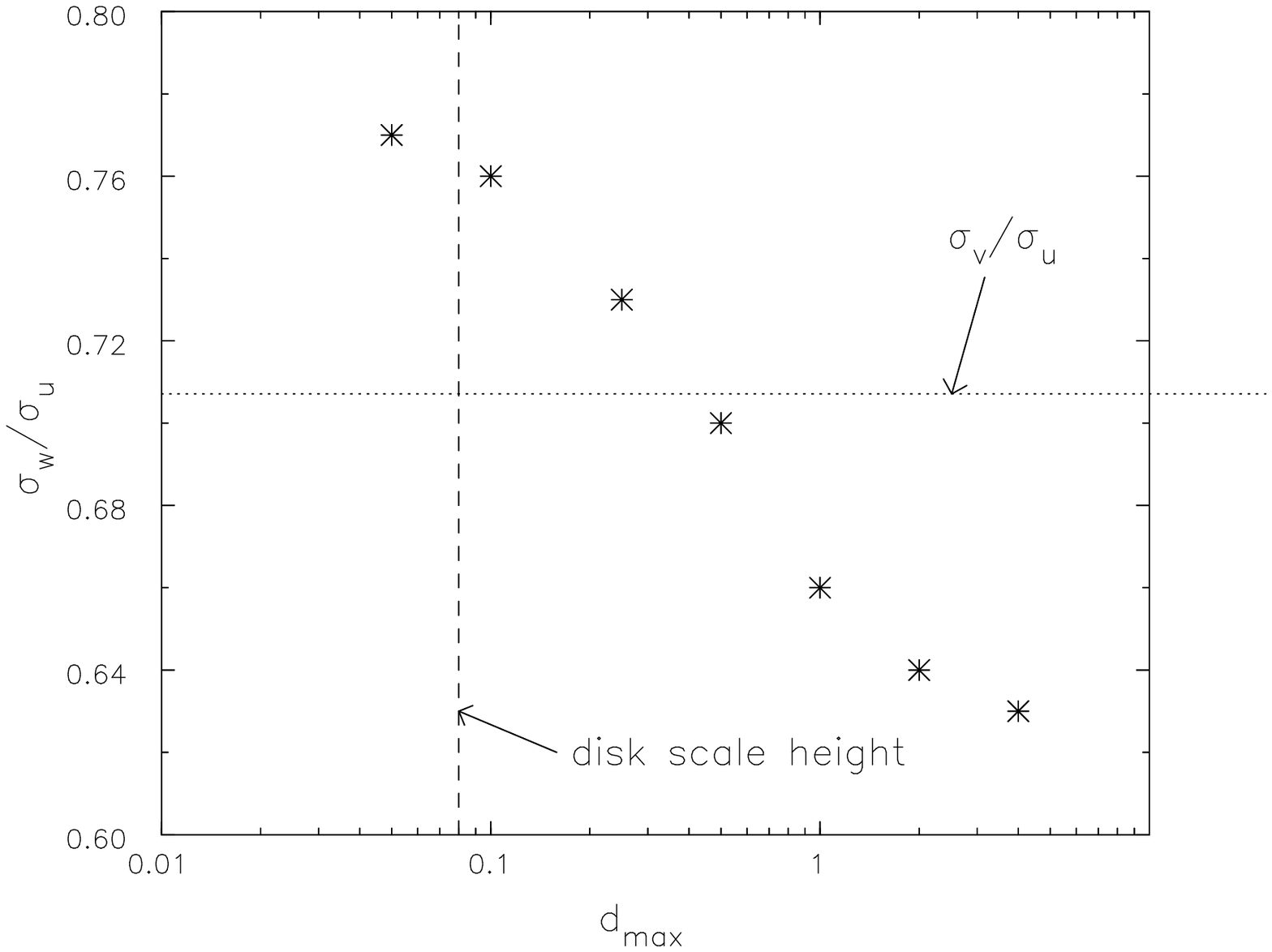}
\end{center}
\caption{The equilibrium axis ratio of the velocity ellipsoid of
  particles plotted as a function of the limiting range of the
  perturbation forces from the heavy particles. See \cite{Sell08b} for
  a description of the calculations.}
\label{fig.axrat}
\end{figure}

The reason for the discrepancy with Lacey's prediction was clarified
by \citet{Sell08b}, who presented local simulations of scattering of
test particles by massive co-orbiting particles.  He artificially
restricted the range of the gravitational forces from the heavy
particles, which vanished at distances greater than some $d_{\rm
  max}$.  Figure~\ref{fig.axrat} shows the equilibrium ratio of the
vertical to radial velocity dispersions $\sigma_W/\sigma_U$ plotted
as a function of the adopted $d_{\rm max}$.  The ratio settles to
something close to the energy equipartition prediction when none but
the closest heavy scatterers perturb the stars but, as the range of
scattering was increased in separate experiments, the equilibrium
ellipsoid gradually flattened and approached the shape predicted by
\citet{Ida93} for no cut off.

The flattened shape of the ellipsoid is determined by the fact that
the perturbing clouds are located within the disk, leading to an
aspherical distribution of impact parameters, with the consequence
that deflections from the more distant clouds preferentially redirect
the in-plane velocity components.  \citet{Lace84} and \citet{BL88}
neglected distant encounters, and therefore missed this effect.
However, the familiar argument that every decade in distance makes an
equal contribution to scattering also ceases to hold in disks
(\S\ref{sec.relax}), and the contribution to scattering by clouds that
are more distant than several disk scale heights drops away rapidly.
Thus it is the clouds at in-plane distances of just a few disk
thicknesses that do most of the redirecting.

\citet{Smit12} confirmed the predicted velocity ellipsoid shape for
the metal rich stars in their data, but reported a rounder ellipsoid
for the hotter, and probably older, metal-poor stars.  Further work is
needed to confirm this metallicity dependence, which may have been
biased by the difficulty of separating thin- from thick-disk stars.
Note that \citet{HNA09} (see Fig.~\ref{fig.avr}) found an ellipsoid
shape that was almost constant with age, and at most only mildly
rounder for the older stars.

\subsubsection{Collective effects}
The preceding calculations of scattering by mass clumps ignored all
collective effects.  Not only are disk galaxies subject to spiral
perturbations, which themselves scatter stars, but the co-orbiting
GMCs induce a collective response from the surrounding stellar disk
\citep{JT66} that substantially enhances their effective mass, a
complication that is ignored in most studies of cloud scattering.  An
exception was provided by \citet{TK91}, who studied scattering by both
a density perturbation and the supporting response from the
surrounding matter.  The density fluctuations in their local
simulations arose from the shot noise of the particles, while the same
particles also took part in the supporting response.  By applying a
radial damping term, they may have unwittingly prevented the growth of
instabilities \cite{Sell12}, making their work a particularly clean
calculation of the heating rate due only to the polarized disk
response to co-orbiting mass clumps.

Since molecular gas is mostly \citep{Niet06}; \citet{Koda09};
\citep{Grat10}; \citep{Efre10}, but not entirely
\citep[\eg][]{Cord08}; \citep{Schi13}, concentrated in spiral arms it
is probably futile to draw a sharp distinction between spiral arms and
the wakes of dense gas clumps, and a correct treatment would be to
calculate the effects of spiral formation and gas dynamics in the
combined star and gas disk.  \citet{BL88} took a step in this
direction, but a full calculation may remain unreachable for some time
if one tries to include a self-consistent treatment of the formation
and dispersal of the gas clumps: molecular gas concentrations probably
grow in the converging gas flow into a spiral arm, and are
subsequently partly dispersed by star formation.  \citet{DVH12} also
showed that massive clumps provoke spiral responses, but the spiral
activity probably included some self-excited collective modes, since
it persisted after the perturbers were removed.

Thus the studies of scattering in disks reviewed earlier make the
simplifying assumption that spirals and mass clumps are distinct
agents.  This assumption at least separates the problem into tractable
pieces.  Perhaps it can be justified if the wakes of cloud complexes
can be lumped with spirals into a single scattering agent that is
distinct from the clouds that caused them.

\subsubsection{Conclusions on scattering}
We now understand that the velocity ellipsoid in the solar
neighborhood is flattened as expected from scattering by GMCs
\citep{Ida93}.  While the clouds efficiently redirect peculiar
velocities to maintain the observed shape of the velocity ellipsoid,
they are not thought to be responsible for much heating.

The magnitude of the peculiar velocities of the intermediate age stars
exceeds what cloud scattering could achieve \citep{Lace91}.  The old
explanation for this, which may not be valid because it neglects the
cumulative effect of intermediate distance encounters, was that the
efficiency of scattering by clouds decreases as stars spend more time
outside the cloud layer that is largely confined to the midplane.
Simulations by \citet{HF02}, that did include distant encounters,
confirmed that GMCs alone are unable to account for the random motions
of the oldest stars.

Thus some other agent, generally assumed to be the spirals, is needed
to boost the rms velocities of intermediate age disk stars to their
observed values.  Even though spirals do not heat the vertical
motions, they drive up in-plane random motions that are efficiently
redirected by GMCs, and the velocity ellipsoid maintains an
approximately constant shape as its size increases, accounting for the
observed trends in Fig.~\ref{fig.avr}.  This picture does not exclude
the possibility that the high peculiar motions of the very oldest disk
stars, also known as the thick disk, have a different dynamical origin
(\S\ref{sec.thick}).

\subsubsection{Heating in simulations}
Note that the behavior in $N$-body simulations needs to be interpreted
with caution.  Section III.B.6 accounted for the observed peculiar
velocities in the solar neighborhood using the combined action of two
distinct mechanisms: heating by spirals with the random motions being
redirected by molecular clouds.  Simulations support spiral patterns
that may resemble those in galaxies and, if collisionless, should not
thicken because they generally omit heavy particles to represent
GMCs.\footnote{\citet{DVH12} included heavy particles, but did not
  discuss their effect on the velocity ellipsoid shape.}  Yet a few
authors \citep[\eg][]{Quin93}; \citep{MD07} have worried that disks
thicken in isolated $N$-body simulations that are heated by spiral
activity.

\citet{Hous11} compared the vertical heating in a simulation with the
solar neighborhood data.  Their simulation, which was probably heated
in part by spirals, included the cosmologically expected infall of
pieces of substructure that could increase the vertical dispersion,
and also modeled the full ``gastrophysics'' of cooling, star formation
and feedback.  However, they employed gas (and star) particles having
the masses of GMCs, which therefore lacked the spatial and mass
resolution to form dense clumps that are crucial to shaping the
velocity ellipsoid.

Collisional relaxation, which is much more rapid in disks \citep[][and
  \S\ref{sec.relax}]{Sell13b}, is a more likely explanation for
redirecting in-plane motions to thicken disks in
simulations.\footnote{\citet{MD07} found that thickening was suppressed
  when the azimuth of every disk particle was randomized after every
  step, in order to suppress the growth of nonaxisymmetric
  disturbances.  However, such a procedure must also largely inhibit
  two-body scattering, as well as all coherent responses from the
  surrounding disk.}  Thus, the simulated vertical heating rate, in
particular, will depend on the number of particles employed, and
comparison with the observed vertical heating of disk stars in the
Milky Way \citep[\eg][]{Hous11} is premature without careful numerical
convergence tests.

\subsection{Radial migration and mixing}
\label{sec.rm}
\begin{figure}[t]
\begin{center}
\includegraphics[width=.9\hsize, angle=270, clip=]{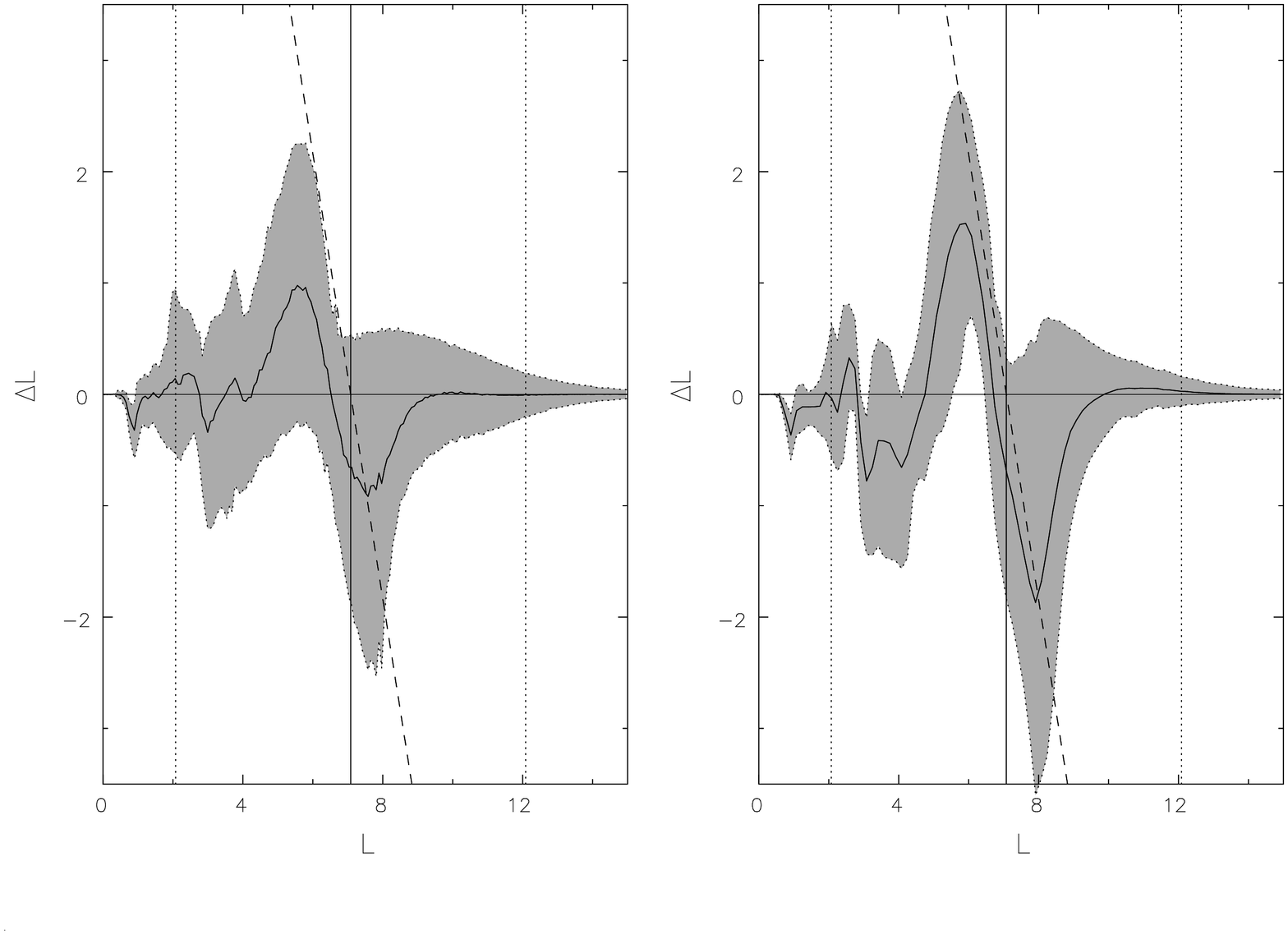}
\end{center}
\caption{Changes in angular momentum, $L$, resulting from a single
  transient spiral mode.  The shaded region includes 90\% of the
  particles and the solid curve shows the mean change.  The vertical
  solid line marks the location of corotation, while the dotted lines
  mark the Lindblad resonances.  The dashed line has slope $-2$.
  From \citet{SB02}.}
\label{fig.dLz}
\end{figure}

For years, the focus of spiral scattering was on heating at Lindblad
resonances, and changes at corotation went unreported.  \citet{SB02}
were therefore surprised to find that a transient spiral mode causes
greater angular momentum changes to stars at corotation than occur at
the Lindblad resonances, as shown in Fig.~\ref{fig.dLz}.  These more
substantial changes had not attracted attention because they do not
heat the disk (Eq.~\ref{eq.Lchange}), and stars largely change places
in a dynamically neutral manner.  However, they do have important
consequences for the distribution of chemical abundances among the
disk stars \citep{Rosk08b}; \citep{SB09a}; \citep{MCM12}.

Changes to the guiding center radii caused by a series of transient
spiral modes with corotation radii scattered over a wide swath of the
disk will cause stars to execute a random walk in radius with a step
size ranging up to $\sim 2\;$kpc.  The resulting strong radial
migration, called {\bf churning}, has implications for abundance
gradients and age-metallicity relations.  The apparent metallicity
gradient is also {\bf blurred} by epicyclic motions, which can readily
be subtracted for an individual star without having to integrate the
orbit \citep[\eg][]{Yu12}, since the guiding center radius of a star is
determined only by its angular momentum (Eq.~\ref{eq.rguide}).

The topic of radial migration is bedeviled by the fact that many
authors have conflated the process first described by \citet{SB02}
with the general redistribution of angular momentum that occurs with
any nonaxisymmetric disturbance.  The fact that spirals, bar
formation, the excitation of responses by satellites, \etc,
redistribute angular momentum across the disk of a galaxy had been
understood for decades.  However, the recently discovered changes at
corotation of a transient spiral mode have the two unique properties
of neither heating the disk nor causing it to spread, as described
next.  All other forms of angular momentum transport do both these
things.  Even processes that are not associated with lasting angular
momentum changes, such as increasing epicycle motions or of trapped
particles that cross and recross a resonance, have been described as
``radial mixing.''  While this phrase may be too deeply embedded to be
redefined, ``radial migration'' and ``churning'' are less widely used
and I suggest these terms be reserved to describe only the changes at
corotation of a transient spiral mode.

While these other processes may play a role in redistributing matter
radially, the low velocity dispersion of disks limits the extent to
which heating at Lindblad resonances can have occurred, as discussed
in \ref{sec.sh}.  Churning by spiral waves over the lifetime of the
disk could, in principle, cause far more mixing with very limited
heating, as found by \citet{SB02} and \citet{SSS12} (see
Fig.~\ref{fig.SSS12}).

\subsubsection{Mechanism of radial migration}
Stars near corotation move slowly with respect to the spiral
perturbation and therefore experience almost steady forcing from the
wave, which allows large changes to build up -- a process that is
analogous both to surfing on ocean waves and to Landau damping in
plasmas, although the consequences differ.  Stars orbiting just behind
the density excess are attracted forward by it and therefore gain
angular momentum.  However, the result of gaining angular momentum is
that the star moves onto an orbit of greater guiding center radius
(Eq.~\ref{eq.rguide}), and its angular frequency about the center
therefore decreases.  If the star were just inside corotation and
therefore gaining on the density excess, the change can cause it to
rise to a radius just outside corotation where it begins to fall
behind.  This behavior is described as a horseshoe orbit.  Conversely,
stars just ahead of the perturbation are pulled back, lose angular
momentum and sink inwards, where they orbit at higher frequency.
Those outside corotation, where the perturbation gains on them, could
lose enough angular momentum to cross corotation and begin to run
ahead of the wave.  As long as the gradient $\partial f/\partial L_z$
is fairly shallow, roughly equal numbers of stars gain as lose, and
they largely change places.  The process affects stars with small
peculiar velocities most strongly, since larger epicyclic motion leads
to less coherent forcing by the spiral potential.

Were the spiral potential to maintain a fixed amplitude, stars on
horseshoe orbits would be described as trapped.  As they are moving
slowly with respect to the wave, it would take them a long time to
reach the next density maximum where the changes just described would
be exactly undone.  However, if the amplitude of a transient spiral
mode has decreased by the time the star reaches the next density peak,
it may no longer be trapped and will continue to move with a lasting
change to its angular momentum.

Adopting variables suited to motion near corotation of a steady
potential perturbation, \citet{SB02} found that the radial extent of
the region where these horseshoe changes occur varies as the square
root of the perturbation amplitude, and therefore widens as a
perturbation grows.  At the same time, the more distant ``trapped''
stars move more rapidly in the frame of the perturbation, and the
shortest period of a trapped star decreases as the inverse square root
of the potential amplitude.  They found the spiral was strong for less
than half the horseshoe period for most trapped stars, which therefore
undergo a single change.

Horseshoe orbits are also responsible for limiting the amplitude of
the spiral.  For a disturbance to grow, the response of the stars to
the growing potential must reinforce the perturbed density, at least
until it saturates.  \citet{SB02} also argued that the maximum
amplitude of a spiral is limited by the widening horseshoe region
where stars are driven away from, instead of toward, the density
maximum.  This change kicks in suddenly because growth is linear in
the disturbance potential, but the horseshoe region grows as its
square root.

\subsubsection{Other radial mixing processes}
Any process that redistributes angular momentum mixes stars and gas
that originated at different radii, and even the transient spiral
modes that churn the disk also transport angular momentum to a much
lesser degree.  However, in contrast to the changes brought about by
churning, redistribution of angular momentum across the disk always
increases random motion and alters the large-scale surface density
profile of the disk.

The role of bars as agents that mix the stars and gas of a disk has
long been recognized \citep[\eg][]{Hohl71}; \citep{FBK94}, and has
received intense recent attention.  The possible effect of resonance
overlap between the bar pattern and the outer spiral was raised by
\citet{Quil03}, see also \citet{MQ06}, and developed by \citet{MF10}
for test particles in assumed nonaxisymmetric potentials of plausible
bars and spirals.  Since they adopted perturbations that grew to
steady amplitudes, they clearly were not exploring the process
described by \citet{SB02}.  Instead they found, as had
\citet[][although without comment]{LC91}, that single perturbations
created regions in which particles were simply trapped to cross and
recross corotation with minimal angular momentum changes at the
Lindblad resonances.  However, simulations with two imposed
disturbances revealed chaotic behavior when resonances of the two
patterns overlapped, which \citet{MF10} described as ``nonlinear''
interaction.  The substantially greater changes in the angular momenta
of the particles were also associated with disk heating \citep{MQ06}.
These studies raise the possibility that additional angular momentum
transport could even occur were galaxies able to support multiple
long-lived nonaxisymmetric structures.  \citet{Quil09} also used the
test-particle technique to show that the orbits of disk particles are
``mixed'' when perturbed by an orbiting satellite.

Results from $N$-body simulations are of greater interest, since the
perturbations that cause the angular momentum changes are generated
self-consistently and have physically reasonable time dependence.
\citet{BC11} calculated diffusion coefficients in disks that form bars
and spiral patterns finding, as seems reasonable, that angular
momentum changes are lower in disks with higher $Q$
(Eq.~\ref{eq.Qdef}).  \citet{Minc11, Minc12} report simulations that
formed bars and spirals, in which they claim evidence of enhanced
angular momentum changes due to resonance overlap.  They also
highlight disk spreading, which is largely due to the angular momentum
changes during bar formation (see \S\ref{sec.barorig}), and also
prolonged changes at the corotation resonance of a bar, which are
likely caused by particles that are trapped to cross and re-cross the
resonance and therefore cannot sensibly be described as something as
irreversible as mixing.

\begin{figure}[t]
\begin{center}
\includegraphics[width=\hsize, clip=]{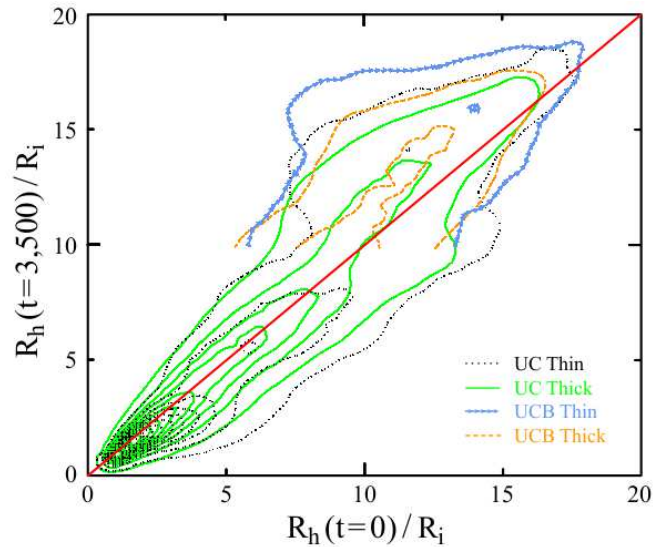}
\end{center}
\caption{Comparison of initial home radii with home radii after $\sim
  10\;$Gyr of evolution for particles in two simulations by
  \citet{SSS12}.  For Milky Way scaling, the radial unit is
  $0.75\;$kpc.  Simulation UC did not form a bar, whereas a bar did
  form in simulation UCB, and contours are drawn separately in
  different colors and line styles for the initially thin- and
  thick-disk populations.  The barred region is omitted because $R_g$
  (Eq.~\ref{eq.rguide}) cannot be defined in a strongly
  nonaxisymmetric potential.}
\label{fig.SSS12}
\end{figure}

\subsubsection{Radial migration in simulations}
\citet{Rosk08a} computed an isolated gas plus stars galaxy model that
tracked star formation and metallicity evolution.  The churning of the
disk by a succession of transient spiral modes caused extensive
migration.  Some heating was caused by the smaller changes at the
Lindblad resonances of spirals in their models, which is unavoidable,
and they reported a change in the gradient of mean stellar age near
the outer edge of the disk, which must have been created by the
outward migration of particles.  \citet{Rosk08b} went on to
demonstrate that migration led naturally to an age-metallicity
relation similar to that in the Milky Way.  A later work
\citep{Loeb11} studied the formation of a thickened disk comprised of
outward migrating stars with enhanced [$\alpha$/Fe] ratios and lower
mean orbital speed.  \citet{Bird11} found that mixing is more
extensive when spiral activity is invigorated by star formation,
although the level of spiral activity in their models depended
strongly on the ``gastrophysical'' prescription adopted.  They showed
that migration persists even for particles with large oscillations
about the midplane, and they determined migration probabilities from
their simulations.

Figure~\ref{fig.SSS12} \citep[from][]{SSS12} shows that the home radii
of stars can migrate either inwards or outwards by many kpc, while the
formation of a bar causes some comparatively mild additional mixing.
Radial migration in the thick disk is only slightly weaker than in the
thin disk, because the spiral potential that drives migration decays
only slowly away from the plane.  In fact, the slightly smaller
average changes of the thick-disk particles were probably caused more
by their larger in-plane random velocities than by their greater
vertical oscillation \citep[][their Fig.~10]{SSS12}.  This is the
likely reason that \citet{Bird13} found that radial migration was less
effective in the hotter, and generally older, particles of the inner
disk of their model.

\citet{Rosk12} presented a detailed study of radial migration in their
simulations.  They identified the locations, relative to the spiral
density maxima, of particles that gained or lost large amounts of
angular momentum and confirmed that even particles that moved rapidly
over a large radial distance remained on near-circular orbits.

\subsubsection{Adiabatic invariants}
One of the advantages of using action-angle variables to describe the
motions of stars (\S\ref{sec.actions}) is that the actions are
adiabatic invariants.  Broadly, this means that they are conserved
quantities when the orbit of the star is subject to slow changes,
except where resonances arise (see \BTii, \S3.6 for a more careful
statement).  For example, \citet{LB63} used the invariance of radial
action of a star to argue that orbital eccentricity would be invariant
during gradual changes to the potential with no change to the angular
momentum of the star.  However, transient spirals change the angular
momentum of stars at corotation without changing the radial action,
and therefore eccentricity is not invariant during these changes.

Nevertherless, the radial action, $J_R$, is a useful adiabatic
invariant during disk evolution except, of course, from the
nonadiabatic changes at Lindblad resonances, where $l=\pm1$ in
Eq.~(\ref{eq.Lchange}).  For nearly circular orbits $J_R \rightarrow
\kappa a^2/2$, with $a$ being the radial excursion of the star
\citep{LBK72}.  For a well-mixed set of stars of fixed $J_R$, we have
$\langle a^2 \rangle = 2\langle v_R^2 \rangle/\kappa^2$ in the
epicycle approximation, and therefore $J_R \approx \langle v_R^2
\rangle /\kappa$.  Thus, during radial migration $\langle v_R^2
\rangle \propto \kappa$ for this group of stars, \ie\ their radial
dispersion decreases during outward migration, as found by
\citet{MFQ12}, and {\it vice versa}.

\citet{SSS12} also showed that vertical action $J_z$, and not vertical
energy, is the conserved quantity when stars migrate.  Their
conclusion stood out far more clearly when vertical action was
calculated exactly (Eq.~\ref{eq.actdef}), than when the simple
epicycle approximation was used \citep[\eg][]{MFQ12}.  A somewhat
puzzling finding by \citet{SSS12} was that $J_z$ was conserved on
average, but not precisely for individual particles; which may have
been due to gradual relaxation that afflicts all $N$-body simulations
of disks \citep{Sell13b}.

Since its vertical action is adiabatically invariant during migration,
the vertical oscillation amplitude of a star varies with the strength
of the vertical restoring force, which in turn changes with disk
surface density.  A group of particles that have migrated outward in a
nongrowing disk of declining surface density must have an increased
scale height and a decreased velocity dispersion \citep{SB12};
\citep{RDL12}.  The vertical thickness would be squeezed by the
increasing mass density that occurrs in a growing disk, but younger
stars formed in the outer disk must still reside in a distinctly
thinner layer than that of the outward migrating stars.

\subsubsection{Tests for radial migration in the Milky Way}
Currently, there has been no decisive test to confirm that radial
migration really does occur in the Milky Way disk or elsewhere.  But
there are a number of strands of indirect evidence to suggest that the
mechanism does occur.

\citet{Hayw08} found evidence for radial migration in a study of the
metallicity distribution of solar neighborhood stars that suggested
some stars were formed elsewhere in the disk.  \citet{Lee11} claim
evidence for radial migration in the thin disk on the grounds that
metallicity is uncorrelated with orbital eccentricity, but found a
decreasing orbital velocity with metallicity in the thicker disk.
\citet{Yu12} reported a decreasing metallicity gradient with age of
Milky Way thin-disk stars, which is the expected consequence of radial
migration.

\citet{Bovy11b} corrected data from SDSS for the selection function of
the survey to determine the properties of the underlying stellar
population.  They found a continuous distribution of
abundance-dependent disk structure with increasing scale height and
decreasing scale length which they argued strongly favors
``inside-out'' disk formation combined with gradual internal evolution
through mechanisms such as radial migration.  Both \citet{Bens11} and
\citet{Chen12}, in separate studies of quite different stellar
populations, also found a short scale length for thick-disk stars,
which are less well mixed because these populations are dynamically
hot.

\citet{Hayw12} concluded that the metallicity distribution of disk
stars seemed consistent with some degree of migration, but drew
attention to a number of puzzling features.  He cited very tentative
evidence of a steplike feature in the radial distribution of
metallicities \citep{Hill12} that, if confirmed, suggests that
migration in the Milky Way may not have been efficient.  However,
\citet{Yong12}, in a study of open star clusters in the disk of the
Milky Way, did not find a discontinuity in the abundance gradient at
the solar radius, but instead found evidence for a change to a
shallower slope at around 12~kpc.

Much stronger tests will emerge from current and future surveys.  The
{\it Gaia\/} mission \citep{Perr01} will yield the kinematics of stars
over a large fraction of the Galaxy in overwhelming detail.  More
detailed kinematics and chemical abundance measurements are being, or
will soon be, collected in surveys such as APOGEE \citep{Alle08},
LAMOST \citep{Deng12} and ARGOS \citep{Free13} -- see also
\citet{RB13}.  Indeed, one of the principal goals of the HERMES survey
\citep{Blan10} is to unravel the history of radial migration in the
Milky Way.

\begin{figure}[t]
\begin{center}
\includegraphics[width=\hsize]{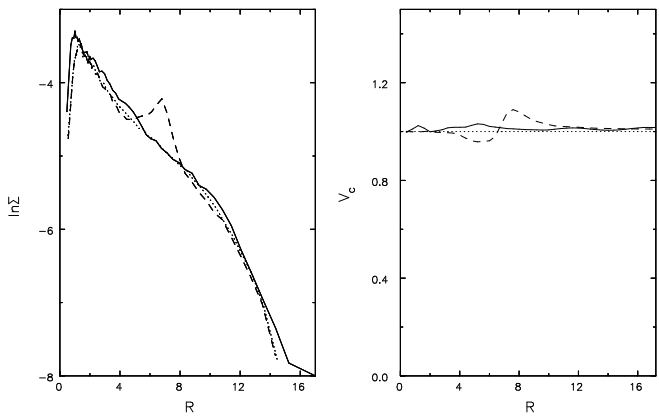}
\end{center}
\caption{The evolution of the surface density (left panel) and the
  total rotation curve (right panel) in the simulation described in
  \S\ref{sec.smooth}.  The dotted lines show the initial unperturbed
  Mestel disk, the dashed lines are drawn after a ring of material is
  added to the disk centered on $R=7$, and the solid lines show the
  distribution five rotation periods later.}
\label{fig.bump}
\end{figure}

\subsection{Smoothing rotation curves}
\label{sec.smooth}
The rotation curve, or circular speed as a function of radius, is
remarkably smooth for most galaxies \citep[see][for a somewhat dated
  review]{SR01}.  There is no feature even where the central
attraction shifts from being baryon-dominated to dark
matter-dominated, which \citet{BC85} described as a ``disk-halo
conspiracy.''  A few authors \citep[\eg][]{Kaln83}; \citep{Kent86};
\citep{PW00} have drawn attention to ``bumps and wiggles'' in
long-slit rotation curves, some of which correspond to photometric
features in the light profile.  While this is undeniable evidence for
significant mass in the disk, the underlying cause of these
small-scale features may be spiral arm streaming rather than
substantial fluctuations in the radial mass profile of the disk.

Spiral instabilities may also be responsible for featureless rotation
curves, as first argued by \citet{LH78}.  While running simulations
with a different purpose, \citet{SM99} noted that as the disk grew in
their models with a dense central mass and a (rigid) cored outer halo,
the mass distribution in the disk rearranged itself such that the
resulting rotation curve was remarkably featureless.  They noted that
they obtained this result with a number of differing rules for the
angular momenta of particles added to the disk.

A more controlled example is illustrated in Fig.~\ref{fig.bump}
(Sellwood, in preparation).  In this simulation, the initial model was
the Mestel disk that has a circular speed independent of radius shown
by the dotted line in the right panel and, in this case, only
one-third of the central attraction is from the disk with remaining
two-thirds due to a rigid halo.  The dashed lines show the
consequences to the surface mass profile and rotation curve of adding,
over a period of less than one disk rotation, an extra ring of matter
composed of live particles to this archetypal featureless model.  The
model quickly developed strong spiral patterns and after just five
rotation periods, the rotation curve and the surface density
distribution became featureless again, as shown by the solid lines.

The spirals that developed in this model were the result of two
unstable modes that were provoked by the density ridge.  Local
stability analysis of a disk with a ridgelike density feature
\citep{SK91} predicts that, for each sectoral harmonic, the normal
modes are wave pairs with corotation on opposite sides of the ridge.
However, only those wavelike distortions to the ridge that can excite
a strong supporting response from the surrounding disk are unstable.
The angular periodicity that excites the strongest supporting response
depends on the $X$ parameter of swing-amplification theory
(Eq.~\ref{eq.Xdef}), and the most rapidly growing pair of modes is for
$m=3$ for the disk mass in this simulation.  As the amplitudes of the
modes rise, horseshoe orbits (\S\ref{sec.rm}.1) develop at both
corotation resonances but, unlike in a featureless disk, the presence
of the ridge causes the resulting $L_z$ changes to be strongly out of
balance at corotation for both modes.  Thus far more particles are
removed from the ridge than are added to it, causing the density
profile of the disk to flatten, as shown by the solid curves in
Fig.~\ref{fig.bump}.

Thus it seems that the distribution of angular momentum in the
baryonic material that makes galaxy disks does not need to be able to
account for the featureless character of most galaxy rotation curves,
and small-scale variations in any reasonable distribution will be
erased by spiral activity.  The experiments of \citet{SM99}, together
with results from more realistic modern simulations \citep{Abad03,
  Ager11} hint that this effect may be substantial enough to control
the overall shape of the rotation curve, although further work is
needed to establish this more interesting conclusion.

\subsection{Angular momentum redistribution}
\label{sec.Lchange}
For a star in a rotationally-supported disk, the angular momentum $L_z
\sim RV_c$ is much greater than the typical radial action $J_R \sim
a\sigma_R$, with $a$ being the radial excursion.  Thus, $J_R \ll L_z$,
and Eq.~(\ref{eq.Lchange}) therefore requires that $\Delta L_z \ll
L_z$ at Lindblad resonances.  The implication is that spirals do not
cause large changes to the distribution of angular momentum among the
stars of a disk, as was borne out in $N$-body simulations
\citep{SJ79}; \citep{Bird13}.  The largest changes to the distribution
of $L_z$ in a disk occur during bar formation (\S\ref{sec.barorig}),
which has long been known \citep{Hohl71} to create a high level of
random motion although, even after this event, $J_R \ll L_z$ in the
outer disk.

This conclusion is not inconsistent with the large changes in $L_z$ at
corotation, that cause stars to diffuse through the disk with little
heating (\S\ref{sec.rm}).  Changes at corotation cause stars largely
to exchange places, with only a very minor net change (if any) to the
large-scale distribution of angular momentum within the disk.  Even
the smoothing of features in the rotation curve (\S\ref{sec.smooth})
causes only a {\it localized\/} smoothing of the angular momentum
distribution.

Note that in growing galaxy disks, the specific angular momentum of
infalling gas is expected to rise over time -- the so-called
``inside-out'' growth of disks \citep[\eg][]{MF89}; \citep{SG03};
\citep{Bird13}.  The distribution of angular momentum among the stars
formed from this material is expected to differ from that of the old
disk.  In this case, the total distribution of angular momentum within
the disk changes for a quite different reason.

Thus the dynamically cool disks of stars in galaxies testify that
their {\it large-scale\/} distribution of angular momentum cannot have
been greatly altered from that at the time the stars formed.  This
constraint from random motion does not apply to redistribution within
the gas component, however, since random motion in the gas is quickly
dissipated.

\section{Disk thickening and survival}
\label{sec.thick}
The disk of the Milky Way contains both a thin layer of young stars
and a thicker layer of old stars.  For a long time they were described
as separate components, with intermediate age stars being included as
part of the ``thin'' disk \citep{GR83}; \citep{LC00}; \citep{Munn04};
\citep{Juri08}; \citep{Ivez08}.  However, \citet{Bovy11a} suggested
that there is no clear distinction between the two populations but
rather a continuous variation in thickness, metallicity and radial
scale-length, with the oldest, most metal-poor and hottest component
having the shortest radial scale length.  Others have challenged this
conclusion \citet[\eg][]{Bens13}, arguing that the thick disk is a
distinct component.  Whichever way this discussion is settled, the
thick and thin terminology remains useful to distinguish the two ends
of the thickness range.

\citet{Burs79}, \citet{Moul05}, and others found evidence for a
thicker layer of older stars in other galaxies, which \citet{YD06}
suggested may be more massive, relative to the thin disk, in lower mass
galaxies.  Furthermore, \citet{Come11} suggested that the thick
components may be more massive than previously believed.

The thin and thick disks of the Milky Way can be distinguished not
only by their scale heights and velocity dispersions, but the thick
disk lags in its net rotational velocity \citep{CB00}, contains older
stars with lower metallicities \citep{Maje93}, its stars have enhanced
[$\alpha$/Fe] ratios \citep{BF05}; \citep{Redd06}; \citep{Fuhr08};
\citep{Ruch11}; \citep{Schl11}; \citep{LvdV12}.  As noted
\citet{Bens11} and \citet{Chen12} suggest a shorter radial scale
length for the $\alpha$-enhanced thick disk, although their estimates
are still quite uncertain. These distinctions are not clear cut, and
the assignment to a population may depend somewhat on whether a
spatial, kinematic, or chemical abundance criterion is applied
\citep{Fuhr08}; \citep{SB09b}; \citep{Loeb11}.

\subsection{Formation of thickened disks}
\label{sec.data}
Whether the thick disk is or is not a separate component has important
implications for its formation.  A distinct component suggests some
event, such as a minor merger (see \S\ref{sec.thin}) in the past,
stirred up the old disk and what is now described as the thin disk
began to form subsequently through gas accretion and star formation,
creating two chemically and dynamically distinct populations.
However, evidence of such an event could well be obscured by one or
more of a number of other mechanisms that may also contribute to the
currently observed properties.

In addition to the minor merger hypothesis, \citet{Abad03} proposed
that the debris accreted from disrupted satellite galaxies could form
part of the thick disk, but chemical analysis of thick-disk stars
\citep{Ruch10, Ruch11} argued against this suggestion.  \citet{Broo05}
and \citet{Bour09} have suggested that stars formed in a thicker gas
layer during galaxy assembly could have given rise to a thick disk.  A
fourth suggestion is that stars migrating outward from the inner
Galaxy would have a thick distribution.  Both the simulations of
\citet{Loeb11} and the semianalytic model for Galactic chemical
evolution that includes radial migration by \citet{SB09a}, showed that
outward radial migration of old stars from the inner disk can create a
thick population of old, metal-poor, stars with enhanced [$\alpha$/Fe]
ratios.  \citet{SB09a, SB09b} and \citet{Scan11} pointed out that it
naturally gives rise to both a thin and a thick disk, under the
assumption that thick-disk stars experience a similar radial churning.
This assumption was validated by \citet{SSS12}.

\citet{Sale09} proposed a test, based on orbit eccentricity, to
distinguish these formation mechanisms, that has not proven decisive
\citep{Dier10}; \citep{Case11}; \citep{Wils11}, although it does
disfavor the accretion scenario.  In this context, it should be noted
that the peculiar velocity components, even of stars having highly
eccentric orbits, could be redirected by GMC scattering
(\S\ref{sec.clouds}), weakening the power of such tests.  Furthermore,
it is likely that more than one of these mechanisms has been at play
as the disk of the Milky Way has built up.

\citet{ST96} suggested that the thick disk was formed by
``levitation'', in which radially eccentric in-plane orbits were
converted to near-circular inclined orbits through resonant trapping
as the potential of the Galaxy became more flattened during disk
growth.  However, the observed orbital eccentricities in the thin and
thick disks today are the other way around.

\subsection{Survival of thin disks}
\label{sec.thin}
The hierarchical model of galaxy assembly (\S\ref{sec.galform}) is
challenged by the thinness of disk galaxies \citep{TO92}, which are
stirred and thickened by the infall of satellite galaxies
\citep{Quin93}; \citep{Walk96}; \citep{VW99}; \citep{Bere03};
\citep{Read08}; \citep{Vill08}; \citep{Kaza09}.  The severity of the
challenge to the current $\Lambda$CDM paradigm involves many questions
that are not easily answered.  \citet{Wyse09} summarizes the evidence
that the thick disk of the Milky Way, and perhaps that of other
galaxies \citep{Moul05}, contains essentially no stars younger than
$\sim 10^{10}$~yr.  If this critical piece of evidence holds up, it
implies that no gravitational disturbance to the disk could have
scattered stars into the thicker layer throughout that time.

The survival of the so-called superthin galaxies \citep[see
  \eg][]{Matt00} presents a similar challenge.  They are believed to
be low-surface-brightness galaxies viewed edge-on that are probably
embedded in a massive halo.  If the low-luminosity density represents
a low disk mass density, then their disks are less coherently held
together by their self-gravity than are normal disks, making them all
the more fragile to gravitational perturbations.  Thus, not only are
these disks remarkably thin, but they would be more easily thickened
by perturbations than would heavier disks.

The expected rate of infall of subhalos as a function of their mass
can be estimated from simulations of the growth of dark matter halos
in the appropriate cosmology \citep[\eg][]{Purc09}.  However the
infalling pieces of substructure can be tidally disrupted, and may
merge into the smooth inner halo \citep{Gao11}.  The Sagittarius
stream \citep[\eg][]{Belo06} provides a clear example of the tidal
stripping of a satellite as it falls into the Milky Way halo.

If the core of a dwarf galaxy is dense enough to survive until it
interacts strongly with the disk, it may deposit some of its orbital
energy into the disk, the remainder being absorbed by the halo through
dynamical friction (\S\ref{sec.dynfr}).  A proper calculation of this
process needs to take into account the damping of the vertical
oscillation by dynamical friction \citep{QG86}, the reorientation of
the disk plane in response to the absorption of misaligned angular
momentum \citep{HC97}, and the excitation of bending waves that can
travel some distance across the disk before depositing their energy
into vertical random motion \citep{SNT98}.  The coherence of the disk
needed to support these last two mechanisms depends both on its
self-gravity and on the degree of random motion \citep{DS99}.

The simulations by \citet{Kaza09} reveal that the disk is
significantly distorted and thickened by the infall of a sequence of
massive subclumps.  The larger clumps which arrived first caused the
most disruption, while the smaller fragments did less damage.  All
three velocity components of the disk particles rose substantially,
while the disk also developed a pronounced flare.

Many \citep[\eg][]{Most10}; \citep{Vill10}; \citep{Puec12} have
pointed out that gas infall subsequent to a minor merger can form a
new thin disk, and that the attraction of the additional mass in the
disk squeezes the thickened layer of older stars.  However, stars
formed prior to the merger remain in a thickened layer \citep[see
  \eg][]{Broo04}; \citep{Robe06}; \citep{Gove09}.  The Milky Way may
have a continuum of disk populations of increasing thickness and age
\citep{Bovy11a}; but if the conclusion of \citet{Wyse09} that the
thickest subcomponent contains no stars with ages $\ltsim 10^{10}\;$yr
holds, then the disk could not have been gravitationally stirred for
all that time.  Such a constraint would present a significant
challenge to current cosmological models.

\subsection{Challenge to radial migration models}
The old age of the thick disk also raises a challenge for radial
migration models \citep{SB09b}.  If some thick-disk stars have
migrated from the inner Milky Way through the action of spirals, why
are they {\it all\/} so old?  One might expect at least a tail of
young stars that have migrated rapidly from the center, although there
were very few in the simulation by \citet{Loeb11}.  Solway and Sellwood
(in preparation) suggest that the formation of the bar in the Milky
Way prevented any stars born in the inner Galaxy from being caught up
by the corotation resonance of spirals and carried to the outer disk.
If this suggestion is correct, then we may be able to date the
formation of the bar in the Milky Way from the oldest stars that have
inner disk metallicities.  Other processes may have added stars to the
thick disk, but there is at least hope that the abundance ratios of
some elements might be unique signatures that the star originated in
the inner Milky Way.

\section{Bars}
\label{sec.bars}
A majority of disk galaxies contain a bar of some kind.  Bars are
clearly visible in some 30\% of galaxies, as judged from SDSS galaxy
images by the Galaxy Zoo project \citep{Mast11}.  A larger bar
fraction is seen in near-infrared images \citep[\eg][]{Eskr00};
\citep{Mene07}, in part at least because bars can be obscured by
star-forming regions in later type galaxies \citep[\eg][]{BW91}.
The bar fraction rises still further when weak oval distortions and
short bars are included \citep[\eg][]{MJ07}; \citep{Rees07}.  Yet
even in these studies, some 30\% of disk galaxies in the local
universe still lack any trace of a bar.

The incidence of bars over cosmic time has been investigated in a
number of studies, which face difficulties of morphological
classification from small images, even with Hubble Space Telescope
resolution, and of band shifting of the light distribution with
redshift.  \citet{Came10}, who reviewed previous work, concluded that the
bar fraction in more massive galaxies has been constant since $z \sim
0.6$, but has increased in lower mass galaxies by about a factor of 2
over the same time interval.  \citet{Shet12} also found that bars are
less common in disturbed galaxies at high redshift.  These findings
seem consistent with a general picture that the bar fraction appears
to be set after galaxies form and settle \citep{KBM12}, which happens
earlier in more massive galaxies.  Bars are therefore believed to be
old, long-lived structures.

Bars are also believed to have a greater extent normal to the disk
plane than does the disk that hosts them, giving them a ``peanut'' shape
when viewed edge on.  Since we cannot see the face-on view in such
cases, the evidence to support this interpretation of box-peanut bulges
is indirect \citep[\eg][]{BA05}.  The inner Milky Way manifests such a
shape \citep[\eg][]{BS91}; \citep{Weil94}; \citep{Stru06}.  In fact,
its peanut shape is so pronounced that it is described as an
``X shape'' \citep{MZ10}; \citep{Nata10}; \citep{Ness12};
\citep{WG13}.

\begin{figure}[t]
\includegraphics[width=\hsize]{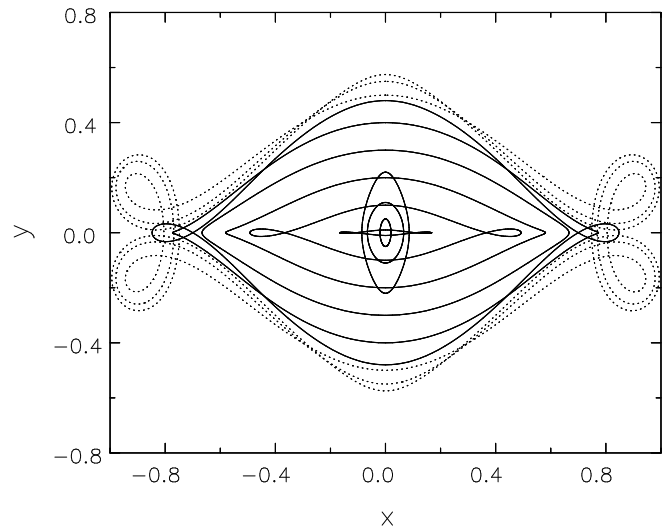}
\caption{The solid lines show examples, in a rotating bar potential, of
  important periodic orbits that close after two radial oscillations
  for every turn about the center: the 2:1 resonant families.  Those
  orbits elongated parallel to the bar axis (horizontal) are members
  of the $x_1$ family.  The $x_2$ orbits are elongated perpendicular
  to the bar.  The dotted lines show three 4:1 resonant orbits (which
  close after four radial oscillations for every turn about the
  center) that may contribute to the somewhat boxy appearance of
  many bars.}
\label{fig.x1x2}
\end{figure}

\subsection{Stellar dynamics of bars}
\label{sec.sdbar}
\citet{SW93} gave a thorough review of barred galaxies.  Although
somewhat dated, I refer the reader to that review for a detailed
account, and give only a brief outline of the basic dynamics of bars
here.

A self-consistent stellar bar has a nonlinear dynamical structure
that is most easily visualized in a frame that rotates with the bar.
A large fraction of the stars in a bar move in the forward sense in
the rotating frame on elongated orbits that are confined to the bar.
Clearly, moving forward in the rotating frame implies, in an inertial
frame, an orbit period about the center that is shorter than the
figure rotation period.

Most of the orbits within the bar occupy regular parts of phase space,
in which the stars are trapped about periodic orbits, and there are a
number of resonant families of such orbits.  Figure~\ref{fig.x1x2}
shows the most important orbit families in the midplane of an
idealized rotating bar, which has the simple effective potential
(\cf\ \BTii\ Eq.~3.103)
\begin{equation}
\Phi_{\rm eff}(x,y) = {\textstyle{1\over2}}v_0^2 \ln \left( 1 + {x^2 +
  y^2/q^2 \over R_c^2} \right) - {\textstyle{1\over2}}\Omega_p^2R^2,
\end{equation}
where $R^2 = x^2 + y^2$, $R_c$ is a core radius inside of which the
potential is approximately harmonic, $q \leq 1$ is the flattening, and
$v_0$ is the circular speed at large $R$ when $q=1$.  As in \BTii, the
values are $v_0 = 1$, $q=0.8$, $R_c=0.03$, and $\Omega_p=1$.

Periodic orbits are described as resonant because they close in the
bar-frame, and the solid curves show 2:1 families that close after two
radial oscillations and one full turn about the center.  The $x_1$
family is described as the ``backbone'' of the bar \citep{Cont80},
because the stars that are trapped around these orbits have a density
distribution that is more elongated than the adopted bar potential.
Since Poisson's equation requires the potential surfaces to be rounder
than the density surfaces, the orbits that make up a self-consistent
bar must be confined to a smaller region than is energetically
accessible to them.  This requires the majority of bar orbits to be
regular, since chaotic orbits fill the volume bounded by their energy
(or Jacobi constant, eq.\ \ref{eq.jacobi}).

Five Lagrange points occur in the bar potential, where a particle
could remain stationary in the rotating frame.  Four lie on the
corotation circle, in an infinitesimal barlike potential: L$_1$ and
L$_2$ lie on the bar major axis, L$_4$ and L$_5$ on the bar
minor-axis, while L$_3$ is at the bar center.  In strong bars, L$_1$
and L$_2$ are closer to the center than are L$_4$ and L$_5$.  The
L$_1$ and L$_2$ Lagrange points in the potential used for
Fig.~\ref{fig.x1x2} are very close to the points $(x, y) = (\pm1, 0)$,

Theorists generally agree that a self-consistent bar structure cannot
extend farther from the center than the major-axis Lagrange points,
because the stellar response outside corotation to forcing by a bar
potential creates a density distribution that is elongated orthogonal
to the bar direction.  In addition, one cause of chaos in phase space
is the overlap of resonances \citep{Chir79}, and the increasing
density of resonances (see ultraharmonic resonances in
\S\ref{sec.resnc}) as corotation is approached led \citet{Cont80} to
suggest that bars should end just before corotation.  \citet{Elme96}
defined the useful dimensionless parameter
\begin{equation}
\label{eq.Rdef}
{\cal R} \equiv R_L/a_B,
\end{equation}
which is the ratio of the distance, $R_L$, of the Lagrange point L$_1$
(or L$_2$) from the galaxy center to the semi-major axis of the bar,
$a_B$, and thus Contopoulos's argument is that ${\cal R}>1$ for all
bars.  While there is no compelling theoretical argument against
${\cal R} \gg 1$, which would be called a ``slow bar'', bar formation
models, and observational evidence from barred disk galaxies (see
\S\ref{sec.omegap}) both indicate a strong preference for ``fast
bars'' that have ${\cal R} \simeq 1.2$, although some exceptions have
been claimed.

A second family of 2:1 orbits, known as the $x_2$ family, is also
illustrated in Fig.~\ref{fig.x1x2}.  These orbits are found only deep
inside the bar and are elongated perpendicular to it.  They generally
appear only in models with dense centers, where the orbital periods
are short.  The forced response of near-circular orbits can be
calculated analytically for an infinitesimal bar perturbation
(\BTii\ eq.\ 3.147), from which it can be seen that the driven orbit
orientation differs by $90^\circ$ on opposite sides of all three major
resonances (\ie\ where $l=0,\pm1$ in Eq.~\ref{eq.res}).  Orbit
integrations are needed in strong bars, where the orientation shift
can be regarded as the generalization of the ILR to large amplitude
perturbations.  The existence of an ILR in the potential of the
azimuthally averaged mass distribution is a necessary, but not
sufficient, condition for the appearance of the $x_2$ family in a bar,
since it can disappear as the bar strength is increased \citep{CP80}.
Even though the influence of the resonance can be recognized from the
orbit structure, it is impossible to identify its location or radius
in a strong bar because some $x_1$ orbits, which align with the bar,
overlap spatially with the perpendicularly oriented $x_2$ orbits
(Fig.~\ref{fig.x1x2}).

The in-plane projections of many bars have distinctly ``boxy'' shapes
\citep{Korm83}; \citep{Atha90}; \citep{Gado11}, suggesting that the
parent orbits should have a somewhat rectangular shape.  (The 3D shape
is discussed later.)  The dotted lines in Fig.~\ref{fig.x1x2} show
three examples of 4:1 resonant orbits that are still elongated along
the bar, whose existence may be related to the boxy shape of bars.
Note that these orbits are found only close to corotation, where the
orbit period in the rotating frame is long enough to allow four radial
oscillations, and they therefore can be populated only in fast bars.

These 2:1, and perhaps also the 4:1, orbit families are the most
important for bar dynamics, but many other less important orbit
families have been found, even when motion is confined to a plane.
The dynamics of motion in the third dimension is considerably richer
\citep{PF91}; \citep{Pats02}; \citep{Skok02}, with multiple
commensurabilities possible between the vertical and in-plane
frequencies.  However, the most important orbits remain those that
resemble the $x_1$ family when seen in projection.  Those that
oscillate about the midplane in either an archlike structure or a
figure of 8 when viewed from the side are 2:2:1 resonant orbits
that complete two radial and two vertical oscillations for every turn
about the center.

Note that the bar pattern speed is equal to the precession rate of the
orbits that support it, and conversely all orbits precess at the bar
pattern speed.  In the absence of the bar potential, every orbit can
still be regarded as a precessing ellipse \cite{Kaln73}, but the
precession rates of the orbits would no longer be equal with the
larger orbits generally precessing at lower rates than the smaller.
Thus the effect of the bar potential is to impose a common precession
rate on orbits that would otherwise prefer to precess at a range of
different rates.  Clearly, the stronger the bar potential, the greater
its ability to trap orbits to precess with it.  Because the bar is a
self-consistent structure, its pattern speed can be regarded as an
average, or ``compromise,'' between the unforced precession rates of
the orbits from which it is built.  Thus bar pattern speeds must rise
if mass accumulates in the bar center, since an increased mean density
raises all orbital frequencies.

\subsection{The origin of bars}
\label{sec.barorig}
\subsubsection{Global bar-forming mode}
It has long been known from both numerical simulations \citep{Hohl71};
\citep{OP73} and global stability studies \citep{Kaln72, Kaln78};
\citep{Jala07}, that simple models of rotationally supported stellar
disks are globally unstable.  In the linear regime, the instability
takes the form of an open two-arm spiral but, as it saturates, the
shape straightens into a bar in the inner disk, while the spiral in
the outer disk winds up and disperses.  \citet{Hohl71} showed that the
instability causes considerable rearrangement of the angular momentum
in the disk,\footnote{Although the instantaneous angular momentum of an
  individual particle in a strongly nonaxisymmetric system changes
  continuously, particles in a settled bar stream round the bar in a
  steady fashion such that the distribution of their instantaneous
  values does not evolve.} and the azimuthally-averaged mass profile is
substantially changed, with the inner disk becoming denser, while
material is also spread far outwards.

The vigorous linear instability that creates the nonlinear bar is the
result of a cavity-type gobal mode, or standing wave, in a massive
disk.  The linear instability can arise only if the combined mass
distribution of the dominant disk, plus contributions from any bulge
and dark matter halo, yields a quasiharmonic potential over the inner
part, so that the rotation curve rises roughly linearly from the
center before flattening around a radius $\Rcore$.  In most
circumstances, the group velocity \citep{Toom69} of trailing spiral
waves is directed away from the corotation radius, while leading waves
propagate toward it, as illustrated for a centrally cusped potential
in Fig.~\ref{fig.dust}.  In that figure, the trailing disturbance was
absorbed at an ILR, but an inwardly propagating trailing spiral can
travel all the way to the galaxy center if it does not encounter this
resonance, and an $m=2$ disturbance easily avoids an ILR when the
potential near the center is quasiharmonic.  In that case, the
incoming trailing wave reflects off the center into an outgoing
leading spiral.  The feedback loop is closed at corotation where the
outgoing leading wave superreflects into an amplified trailing wave.
At the same time, an outwardly propagating trailing wave is excited
outside corotation that satisfies wave action conservation
requirements.  The mode is unstable because the wave train is
amplified at corotation, and the instability typically exponentiates
on the time scale of an orbital period.

The bar that results from this instability generally extends to a
radius that is perhaps 20\% -- 30\% greater than $\Rcore$.  The
initial bar has a pattern speed that is often slightly lower than that
of the eigenmode that caused it, and it almost fills its corotation
circle.  Thus bars are shorter in models with smaller harmonic cores
\citep{Sell81}, although something else can happen (see
\S\ref{sec.barnl}) if the core is very small or absent.

\subsubsection{Stabilizing mechanisms}
Because it grows through swing-amplification, this mode is highly
unstable in massive disks with low velocity dispersion, whenever the
feedback loop is open.  If the disk is massive enough to contribute
most of the central attraction, then $1 \ltsim X \ltsim 3$ for $m=2$
disturbances (Eq.~\ref{eq.Xdef}), and strong amplification occurs
unless $Q \gtsim 2$.

In mass distributions with quasiharmonic cores, \ie\ those that allow
ingoing waves to reflect off the center, the instability can be
quelled either by a high degree of random motion \citep{AS86} or by
making the disk unresponsive to $m=2$ disturbances, by making $X
\gtsim 3$ (Eq.~\ref{eq.Xdef}).  This latter solution is that favored
by \citet{OP73}, \citet{ELN82}, and by \citet{Chri95} who correctly
argue that bar stability can be achieved if a large fraction of the
central attraction over most of the inner disk comes from unresponsive
spherically distributed matter (bulge and halo).  However, this is not
the only, or even the most realistic, way to stabilize a dynamically
cool disk.

After having elucidated the mode mechanism, \citet{Toom81} predicted
that the formation of a bar could be prevented if the in-going $m=2$
wave were unable to reflect off the center of the disk.  The easiest
way to prevent feedback through the center is to ensure that the wave
encounters an ILR, where it will be absorbed  as illustrated in
Fig.~\ref{fig.dust}.

Simulations of models with centers dense enough to force an ILR for
most patterns \citep{Sell85}; \citep{SM99}; \citep{SE01} confirm that
Toomre's proposed mechanism can indeed stabilize a disk in
high-quality numerical work.  These globally stable galaxy models have
massive disks with realistic rotation curves and a moderate degree of
random motion.

\subsubsection{Bar formation through nonlinear trapping}
\label{sec.barnl}
However, \citet{ELN82} reported that bars formed in their simulations
with equal vigor no matter how dense they made the central bulge, in
apparent contradiction with Toomre's prediction.  \citet{Sell89b}
confirmed that bars formed in his similar simulations with dense
bulges when the initial arrangement of the particles was random.  But
he also found that the {\it same\/} models did not form bars when the
particles were uniformly spaced in azimuth around rings to reduce the
initial level of shot noise in the low-order sectoral harmonics -- a
{\bf quiet start}.  This different behavior arose because the
absorption of in-going spiral waves at an ILR is predicted to occur at
small amplitudes only.  The higher level of shot noise from randomly
placed particles seeds larger amplitude disturbances that can
overwhelm the ability of the ILR to absorb them.  The resulting
non-linear trapping of particles causes a bar to form that is
superficially similar to that formed through the global linear
instability.  Had \citet{ELN82} employed a much larger number of
particles, a less subtle way to beat down the level of shot noise,
they should have found that dense bulges can indeed stabilize a disk,
as Toomre predicted.  Thus the bulge and halo masses required by the
popular stability criteria proposed by \citet{OP73}, \citet{ELN82} and
\citet{Chri95} simply do not apply to galaxies with dense centers, or
to high-quality simulations of models with this property.

Shot noise from the $\sim 10^{10}$ disk stars in real galaxies must
have a low amplitude.  Larger density fluctuations caused by star
clusters and GMCs still seem unlikely to trigger nonlinear bar
formation.  However, this nonlinear method of making bars in globally
stable disks could still occur, for example, by a large amplitude
perturbation caused by a tidal encounter or minor merger (see
\S\ref{sec.barorig}.5).

\subsubsection{Slow trapping of orbits}
\citet{LB79}, in an elegant piece of dynamical analysis, proposed a
mechanism for the secular growth of bars in galaxies.  He showed that
eccentric stellar orbits could, under certain reasonable conditions,
gradually become trapped into a rather slowly rotating bar structure.
Orbits tend to align in the inner parts of galaxies, thereby
reinforcing the bar, when the overall density distribution is not too
sharply peaked toward the center.  \citet{LB79} envisaged that the
angular momentum loss from the inner part of the galaxy would be
mediated by spiral patterns, as may have happened in the simulations
of \citet{JS78}.  This mechanism can be important for the secular
growth of bars discussed below (\S\ref{sec.bargrow}).

Even though the mechanism was originally envisaged as a slow trapping
process, \citet{Poly04, Poly13} argued it may also form slow bars on a
dynamical timescale.  He argued that this was the cause of the weak,
slow bars that formed in the simulations described in the appendix of
\citet{AS86}, which had velocity distributions that were strongly
radially biased.  The mechanism has also been identified
\citep{Palm90} as the root cause of the radial orbit instability in
spheroidal stellar systems with radially biased DFs.

\subsubsection{Bar formation through tidal encounters}
A number of studies of tidal interactions of satellite galaxies with
disks have shown that bars are often triggered by the encounter
\citep{Byrd86}; \citep{Nogu87}; \citep{Geri90}; \citep{Salo91};
\citep{MW04}; \citep{Roma08b}.  If the unperturbed disk were stable,
bar formation could still occur through the nonlinear trapping
mechanism described in \S\ref{sec.barnl}.

\citet{MN98} suggest that tidally induced bars might be slower, in the
sense that ${\cal R} \gg 1$, than those formed through the usual bar
instability.  \citet{Curi06} also report that bars seemed to form more
readily in cosmologically formed halos with moving substructures than
in cases where the halo is smooth and nonevolving.  \citet{Bere07}
reported that interacting galaxy models with large gas fractions
appeared to be less susceptible to bar formation than their gas-free
counterparts, although the high numerical viscosity of the SPH method
(\S\ref{sec.flows}) may have had an undue influence on this
conclusion.  The variety of possible galaxy mass ratios, orbits, spin
directions, gas fractions, \etc, implies that the comparatively few
simulations so far reported have barely scratched the surface of this
vast multidimensional, parameter space.

Despite a number of studies to attempt to determine whether galaxies
with nearby companions or those in dense environments are more likely
to be barred, the results have generally been inconclusive.
\citet{Skib12} review this work, and present a much larger study of
their own, based on barred classifications from the Galaxy Zoo
project, that appears to find a significant excess of bars in galaxies
with moderately distant companions.  Much more theoretical work and
further observational studies \citep[\eg][]{Mend12} are required to
determine the extent to which bars in real galaxies could be caused by
interactions.

\subsubsection{Recurring bars?}
The changes to the distribution of both mass and angular momentum that
result from this global instability are the largest that occur in an
isolated disk \citep{Hohl71}; \citep{Deba06}.  However, for reasons
given below, bar formation through a global instability is widely
believed to happen just once in the life of most disk galaxies, and
the associated large structural changes are not expected to recur.

\citet{BC02} and \citet{Comb08} offer a dissenting view, and found
that the instability can recur in their simulations.  The process of
bar formation and dissolution (\S\ref{sec.dissol}) creates much random
motion, leaving the original stellar disk dynamically ``hot'' and
unresponsive.  However, gas settling onto circular orbits and forming
new stars creates a new, dynamically responsive, component.  Thus,
they invoke a high gas accretion rate in order that the whole disk may
again become bar unstable.  Some of their simulations did form a
second bar, in \citet{BC02} after the disk mass had roughly doubled,
and \citet{SM99} presented an additional case.

However, it is hard to see how such behavior could recur repeatedly,
since each cycle adds mass to the hot unresponsive disk population.
Also, angular momentum changes associated with earlier bar formation
will have made the disk more centrally concentrated, which is
stabilizing as \citet{SM99} demonstrated.  Furthermore, inside-out
disk growth suggests that fresh gas is expected to be accreted less in
the center, where it would most be needed, and more in the outer disk.

\subsection{Continued growth of bars}
\label{sec.bargrow}
A globally unstable bar-forming mode generally has an open spiral
form, which causes a large-scale rearrangement of the angular momentum
in the disk.  However, the spiral soon fades, leaving the bar as the
only persistent feature and the outer disk dynamically much hotter.
If the disk does not extend far beyond the bar, and the simulation
does not include a live halo or any dissipative component, then very
little further happens.

A more extended disk can support continuing spiral activity for a
period, which generally has a lower pattern speed than that of the bar
\citep{SS88}, and the duration of that activity can be extended,
perhaps indefinitely, by mimicking dissipation.  The apparent
connection between the bar and the spiral pattern in such cases must
be transitory, although \citet{SS88} showed that contours of the
nonaxisymmetric density distribution appear to join the bar to the
spiral for a significant fraction of the beat period.  Whether spirals
in real barred galaxies are driven responses to the bar or distinct
dynamical entities has proven harder to establish \citep{Buta09};
\citep{MRM09}.

Although spiral patterns in simulations generally have different
pattern speeds from that of the bar, the two nonaxisymmetric
structures do interact.  Generally, it is found that spiral activity
is associated with bar growth, since spirals remove angular momentum
from particles, allowing them to become trapped into the bar
\citep{LB79}; \citep{Sell81}; \citep{SSS12}.  This process causes the
bar to slow as it grows, because it adds material to the bar that has
a lower natural precession rate (see \S\ref{sec.sdbar}), and the
requirement ${\cal R} > 1$ continues to hold.  Note that spirals do
not always cause bars to strengthen and can sometimes cause them to
weaken, as described in \S\ref{sec.dissol}.

\citet{Tagg87} and \citet{MT97} argue that bars can drive spirals
through nonlinear resonance coupling.  In their picture, the location
of corotation of the bar coincides with the ILR of the outer $m=2$
spiral, which has a lower pattern speed, and the coupling is mediated
by a third mode, which may be axisymmetric ($m=0$) or $m=4$.  Similar
ideas were proposed by \citet{Fuch05}. There is no doubt that many
spirals in simulations have pattern speeds of approximately the
angular frequency for this to be a possible explanation, but the
evidence for the third mode that would confirm it has proved more
elusive.

Bars also grow in length due to dynamical friction with the halo
(\S\ref{sec.dynfr}), and growth by this process can be extreme
\citep{AM02}; \citep{Mart06}; \citep{VSH09}.  \citet{AM02} show that a
bar in a moderately dense halo continues to grow until it is as large
as the disk from which it formed!  Perhaps the mechanism proposed by
\citet{LB79} operates in this context also, with secular bar growth
caused by loss of angular momentum to the halo instead of to the outer
disk.  Of course, the bar slows as it grows in these cases also.
Since we do not observe bars of this size, in relation to their disks
\citep[\eg][]{Erwi05}, it seems reasonable to conclude that halo
friction, which is determined by the {\it inner\/} halo density
(\S\ref{sec.dynfr}), is too mild for excessive bar growth to occur in
nature.

\subsection{Buckling instability}
\label{sec.buckle}
After a bar has formed and settled, it generally experiences a second
instability that causes it to thicken out of the plane into a
pronounced peanut shape, as first reported by \citet{CS81}.
\citet{Comb90} and \citet{PF91} suggested that thickening is caused by
a vertical resonance, since gradual thickening also occurs in
simulations in which the buckling mode is suppressed by forcing
vertical symmetry of the potential about the midplane, but
\citet{FP90} concede that thickening is more rapid when buckling is
allowed.  The asymmetric bending of the bar when viewed edge-on in
many simulations \citep{Raha91}; \citep{OD03}; \citep{MS04};
\citep{Mart06} is a clear indication that a dynamical buckling
instability is the principal cause of the peanut-like shape.  The
buckling instability of a bar is believed to have formed the peanut
shape of the Milky Way bulge \citep[\eg][]{Shen10}; \citep{LS12};
\citep{GM12}, and additional kinematic data \citep{Vasq13} seem
to support this picture.

Buckling instabilities had been predicted for a stellar system with an
excessively flattened velocity dispersion ellipsoid \citep{Toom66};
\citep{Kuls71}; \citep{FP84}.  \citet{Arak85} showed that the
instability is present in a uniform stellar sheet with a
$\hbox{sech}^2(z/z_0)$ vertical profile provided $\sigma_z <
0.3\sigma_x$, and this criterion appeared to be roughly correct in a
global axisymmetric model \citep{Sell96}.  Simulations of the
nonlinear evolution of the instability in strongly prolate systems
\citep{MH91}, disks \citep{SM94}, and the rotating bars of interest
here, reveal that the flattened system develops an increasing bend in
the vertical direction until self-gravity is no longer able to confine
the particles to the bending layer; the nonlinear evolution is a
puffier system with less extreme velocity anisotropy.  In the case of
rotating bars, the disk in which the bar formed may have been quite
stable to buckling when axisymmetric, but the formation of the bar
creates an elliptical flow, with substantial streaming motion in the
radial direction that has the same destabilizing effect on the bending
dynamics as does random motion.

It is clear that the 2:2:1 resonant orbit family invoked by
\citet{PF91} is the reason that the bar takes on a peanut shape.
Orbits of this family dominate in rapidly rotating 3D bars
\citep[\eg][]{PF91}; \citep{Mart06}; in the rotating frame, they close
after two radial oscillations (as do the $x_1$ family in 2D) and two
vertical oscillations, with the vertical excursions peaking when the
particle is far from the center.

The buckling instability weakens the bar \citep{Raha91}; \citep{MS04};
\citep{Deba04, Deba06}, and causes it to become slightly more
centrally concentrated, as energy added to the vertical motions is
removed from the horizontal.  The peanut shape of the bar after the
instability may also be affected by the degree concentration of the
central mass: the central waist is more pronounced in models with a
quasiuniform inner density distribution, while the thickness is more
uniform when the central density is strongly peaked \cite{Bere07}.

The peanut shape generally does not encompass the full extent of the
bar, \ie\ there is some flat bar outside the buckled inner part
\citep{Lutt00}; \citep{Atha05}; \citep{Gado07}; \citep{ED13}, as has
also been claimed for the Milky Way \citep{MVG11}, with the ``long
bar'' seen in counts of the mid-IR sources \citep{Benj05} and variable
stars \citep{Gonz12}.  \citet{ED13} also argue that not every bar
thickens vertically and estimated that at least 13\% of bars in
galaxies have not buckled.

Unfortunately, the nature of the buckling instability in simulations
depends on spatial resolution (or the gravity softening length) used
in the $N$-body code.  Codes, such as that used by \citet{Raha91},
which do not have many zones or softening lengths within the vertical
thickness of the disk yield restoring forces to the midplane that are
not as sharp as they should be.  A soggy restoring force increases the
spatial scale of the instability, leading to the simple low-order
buckling mode reported by \citet{Raha91}.  Subsequent models with
better spatial resolution found that bars still thicken, but the
buckling occurs on shorter length scales, causing less pronounced
bends before the mode saturates.

The inclusion of a rigid mass component, especially a bulge or central
mass concentration that is held fixed, also compromises the proper
representation of the buckling instability.  Such models provide an
additional restoring force to the fixed center, whereas a fully mobile
mass distribution should move in response to the bend in the thin
component \citep{Bere07}.

\citet{Mart06} show that a bar that grows substantially in length may
undergo a second buckling instability.  \citet{AM02} also found that
the extent of peanut appearance grew significantly as the size of
bar continued to increase.

\subsection{Gas response to bar forcing}
\label{sec.flows}
As described in \S\ref{sec.ISM}, the interstellar medium (ISM) in
galaxies is not a simple fluid with a well-defined equation of state.
Thus before running simulations to model the gas flow, one must first
decide how best to approximate the dynamical behavior of the ISM.

One approach \citep[\eg][]{vAR81}; \citep{Pine95}; \citep{Kim12} is to
use a standard Eulerian hydrodynamic code with an isothermal equation
of state, adopting a sound speed that is representative of the
velocity spread of the clouds, typically between 5 and 10~km~s$^{-1}$
rather than the much lower thermal speed.  These well-developed
methods have the advantages of optional adaptive grid refinement
\citep[\eg][]{Krav02} and a low numerical viscosity, but they also
attribute a pressure that resists compression in a converging flow
where the physical properties of the gas suggest that we should expect
strong dissipation through some kind of bulk viscosity.

Lagrangian methods have also been applied, the most popular of which
is smooth particle hydrodynamics \citep[hereafter SPH, see][for a
  review]{Spri10a}.  The advantage of these methods is that they
concentrate numerical resources in the interesting regions of high
density, and self-gravity of the gas can readily be combined with the
scheme used for the stellar particles.  A known weakness of SPH is its
inability to support some standard fluid instabilities, especially the
Kelvin-Helmholtz instability \citep{Ager07}, but fix-ups have been
developed \citep{RH12}; \citep{Hopk13}.  Another weakness of all
Lagrangian methods, such as ``sticky'' particle and ``colliding''
particle schemes as well as SPH, is the high numerical viscosity due
to the finite radius (or kernel width) of the particles.  This is of
particular importance in spatially separated, but nearby
counter streaming flows, such as can occur in strongly
nonaxisymmetric potentials.  Two nearby streams of oppositely flowing
particles whose interpolation kernels overlap will clearly drag on
each other, causing viscous dissipation that may be greatly
over-estimated.

\citet{Spri10b} described a promising new hybrid Eulerian-Lagrangian
method that adjusts the grid cell boundaries as the fluid flows.
However, the number of published examples is so far rather small, and
most are applied to galaxy formation, rather than to galaxy evolution.

Since no one method perfectly mimics the dynamics of the ISM, it is
good to compare the behavior in any one problem using a variety of
techniques.  One has greater confidence in behavior that is
reproducible by more than a single method.

\subsubsection{Flows in two dimensions}
Many have reported simulations of a massless gas component flowing in
a rigidly-rotating bar potential \citep[a partial list of some the
  more important papers is:][]{SH76}; \citep{ST80}; \citep{MT80};
\citep{Schw81}; \citep{vAR81}; \citep{Atha92}; \citep{Kim12}.  These
simulations used a variety of approximations to model the gas, but
generally they found that gas within the bar region is driven inward
toward the galaxy center, where it accumulates, while gas in the
region outside corotation is driven outward.

Because the gas is moving highly supersonically, pressure is
negligible and, except where shocks arise, the motion of a fluid
element follows a ballistic orbit.  Therefore, were shocks absent,
mild dissipation would drive gas onto streamlines corresponding to
periodic orbits in the potential.  However, in most bar flows, the
periodic orbits do not nest without intersecting others or themselves,
as exemplified in Fig.~\ref{fig.x1x2}, and shocks must form.  Shocks
form along the leading edges of the bar as it rotates, causing the gas
to lose both energy and angular momentum.  The loss of angular
momentum occurs because the shocks skew the flow pattern with respect
to the axis of the bar, and therefore the gas spends more than half
its time on the leading side of the bar, where the nonaxisymmetric
part of the bar potential applies a retarding acceleration.  In all
models except those that lack a central mass concentration, the inflow
stalls at some distance from the center, which happens where the $x_2$
orbit family appears.

Observational evidence from barred galaxies \citep[reviewed in][]{KK04}
suggests that something like the behavior just described also happens
in nature.  \citet{Pren62} appears to have been the first to associate
shocks with the dust lanes that are generally seen on the leading
edges of the bar, assuming the outer spiral to trail.  Physically, a
shock in simulations of the idealized ISM implies, in real galaxies,
locations where streams of gas clouds undergo more frequent
collisions, causing a change in momentum, and a large increase in
density that gives rise to the dust lane.  Steep velocity gradients
across dust lanes can be detected in high resolution velocity maps
\citep[\eg][]{Wein01}; \citep{Hern05}; \citep{ZSS08} and massive gas
concentrations are observed in the centers of barred galaxies
\citep{Shet05}; \citep{Rega06}; \citep{Geri88}; \citep{GB91a}.  Large
accumulations of gas, presumably having been driven inwards by the
bar, are often found in circumnuclear rings \citep{GB91b} (see
\S\ref{sec.rings}).

Extracting a reliable estimate of the inflow rate of gas from
simulations is fraught with difficulties, however.  The high numerical
viscosity of some methods may enhance the inflow rate \citep{Pren83}
but, even more insidious, is that the precise position of the shock,
and therefore the magnitude of the gravity torque on the gas, is
strongly affected by the choice of numerical scheme and parameters.
\citet{Quil95} imaginatively took an observational approach to avoid
these pitfalls, although other difficulties arise associated with
accounting for all phases of the gas.

A small fraction of the radial flow may continue inward
\citep{Wada04}; \citep{Kim12}, perhaps driven by weak spirals that are
particularly prominent in dust \citep{Caro98}; \citep{Mart03}.  Again,
the inflow rate in a simulation depends strongly on the numerical
scheme and parameters \citep{Kim12} and \citet{KS12} found that
including magnetohydrodynamics substantially increases the inflow rate
in this region.  However, it is clear from the observed build up of
gas in the nuclear rings of real barred galaxies, that the inward mass
flux interior to the ring must be lower than that which flows down the
bar into the ring.

\citet{WK01} included self-gravity of the gas, as well as heating and
cooling.  But the more important limitation of most simulations
mentioned in this section is the neglect of the bar response to the
angular momentum gained or lost by the gas and the evolution of the
gravitational potential as mass accumulates in the center.

\subsubsection{Flows in 3D}
Most 3D studies of gas flows in bars have employed the SPH method with
an isothermal equation of state.  Since there are few results from
other 3D methods with which to compare, it makes sense to compare with
the 2D behavior, especially as no dramatically new features have been
reported that arise specifically from the freedom of motion in 3D.
Indeed, \citet{Pere08} showed that the flow velocities obtained by SPH
compared well with those from a 2D Eulerian method.

Another feature of added realism in most studies is that the
self-gravitating evolution of the SPH particles is combined with that
of stellar particles \citep[\eg][]{Bere98}; \citep{Fux99}.  This
aspect therefore implies that the simulations capture both the angular
momentum loss to the bar, and the changing gravitational potential as
the gas accumulates in the center.  However, while the flow patterns
are broadly similar to those seen in 2D models, a characteristic new
feature of many of these Lagrangian models is a high inflow rate of
isothermal gas to the center.\footnote{\citet{Deba06} found reduced
  inflow with an adiabatic equation of state because the gas is then
  more resistant to compression.  However, the pressure of even an
  isothermal gas may be unrealistic (see \S\ref{sec.ISM}).}  This
finding raises a concern that the quantitative inflow rate may be
substantially over-estimated because of the numerical viscosity
inherent in the SPH method.  An artifact of this kind will cause gas
to accumulate in the center too quickly, and the effects of the
central mass build-up, especially in gas-rich models, may occur too
rapidly.  Note that this concern is not over the physical process,
which surely does happen, but over the rate at which it happens in the
simulations.

\citet{Bere07} found that the build-up of a central mass concentration
due to gas inflow caused a slight increase in the bar pattern speed,
probably because the increased mass in the center raises orbit
frequencies.  They also found a reduced slow-down rate of the bar,
since the bar must take up the angular momentum lost by the gas as it
is driven inwards.  Another finding was that a significant gas
fraction altered the buckling behavior, which changed the final
strength and 3D shape of the bar.  In a follow-up study, \citet{VSH10}
found that moderate fractions of gas ($\ltsim 5\%$ of the disk mass)
have little effect on the behavior; the bar grew, slowed and buckled
pretty much as in a comparison stars only case.

\subsection{Bar dissolution}
\label{sec.dissol}
Bars in $N$-body simulations that do not include any dissipative
component or a live halo are long-lived structures \citep{MS79} that
are also quite robust \citep{SS87}.  But it has long been recognized
\citep[\eg][]{PN90} that the build-up of a central mass concentration
(CMC) at the center of the bar can change its dynamical structure.
Studies to determine the response of a bar to an imposed central mass
\citep{Norm96}; \citep{Shen04}; \citep{ALD05}; \citep{Deba06} have
generally found that the bar is weakened, but not completely destroyed
by a central mass as large as a few percent of the disk mass, and
still larger masses are needed to cause the bar to dissolve entirely.
The destructive power of a given mass is also increased by making it
more dense.  Note that a high central density requires that simulation
particle orbits near the center have short time steps \citep{Shen04},
and numerical errors in this regime can accelerate bar dissolution.

The CMC alters the gravitational potential of the bar, which in turn
requires the orbital structure to adjust.  \citet{Shen04} found that
the massive compact CMC in their model made large parts of phase space
chaotic,\footnote{\citet{GB85} predicted this consequence for
  nonrotating ellipsoidal galaxies.} causing an abrupt dissolution of
the bar.  Lower mass CMCs also caused some orbits to become chaotic,
weakening the bar after which the weakened bar continued to adjust
more gradually toward a new structure in the presence of the CMC.

A complete dissolution of the bar leaves the disk dynamically hot,
since the highly eccentric orbits of the bar no longer remain aligned
in a coherent streaming flow, but become randomly oriented.  The
process can be very rapid because of collective effects; the coherent
alignment of the bar orbits is maintained by the bar potential and, as
the bar weakens, the orbits of remaining stars are less strongly
constrained to precess at the original common rate.  Since the bar has
usually buckled by this time, the hot inner disk formed this way is
also quite thick.  Finally, a very dense central mass can scatter
orbits in any direction, and the stars could take up a spheroidal
shape, perhaps flattened slightly by the potential of the surrounding
disk \citep{Norm96}.  There should be observable consequences from
this sequence of events that could test the predictions of the
simulations.

\citet{BCS05} and \citet{Comb08}, who use sticky particles to
mimic gas, claim that the back reaction of the torque between the bar
and the gas can be strong enough to dissolve the bar.  They correctly
point out that gas inflow must add angular momentum to the bar which
should weaken it \citep{LBK72}.  However, the angular momentum
required to dissolve the bar should be at least roughly equal to that
it lost to the outer disk when it was formed, perhaps more if the bar
has been intensified through spiral activity or halo friction
(\S\ref{sec.bargrow}).  Thus gas inflow through the comparatively
small lever arm of the bar, in comparison to the outer disk, would
indeed need to be prodigious to supply the angular momentum to unbind
the bar.  \citet{Bere07}, for example, did find that the bar weakened
earlier as the gas mass fraction was increased, but they argued this
behavior was caused by the accumulation of mass into the center rather
than a back-reaction of the torque between the bar and the gas.

A third possible internal method to weaken a bar is an interaction
with an exceptionally strong spiral, which has occurred in a few
simulations \citep[\eg][]{SM99, SSS12}. \citet{LBK72} derived
Eq.~(\ref{eq.LBK}) by averaging over all phases, which they assumed to
be uniformly populated, leading them to the widely cited conclusion
that spirals remove angular momentum from the inner disk.  However,
the stars are far from uniformly distributed in azimuth near the end
of a bar and the fact that they are trapped in the bar further
invalidates, in this context, the assumptions that underlie the
derivation of Eq.~(\ref{eq.LBK}).  The behavior in this more complex
situation seems to depend on the relative phase of the spiral arm and
the bar.  For most of the cases when the bar leads, or is close to the
same phase, as the inner end of the spiral, the spiral can remove
angular momentum from nonbar stars which may allow them to become
trapped into the bar, thereby increasing the bar strength, as
described in \S\ref{sec.bargrow}.  On the other hand, when the spiral
density maximum significantly leads the bar, their mutual attraction
adds angular momentum to stars in the bar, which weakens it.  This
behavior has not been studied in detail, and further work is required
to understand it and quantify its likelihood.

All three bar weakening mechanisms are discussed further in \S\ref{sec.korm}.

\subsection{Discussion of bar fraction}
\label{sec.barfrac}
None of the proposed methods to form bars, or of preventing their
formation, seems able to give a convincing explanation for the
observed fraction of bars in galaxies.  Furthermore, \citet{Bosm96},
\citet{Cour03}, and others have pointed out that barred galaxies seem
little different from their unbarred cousins in most respects --
\eg\ they lie on the same Tully-Fisher relation.  \citet{SG12} did find
significant differences between barred and unbarred galaxies in
photometric parameters, which they attribute to evolution {\it caused
  by\/} the bar.

\citet{Barr08} report an anticorrelation of bar frequency with the
bulge light fraction and \citet{Buta10a} found a decreased frequency
of strong bars in S0 galaxies, which have dense and massive bulges.
Both these studies offer weak support for the stabilizing mechanism
proposed by \citet{Toom81}.  But this cannot be the whole story
because some near-bulgeless disks lack a strong bar (\eg\ M33) while
other barred disks have massive bulges.

The inability of theory or data to find a clear predictor for the
incidence of a bar in a particular galaxy suggests that whether a
particular galaxy is or is not barred may depend on unobservable
factors such as its formation history \citep[see also][]{Shet12}.

\citet[][see also Hoyle \etal\ 2011]{Erwi05} found that bars in
early-type galaxies are larger than those in late-type galaxies, both
in absolute size and in terms of the scale length of the disk light.  He
also noted that bars in many real galaxies, especially of late Hubble
type, are shorter than those in simulations, which is another reason
to think that our understanding of bar formation in real galaxies
remains incomplete.

\subsection{Bar pattern speeds}
\label{sec.omegap}
\citet{TW84a} devised a method to measure the pattern speed of a bar
directly from observations of a tracer component, which must obey the
equation of continuity.  Their original method assumes that the galaxy
has but a single pattern, and would yield a misleading result were
there more than one pattern, each rotating at a different angular
rate.

The stellar light distribution of early-type barred galaxies is
believed to obey the equation of continuity because these galaxies
have little dust obscuration and no star formation.  They also rarely
possess prominent spirals in the outer disk.  Results of many studies
using this method for early-type barred galaxies were summarized by
\citet{Cors08}.  While some individual measurements are quite
uncertain, the data seem to favor $1 < {\cal R} \ltsim 1.4$.
\citet{CH09} found a counter-example in a low-luminosity galaxy.

\citet{Fath09} and \citet{MRM09} applied the method of \citet{TW84a}
to ionized and to molecular gas, respectively.  Both groups argue that
this is valid, even though the separate gas components do not obey the
continuity equation that underlies the method.  \citet{Fath09} generally
found fast bars.  \citet{Meid08} generalized the method to attempt to
measure radial variations in the pattern speed and \citet{MRM09} found
suggestions of pattern speeds that are lower at large radii than those
near the center.

Other methods can yield indirect estimates of bar pattern speeds.
Fits of models of the gas flow (\S\ref{sec.flows}) have been reported
for a few galaxies \citep{Lind96}; \citep{Wein01}; \citep{Pere04};
\citep{ZSS08}, finding ${\cal R} \sim 1.2$ in all cases.
\citet{Atha92} argued that the shapes and locations of dust lanes in
bars also seem to suggest that ${\cal R} \simeq 1.2$.  If the 4:1
resonant orbit family (dotted curves in Fig.~\ref{fig.x1x2}) gives
rise to the ``boxy'' appearance of a bar, then that bar must be fast,
as the orbit family cannot be populated in slow bars.  Identifying a
ring in a barred galaxy as the location of a major resonance with the
bar \citep{BC96} yields, with kinematic information, an estimate of
the pattern speed.

\citet{RSL08} computed models of the stellar and gas (using sticky
particles) responses to forcing by photometric models of 38 barred
galaxies, in which they assumed that the entire nonaxisymmetric
structure rotated at the same pattern speed.  They attempted to match
the model to the visual morphology of the galaxy, and found a range of
values for $\cal R$.  However, in most cases where ${\cal R} \gg 1$,
the fit is dominated by the outer spiral, which may have a lower
angular speed than does the bar.

\subsection{Bars within bars}
\label{sec.twobar}
The nuclear regions of many barred galaxies show isophote twists
\citep[\eg][]{Shaw93} that are interpreted as inner {\bf secondary}
bars within large-scale {\bf primary} bars.  \citet{ES02} identified
secondary bars in $>25\%$ of barred galaxies and reported that they
have a length some $\sim 12\%$ of that of the primary bar.  The
deprojected angles between the principal axes of the two bars appeared
to be randomly distributed, suggesting that the two bars may tumble at
differing rates.  This inference was supported by \citet{CDA03}, who
used the \citet{TW84a} method to show that the two bars in NGC~2950
could not have the same rotation rates; \citet{Maci06} used the same
data to argue that the secondary bar has a large retrograde pattern
speed.  \citet{Fath07} infer an angular speed for the secondary bar
that is higher than that of the primary in NGC~6946.

The theoretical challenge presented by these facts is substantial, and
progress toward understanding the dynamics has been slow.
\citet{MS00} studied the orbital structure in a potential containing
two nonaxisymmetric components rotating at differing rates.  However,
a self-consistent secondary bar can neither rotate at a uniform rate
\citep{LG88} nor can it maintain the same shape at all relative phases
to the primary.

\citet{FM93} argued that gas was essential to forming secondary bars
\citep[see also \eg][]{HSE01}; \citep{ES04}.  However, some of the
collisionless simulations reported by \citet{RS99} and \citet{RSL02}
manifested dynamically decoupled inner structures when the inner disk
had high orbital frequencies due to a dense bulge.  The structure was
more spiral-like in some models, but others appeared to show inner
bars that rotated more rapidly than the main bar.

\citet{DS07} created long-lived, double-barred galaxy models in
collisionless $N$-body simulations having dense inner disks, which
they described as pseudobulges.  They followed up with a more detailed
study \citep{ShD09} that also made some predictions for observational
tests.  The secondary bars in their models indeed rotated at
nonuniform rates, with a shape that also varied systematically with
phase relative to that of the primary.

These models prove that purely collisionless dynamical systems can
support this behavior.  However, it remains unclear what initial
conditions have given rise to double-barred galaxies in nature.

The possible consequence of gas inflow in these galaxies has attracted
a lot of attention.  \citet{SFB89} speculated that bars within bars
might lead to gas inflow over a wide range of scales, from global to
the parsec scale where accretion onto a black hole might cause AGN
activity.  While inflows may have been observed \citep[\eg][]{Haan09};
\citep{vF10}, understanding of gas flow in these nonsteady potentials
remains rather preliminary \citep{Maci02}; \citep{Hell07}.

\subsection{Fueling of AGN by bars?}
Many papers \citep[a partial list is:][]{Knap00}; \citep{Lain02};
\citep{Laur04}; \citep{Hao09}; \citep{Lee12} have discussed the vexing
question of whether there is, or is not, an excess of active galactic
nuclei (AGN) in barred galaxies.  Even the observational question is
hard to answer, because a low level of AGN activity can be confused by
a high rate of star formation, and low-ionization nuclear emission
regions (LINERs).  It is also necessary to ensure that the barred and
unbarred galaxy samples to be compared have similar distributions of
luminosities, colors, \etc

Emphatically one can answer that a single large-scale bar in a galaxy
cannot drive gas close enough to the black hole to be accreted, and
therefore produce an enhanced level of activity.  Torques on the gas
from the bar are able to reduce its angular momentum by about a single
order of magnitude, leaving it orbiting the nucleus at speeds $\gtsim
100\;$km~s$^{-1}$ at a distance $\gtsim 200\;$pc.  Its angular
momentum must be reduced by at least a further 2 orders of magnitude
before the gas could join even the dusty torus that is thought to
surround the accretion disk in a typical AGN \citep{Krol99}.  Thus the
essence of the debate is whether secondary (or even multiple) bars,
nuclear spirals, magnetic fields, \etc\ can bridge this gap and
deliver to the accretion disk some of the larger supply of
circum-nuclear gas that resides in barred galaxies.

Since the argument over the observational evidence continues, with
perhaps the nay sayers in the ascendant at the present time, one
concludes that there is no clear, direct connection between
large-scale bar inflow and AGN activity, and there may be none at all.

\section{Dynamical friction}
\label{sec.dynfr}
\citet{Chan43} pointed out that a massive object moving through a
background sea of light particles would experience a drag force, known
as dynamical friction.  It is believed to affect globular clusters,
satellite galaxies, and bars as they move or rotate inside dark matter
halos.  Orbital decay of satellites, or the slow-down of bars,
together with the gain of energy by the halo, are important aspects
of secular evolution.

\subsection{Mechanism}
Each particle in the background sea experiences an attractive
gravitational impulse as it is passed by the advancing massive body.
Since the attracted particles converge behind the perturber as it
moves forward, the perturber in effect ``focuses'' the background
particles into a trailing density excess, or wake.  The gravitational
attraction between the wake and the perturber gives rise to an
apparent ``frictional'' drag that slows the motion of the perturber.
The kinetic energy lost by the perturber is added to the random motion
of the background particles.

\begin{figure}[t]
\begin{center}
\includegraphics[width=.8\hsize]{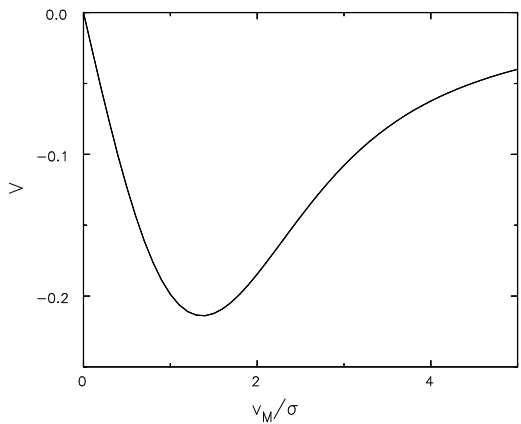}
\end{center}
\caption{The dimensionless acceleration function $V$ defined in
  Eq.~(\ref{eq.chand}) for the case of a Maxwellian distribution of
  velocities among the background particles.  The function is negative
  because the acceleration is directed oppositely to the velocity.}
\label{fig.chand}
\end{figure}

The deceleration of a mass $M$, moving at speed $v_M$, through a
background sea of particles of uniform density $\rho$ with an
isotropic velocity spread $\sigma$ may be written (\BTii, \S8.1)
\begin{equation}
{dv_M \over dt} = 4\pi \ln\Lambda G^2 {M \rho \over \sigma^2} V\left(
{v_M \over \sigma} \right), 
\label{eq.chand}
\end{equation}
where $\ln\Lambda$ is the Coulomb logarithm introduced in
\S\ref{sec.relax}.  The function $V(x)$ describes how the retarding
acceleration varies with the speed of the perturber relative to the
random motions of the stars, and is illustrated for Maxwellian
velocity distribution in Fig.~\ref{fig.chand}.  A rapidly moving
perturber gives weak impulses to the background particles that create
a mild wake far downstream behind the perturber, and the drag
therefore dies away as $V_M \rightarrow \infty$.  The drag on a
perturber moving slowly with respect to the rms motions of the
particles is also mild, because there is only a small excess of
particles dragging it backward over those urging it forward, and the
acceleration must also vanish as $V_M \rightarrow 0$.  For a
Maxwellian velocity distribution, the drag force peaks when the
perturber moves somewhat faster than the 1D dispersion $\sigma$,
\ie\ the rms velocity in any coordinate direction.

\subsubsection{Improved treatment}
There are a number of conceptual problems with Chandrasekhar's
analysis.  Foremost is that fact that the hypothesized infinite sea of
uniform density cannot be realized; away from the center of any finite
distribution of matter, the motion of the perturber would not be a
straight line but a curved orbit within the host mass distribution.
Second, each particle in the sea is supposed to have a single
interaction with the perturber, but particles must be bound to any
finite system (otherwise it would disperse), and therefore repeatedly
interact with a bound perturber.  Choosing values for the density
$\rho$ and velocity dispersion $\sigma$ appropriate for use in
Eq.~(\ref{eq.chand}) also presents difficulties, especially when the
mass distribution is highly inhomogeneous \citep{AB07}.

These problems were all overcome in a seminal paper by \citet{TW84b},
who reformulated the analysis in action-angle variables.  They found
that the rate at which angular momentum is lost from a disturbing mass
orbiting in a spherical system of background particles, hereafter a
``halo,'' is given by an expression identical in form to that derived
by \citet{LBK72} for a disk (Eq.~\ref{eq.LBK} in \S\ref{sec.lzchange})
except that the DF in the spherical case expresses the density in the
6D phase space.  The number of possible resonances is substantially
increased, since they arise wherever combinations of the three
separate frequencies (orbital, radial and vertical) for the
unperturbed motion of background particles match the angular frequency
of the perturbation.  In a follow-up paper, \citet{Wein85} evaluated
the frictional drag expected from a halo, modeled as a singular
isothermal sphere, on a rotating bar, modeled a homogeneous prolate
spheroid rotating about an axis in its equatorial plane.  Assuming the
bar also had the moment of inertia of the rigid spheroid, Weinberg
found that in a dense halo, the pattern speed of a strong bar would
decay with a half-life of a few of its initial rotation periods.

While the complicated LBK torque expression (Eq.~\ref{eq.LBK}) is
daunting, its physical meaning is very similar to that of
Chandrasekhar's formula (Eq.~\ref{eq.chand}).  The drag force arises
because, to second order, the halo builds up a density excess, or
wake, behind the perturber, leading to a gain, on average, of angular
momentum by the halo particles and a corresponding loss by the
perturber.  The lagging wake was illustrated for the case of a bar by
\citet[][their Fig.~1]{WK07}.

As \citet{LBK72} showed for spiral perturbations, halo particles that
are not in resonance also experience changes that average to zero and
therefore the drag is caused only by resonant particles.  Referring
back to Eq.~(\ref{eq.LBK}) again, the contribution to the drag from
each resonance depends on the gradient of the halo particle DF across
that resonance, and there is generally an excess of gainers over
losers.  But the gainers will depopulate the high-density side, and
cross the resonance to the previously low-density side.  Thus, were
the pattern speed of the perturber to remain constant, the local
gradient across the resonance would be reduced, and the system would
adjust toward a balanced equilibrium in which the gradient in the DF
in the immediate vicinity of the resonance had flattened.  Then the
imbalance of gainers over losers would be erased and friction would
die away.  This effect was explicitly demonstrated by \citet{LT83} for
the case of an orbiting satellite, and will turn out to be important
in \S\ref{sec.awf}.

However, the motion of the perturber {\it is\/} affected by its loss
of angular momentum.  An orbiting mass clearly sinks inward, making
its angular speed about the center rise.  Because a bar is not a solid
object with a well-defined moment of inertia, \citet{TW84b} were less
certain about whether its pattern speed would increase or decrease as
a result of angular momentum loss.  However, simulations
\citep{Sell80}; \citep{DS98, DS00}; \citep{Sell03}; \citep{Atha03};
\citep{VK03}; \citep{HWK05}; \citep{Mart06}; \citep{VSH09};
\citep{Minc12} have always found that bars slow as they lose angular
momentum.

The change in pattern speed of the perturber causes the resonances to
sweep through phase space \citep{SD06}; \citep{DBS09}, and therefore the
gradients across resonances do not have time to flatten and, in
general, friction persists as the perturber loses angular momentum,
although exceptions have occurred (see \S\ref{sec.awf}).

\citet{LT83}, \citet{BvA87}, \citet{AB07} and \citet{JS09} for an
orbiting satellite, and \citet{Sell06} for a rotating bar, showed that
the dependence of the friction term on the perturber mass, its angular
speed, and both the dispersion and density of the halo, were all as
expected from Eq.~(\ref{eq.chand}).  The inclusion of self-gravity of
the halo causes a slightly larger density response, with a
corresponding increase in the drag force, but does not appear to
change the scaling.

\subsubsection{Bar-halo friction}
The case of a bar rotating in a halo has received a great deal of
recent attention, as friction from a moderately dense halo slows the
bar on a time-scale of a few rotations \citep{Sell80}; \citep{Wein85}.
Bars also grow in length and strength (\ie\ the quadrupole moment
rises) as they are braked by friction.

While many resonances exist between the bar and halo orbits,
\citet{WK07} stressed that frictional drag was dominated by a small
number of resonances, of which the ILR was by far the most important.
The consequence of flattening the gradient in the halo DF at this
resonance has also affected simulation results (see \S\ref{sec.awf}).

The phase lag angle between the axis of the halo response and
principal axis of a bar varies with the drag force \citep{DS00};
\citep{Sell06}.  Friction is weak at high angular speeds because the
halo response is almost exactly perpendicular to the bar, it
strengthens as the pattern slows reaching a maximum when the response
is $\sim 45^\circ$ to the bar axis, and decreases to zero as the
response becomes aligned with the bar, which generally happens before
the bar is brought to a complete stop.  Thus the system reaches a
steady state in which the corotating halo response, which has been
described as a halo bar \citep{Atha07}, is aligned with the bar in
the disk.  Since the response of orbits to a steadily rotating bar
potential is a forced precession (see \S\ref{sec.sdbar}), it is no
surprise that studies of halo orbits in barred simulations
\citep{Atha02, Atha03}; \citep{CK07} have found a large fraction of
resonant orbits.

It is possible that such trapped orbits are responsible for the
asymmetric distribution of thick-disk and/or luminous halo stars in
the Milky Way.  A density excess of such stars that are only a few
kiloparsec from the Sun in the first Galactic quadrant, with no
counterpart in the fourth quadrant, has been reported most recently by
\citet{Lars08}.  A similar asymmetry, that could be aligned with the
bar in the disk, was also found by the same group \citep{Park04} in
stellar kinematics.

\citet{DS00} experimented with rotating halos, finding that the
frictional drag varies with the degree and sense of halo rotation.  It
is little changed in halos having mild rotation in the same sense as
the disk, and is significantly weakened only by an unrealistic degree
of halo rotation.

The loss of angular momentum from the bar to the halo slows the bar
and allows it to continue to grow \citep{DS00}.  Continued evolution
of bars in moderately dense halos \citep{AM02}; \citep{Atha03};
\citep{Mart06} can cause the bar to grow until it fills almost the
entire disk, at which point corotation can be outside the disk.  The
unreasonably large bars, relative to the disk size, obtained in these
simulations would appear to argue against even moderately dense {\it
  inner\/} halos for real galaxy disks.

The halo that absorbs the angular momentum, need not be just the dark
matter halo; visible spheroidal components are also angular momentum
sinks.  \citet{Saha12} found that an initially nonrotating classical
bulge acquires angular momentum from a strong bar and becomes
triaxial, corotating with the bar when angular momentum transfer is
complete.  The bulge streaming motions they report are quite large.

\subsection{Constraint on halo density}
\label{sec.barfr}
While the above theory leads to the expectation that a halo must exert
a strong frictional drag on a massive bar, fully self-consistent
simulations were needed to show that a bar in a disk embedded in a
dense halo slows to the point that ${\cal R} > 1.4$ \citep{DS00};
\citep{WK02}; \citep{Sell03}; \citep{VK03}; \citep{OD03};
\citep{HWK05}; \citep{VSH09}, which is inconsistent with most observed
values.

The pattern speeds of strong bars are generally fast, in that ${\cal
  R} \sim 1.2$ (Eq.~\ref{eq.Rdef}); the evidence is reviewed in
\S\ref{sec.omegap}.  Furthermore, the slow bar reported by
\citet{CH09} is in the low-surface-brightness galaxy UGC 628, which is
believed to have a large DM fraction that should have slowed a strong
bar.

The observationally accessible ratio $\cal R$ is not a direct measure
of friction, since both the bar length and corotation radius can
change independently, but halo friction has driven this ratio to a
high value in all collisionless simulations in which a strong bar
rotated in a dense halo.  Since no convincing counter-examples of
strong bars in dense halos remaining fast for long periods have been
found (see \S\ref{sec.awf}), the simulations clearly predict low
density halos for the majority of strongly barred galaxies.  This
result is little changed in simulations with moderate fractions of gas
\citep{VSH10}, but friction is weaker because bars are smaller and
weaker in simulations with unreasonably large gas fractions.
Obviously, the magnitude of the friction force varies with bar
strength, and weak bars therefore experience little friction,
\eg\ simulation MHH2 by \citet{Atha03}.  Unfortunately, this author
does not give the crucial value of $\cal R$ for any of her
simulations, limiting further interpretation of her many results.

Since strong bars experience fierce braking from moderately dense
halos, \citet{DS98, DS00} argued that bars in real galaxies can remain
fast only if the central dark matter density is much lower than was
predicted by galaxy formation models.  The implication of their result
led to the conclusion being examined very closely by others, but even
though two serious issues were raised (see below), their conclusion
still holds.

\subsubsection{Anomalously weak friction}
\label{sec.awf}
\citet{VK03} reported that the bar in a disk embedded in a cuspy halo
rotated at almost its initial speed for quite some time, and did not
experience strong braking, which was inconsistent with all other
numerical results and with theory!  \citet{SD06} reproduced their
anomalous result when they reran the exact same model, and were able
to show that the reason for the absence of friction stemmed from a
flattened gradient in the halo DF at the crucial ILR.  While the halo
DF had a density gradient $\partial f/\partial L_z$ that was negative
everywhere initially, the halo DF was changed by interactions with the
disk bar as it formed.  As the disk bar buckled, its pattern speed
increased, with the implication that the resonances returned to parts
of phase space where the previous changes at those resonances had
flattened the gradient in the DF.  Since friction depends upon there
being an excess of gainers over losers caused by a decreasing density
across the resonance, the absence of a gradient allowed the bar to
rotate without friction -- exactly as \citet{LT83} had shown for a
sinking satellite.  Strong friction resumed after a while, probably
because mild braking at minor resonances eventually moved the ILR
resonance into a region where the DF gradient was normal.
\citet{SD06} were also able to show that the near-frictionless state
was fragile, and could not be expected to arise in nature, since tiny
perturbations by passing low-mass satellites were sufficient to shake
the halo out of its ``meta-stable'' state and to cause fierce friction
to resume.

This type of behavior was the root cause of the stochastic variations
found by \citet{SD09}.  They reported that the evolution of the bar
pattern speed in many different random realizations of the same
(isolated) model galaxy varied widely because the halo DF was
sculptured by the early disk evolution, which was stochastic.  The
chance absence of gradients in the DF at the dominant resonance
inhibited friction for periods that varied widely from realization to
realization.  Stochastic behavior was unaffected by changes to the
numerical parameters, even the particle number, and persisted when
they used a tree code instead of a grid code.  Thus a simulation of an
isolated galaxy that shows weak friction between a strong bar and a
dense halo for a short period \citep[\eg][]{Klyp09} is inconclusive;
firm evidence for weak friction that contradicts all the theory and
experimental evidence summarized above would require longer
simulations that are repeated with different random realizations.

\subsubsection{Particle number}
\citet{WK07} claimed that simulations require immense numbers of
particles to reproduce the correct frictional drag.  Their argument
stemmed from their recognition that most of the frictional drag arises
from the gradient in the halo DF across a single resonance, the ILR.
They argued that too few particles in this small region of phase space
would prevent the bar from experiencing the correct drag.  Their
predicted symptoms of this deficiency would be stochastic variations
in the drag force resulting from shot noise in the local distribution
of particles that happened to be in the vicinity of the resonance.
(It should be emphasized that they were discussing friction with an
imposed rigid bar, which is separate from the possible stochastic
variations due to disk evolution that were discussed in the previous
paragraph.)  They went on to estimate that simulations needed a
hundred-fold increase in the number of particles, from $\sim10^6$ to
$\sim 10^8$, before this problem would be brought under control.

\citet{Sell08b} conducted a series of simulations of exactly the kind
that \citet{WK07} envisaged (a rigid bar in a halo of massive
particles) and found that only very small numbers of particles
($\ltsim 10^4$) behaved stochastically in the manner predicted by
\citet{WK07}.  Once $N \gtsim 10^5$, the evolution of both the pattern
speed and the halo mass profile was independent of $N$, and no new
behavior emerged when the number of particles was increased to
$N=10^8$.

The reason for the discrepancy with the predictions of \citet{WK07} is
that the resonances are broadened by the time dependence of the bar
pattern speed.  Their estimates of particle number requirements used
only the intrinsic width of the resonance due to the finite amplitude
of the perturbation -- in effect, they assumed a fixed pattern speed.
A much larger fraction of the particles than these authors expected
contribute to the friction when the pattern speed decreases rapidly
due to frictional drag from the halo, and friction can be reliably
reproduced in simulations of strong bars with $10^5$ to $10^6$ halo
particles.  \citet{CK07} and \citet{DBS09} came to similar
conclusions.

\subsection{Change in halo density}
\label{sec.hdc}
As noted in \S\ref{sec.galform}, all collisionless cold dark matter
models of halo formation predict a steep inner gradient to the halo
density profile, whereas shallower density profiles seem to be
required \citep[\eg][]{Sell09}; \citep{KS11}.  Dynamical friction
between the baryons and the dark matter halo has been proposed as a
possible solution to this discrepancy.

\citet{WK02} and \citet{ElZa01} argue that dynamical friction between
the halo and bars or gas clumps respectively could transfer enough
energy to the dark matter halo to reduce its density in the inner
parts of galaxies.  Note that the mechanism invoked in these papers is
somewhat distinct from galaxy formation models \citep{RG05};
\citep{Gove10}; \citep{PG12} that invoke repeated changes to the
gravitational potential as gas disks collect and then evaporate; since
this latter mechanism is essentially part of current ideas for galaxy
formation, I do not review it here.

\subsubsection{Halo density reduction by bars}
A number of authors have reported a mild reduction in halo density
resulting from bar friction \citep{HW92}; \citep{DS00};
\citep{Atha03}; \citep{MD05}, and both \citet{HWK05} and
\citet{Sell08a} verified that the dominant changes occurred at the ILR.
Simulations with rigid bars that are pinned to a center can suffer
from a numerical artifact if the halo is allowed to become lop-sided
\citep{Sell03}; \citep{MD05}, resulting in an erroneously large
angular momentum transfer.

A convenient measure of halo density is its mean value inside the
radius where the circular speed has risen to half its peak value
\citep{Alam02}.  \citet{Sell08a} found that moderate, rigid bars
reduced this quantity by a few percent, as already noted, but an order
of magnitude reduction required a bar of length $\sim 12$--20~kpc, an
axis ratio $a/b\gtsim3$, and a {\it bar\/} mass $\gtsim30$\% of the
enclosed halo mass.  In fact, the angular momentum given up by the
bar in order to achieve this density reduction exceeded the likely
store of angular momentum in a galaxy disk \citep{MD05};
\citep{Sell08a}.

Furthermore, a real bar, formed say through instabilities in the disk,
contracts as it loses angular momentum to the halo, since the sizes
of the stellar orbits that make up the bar themselves shrink.  The
resulting increase in the central attraction causes the halo density
to rise, an effect that can overwhelm the density reduction due to
angular momentum changes \citep{Sell03}; \citep{CVK06}; \citep{DBS09}.

\subsubsection{Halo density reduction by moving mass clumps}
\citet{ElZa01}, and later \citet{Toni06}, proposed that moving clumps
of dense gas will also transfer energy to the DM halo through
dynamical friction and lower its density.  They envisaged that baryons
would collect into clumps through the Jeans instability as galaxies
are assembled and present somewhat simplified calculations of the
consequences of energy loss to the halo through dynamical friction.
The dynamical process is that of mass segregation, which is well known
in other contexts, such as in globular clusters \citep[\eg][]{Merr04},
but requires much larger mass differences for evolution on an
interesting time scale.

The idea has been tested in idealized $N$-body simulations in which
the heavy mass clumps were modeled as softened point masses
\citep[\eg][]{JS09}; \citep{Goer10}; \citep{Cole11}.  Generally,
low-mass clumps were found to be ineffective, because dynamical
friction is too weak.  However, the orbital decay of a really massive
clump, about 1\% of the virial mass of the halo, does transfer enough
energy to the halo particles to effect a substantial reduction in its
density, even as the heavy clump itself deepens the gravitational
potential.

\citet{MBK04} used clumps which were themselves composed of particles
that could therefore suffer tidal stripping, \etc\ They reported that
the stripped particles remained roughly at the radii at which they
were detached, and also found that significant reduction in the halo
density was caused by only the heaviest clumps.

The proposed mechanism faces a number of challenges, however.  The
settling gas clumps are assumed to maintain their coherence for many
dynamical crossing times without colliding with other clumps or being
disrupted by tidal fields and/or star formation, for example.
\citet{Roma08a} claimed that baryonic physics had precisely this
effect in their galaxy formation simulations, but calculations
\citep[\eg][]{Kauf06} of the masses of the condensing gas clumps
suggest they range up to only $\sim 10^6\;M_\odot$, which is too small
to experience strong friction.  Larger clumps will probably reside in
subhalos, which may get dragged in, but simulations with subclumps
composed of particles \citep[\eg][]{MBK04} indicate that the DM halos
of the subclumps will be stripped, and the stripped matter largely
replaces any DM moved outwards in the main halo.  Thus if dynamical
friction is to accentuate the separation of the baryons from dark
matter before the baryonic mass clumps in subhalos settle to the
center, they must somehow be stripped efficiently of their dark
matter without dissolving the gas clumps themselves.

\citet{MCW06, MCW07} argue that the energy input to the halo, mediated
by the motion of the mass clumps, can be boosted if the gas is stirred
by the usual feedback from stellar winds and supernovae.  A challenge
for this mechanism is the difficulty of accelerating such massive gas
clumps into coherent motion, since the high-pressure material from the
postulated energetic events will vent more easily along low-density
paths, thereby relieving the pressure before the dense clumps gain
much momentum \citep[\eg][]{MF99}.

\section{Rings and outer light profiles}
Gas in nonaxisymmetric galaxies is driven inward inside corotation,
and outwards at larger radii.  This behavior contrasts with that of
the stars (\S\ref{sec.Lchange}); dissipation allows the gas to stay
dynamically cool while experiencing large changes in $L_z$.  Secular
evolution of this type is believed to be responsible for the formation
of most rings observed in galaxies.  The faint outer light profiles of
galaxies also manifest features, but their origin is less clearly
attributable to secular evolution.

Encounters between galaxies are invoked to explain other types of
galaxy rings \citep{LT76}; \citep{Stru10}; \citep{Elic11} or polar
rings \citep[\eg][]{Spar08}, which are not, therefore, the result of
secular evolution.

\subsection{Rings}
\label{sec.rings}
Long-lived perturbations, such as bars, can drive gas radially until
the flow stalls at resonances where rings of star-forming gas build
up.  \citet{Buta95} identifies three types of ring: {\bf outer} rings,
{\bf inner} rings, and {\bf nuclear} rings, all of which are commonly
found in barred galaxies, but some are known in unbarred galaxies
also.  Outer rings, which are divided into two subtypes depending on
their elongation relative to the bar, are generally believed to occur
at the OLR of the bar.  Inner rings have mean radii that are about as
large as the bar semi-major axis, while nuclear rings are deep inside
the bar.  The rings are thought to depart from circles because of the
quadrupole field of the bar, and the distortion is enhanced by being
located at a resonance.

Outer rings have been identified in the light profiles of 66
early-type barred galaxies by \citet{Erwi08}, who found an occurrence
rate (or a feature at the expected radius) in 35\% of the cases.
\citet{Buta10b} are conducting an on-going search using their deep
3.6$\mu$m survey with {\it Spitzer\/} (dubbed S$^4$G) for additional
outer rings, but are finding few new cases, perhaps because these
features tend to be quite blue.  

The outer, inner, and nuclear rings are widely believed to form though
secular evolution in, mostly barred, galaxies.  \citet{BC96} give a
thorough review of rings and the theory of secular formation of rings,
and though written some years ago, remains reasonably up-to-date as
the subject has not advanced much since.  A more recent review of the
properties of such rings and their formation mechanisms was included
in \citet{KK04}.

The gas in a nonaxisymmetric potential must shock when periodic
orbits cross (see \S\ref{sec.flows}), causing an irreversible change
to the orbital motion.  The shock is generally offset from the
potential minimum, resulting in an angular momentum exchange between
the gas and the bar or spiral.  The position of the shock relative to
the potential minimum determines the sign of the exchange: gas
loses angular momentum inside corotation, whereas it gains outside
this resonance, since the gas flow relative to the wave is in the
opposite sense.  Thus gas is driven away from corotation until the flow
stalls, at an OLR, or where the dominant orbit family switches
orientation in the nuclear region of a strong bar.  Two orbit families
can support rings at the OLR; just inside the resonance, orbits are
elongated perpendicular to the bar, whereas the elongation is parallel
to the bar just outside that resonance.  The early simulations by
\citet{Schw81}, which employed sticky particles, were able to produce
rings of both orientations, and there is evidence for both types in
real galaxies \citep[see][for examples]{BC96}.  More recent models
for the formation of outer rings were presented by \citet{Bagl09} as
the response of collisionless test particles to bar forcing, and they
also compare their models with rear-infrared images of galaxies.

Inner rings are believed to be located at the ultraharmonic resonance
(UHR, see \S\ref{sec.orbs}) of the bar, where the potential supports
4:1 orbits (dotted in Fig.~\ref{fig.x1x2}).  \citet{BC96} suggest that
inflow from corotation stalls at the UHR to make this ring, which is
perhaps consistent with the behavior also found in Schwarz's work
\citep{Simk80}.  There the ring is simply a pointy oval, a shape that
is often found in real galaxies.  However, if the 4:1 resonant family
is responsible, it is somewhat surprising that such rings are not more
boxy; perhaps the theoretical interpretation of inner ring formation
deserves further study.

Subsequent to the review by \citet{BC96}, most attention has focused
on nuclear rings.  Bars appear to be efficient at driving gas inwards
until the flow stalls in a nuclear ring, as described in
\S\ref{sec.flows}.  The gas concentrations in these nuclear rings
appear to be forming stars at a prodigious rate \citep{Hawa86};
\citep{Maoz01}; \citep{Bene02}; \citep{Mazz08, Mazz11}.

\subsection{Outer light profiles}
\label{sec.OLP}
While galaxy disks are frequently described as exponentials, few
galaxies have light profiles that can be fitted with a single
exponential over several length scales.  The light profiles reported
by \citet{Free70} did not extend to very faint light levels, by the
standards of today.  Yet he identified both type I profiles, which
were good exponentials over the limited dynamic range of his data, and
type II, in which the surface brightness of the inner disk rises less
rapidly than the inward extrapolation of the outer exponential.  Both
these types have been found in modern, much deeper photometry
\citep{Pohl02}; \citep{BH05}; \citep{EBP05, Erwi08}; \citep{HE06};
\citep{PT06}, which also revealed type III, in which the light
profile at large radii declines less steeply than the inner
exponential.  \citet{EPGB} found that type II profiles are more common
in barred galaxies.  The fraction having type III profiles rises
to late Hubble types, but galaxy interactions also appear to play a
role \citep{EBP05}.

The origin and significance of this variety of behavior is still not
fully understood, and may be related to galaxy formation, environment,
or star-formation efficiency \citep{Sanc09}; \citep{Mart09}.  However,
some aspects may be due to internal disk evolution \citep{Deba06};
\citep{Foyl08}; \citep{Minc12}.  \citet{Mart12} proposed that breaks
might be phenomena related to a threshold in the star formation, while
truncations are more likely a real drop in the stellar mass density of
the disk associated with the maximum angular momentum of the stars.
On the other hand \citet{Rosk08a} and \citet{Muno13} suggested internal
secular evolution may be the cause.

While \citet{Truj09} assert that the extended type III disk in M94 is
not a ring, they nevertheless suggest it could be formed by an outflow
in the disk that was driven by a rotating oval distortion in the inner
part of the disk.  Also noteworthy is the suggestion by
\citet{Rosk08a} that the radial decline in the mean ages of disk
stars, caused by inside-out disk formation, could be reversed in the
far outer disk by the outward migration of older stars.  An attempt to
verify this prediction \citep{Yoac12} met with mixed results, however.

\section{Pseudobulges and lenses}
\label{sec.korm}
Classical bulges, which have $R^{1/4}$ light profiles, are not
strongly flattened, and rotate rather slowly, are believed to have
been formed from violent mergers of protogalactic fragments in the
early stages of galaxy formation, as described in \S\ref{sec.galform}.
Galaxy disks with an embedded classical bulge are presumed to have
built-up subsequently through the usual process of dissipative gaseous
in-fall.

However, it has become clear that many galaxies host bulges having
quite different properties that are now described as a {\bf
  pseudobulge}s.  They have more nearly exponential light
distributions \citep{AS94}; \citep{FD08}, exhibit quite a high degree
of rotation that has a roughly cylindrical flow pattern in 3D
\citep{KI82}, and are generally flatter than are classical bulges.
\citet{KK04}, updated in \citet{Korm12}, gave a more detailed
description of how a pseudobulge can be distinguished from a
classical bulge.

The observed properties of pseudobulges strongly suggest a different
formation mechanism and it seems highly likely that they formed
through internal evolution from the disk \citep{KK04}, and that this
evolution is more rapid in galaxies with a higher gas fraction
\citep{Korm12}.  Their basic idea is that pseudobulge formation
is mediated by a bar, which first forms and buckles, as described in
\S\ref{sec.bars}, and then dissolves into a dynamically hot, but
flattened and rotationally supported bulgelike structure.  

\citet{Korm12} proposed that a CMC of both stars and gas causes the
bar to dissolve and create a pseudobulge.  Gas is indeed driven inward
by bars (\S\ref{sec.flows}) and simulations (\S\ref{sec.dissol}) show
that bars dissolved by massive CMCs do indeed form thickened,
rotationally supported, near axisymmetric structures that resemble
pseudobulges.  The mass fraction in the dense central concentrations
required to cause the bar to dissolve entirely is very high
\citep[\eg][their simulation NG5]{Deba06}.  The more concentrated the
mass the more efficiently it destroys the bar \citep{Shen04}, but the
mass required is far larger than that suggested for any supermassive
black hole \citep[\eg][]{Gult09}.  Large gas concentrations spread
over an area of few hundred parsecs in radius are observed
\citep[\eg][]{Saka99}; \citep{Shet05}, but again are nowhere near
massive enough.

However, the gas in the nuclear region forms stars at a vigorous rate
(\S\ref{sec.rings}), with presumably a significant fraction of the
mass being locked into long-lived stars that are gravitationally bound
to the region where they formed.  \citet{Korm12} therefore proposed
that the stars built up in the nuclear region over a protracted
period, together with the gas, eventually reach the combined mass
required to dissolve the bar.  Kormendy developed this proposal at
length in his review, to which I refer the intersted reader for the
full picture.  If his plausible idea is correct, it once again implies
that significant secular evolution is mediated by the behavior of gas.

No simulation has yet tested this suggestion, however.  Previous
studies of bar dissolution have created the central mass rather
quickly, giving the bar little time to adjust as the mass grows, and
further simulations of more gradual growth are needed to confirm that
dissolution can eventually occur.  The numerical task is particularly
challenging for several reasons: (a) The evolution must be followed
for a long period while the orbit time-scales in the very center are
short. (b) Gas would have to be accreted continuously to the bar
region, and the subsequent inflow rate should not be exaggerated by
numerical viscosity (\S\ref{sec.flows}).  (c) The halo would need to
modeled self-consistently to follow bar growth through dynamical
friction (\S\ref{sec.dynfr}).

Two other methods that might dissolve a bar were discussed in
\S\ref{sec.dissol}: \citet{BCS05} and \citet{Comb08} suggest that the
angular momentum added to bars as they drive gas inwards can weaken or
destroy them.  While more work on this scenario is needed, the inflow
requirements are severe, and the consequence would not be so different
from the build-up of a CMC.  It is also noteworthy that some possible
interactions between a bar and a strong spiral can weaken or destroy
the bar.  The mechanism and the conditions under which this behavior
can occur also require further study, but the process may prove useful
in this context, especially as strong spirals are most likely to arise
in gas-rich outer disks.

Bars could also be destroyed in minor mergers, of course.  But to make
a pseudobulge, the perturber would have to be dense enough to not be
tidally disrupted before reaching the bar, but not so massive as to
destroy the cylindrical flow pattern and/or shallow inner radial light
profile.  This degree of fine-tuning makes the explanation seem
untenable to account for the observed high frequency of galaxies that
seem to host pseudobulges \citep{Korm10}; \citep{FD11}.

Other mechanisms for pseudobulge formation have been proposed.
\citet{Gued12} found that the pseudobulge in their simulations was
formed at an early stage through mergers, although its subsequent
development was still mediated by a bar.  \citet{Okam13} argued for an
early starburst origin.  However, these ideas may be inconsistent with
a broad range of ages among the stars of pseudobulges \citep{Fish09}.

\citet{Korm12} also highlighted the lens component seen in some barred
galaxies, which he argues is the intermediate case in which the bar is
dissolving, while a lens in an unbarred galaxy is a fully dissolved
bar \citep[see also][]{Comb08}.  He therefore suggested that bar
dissolution could be gradual, else we would not observe many
transition cases.  More moderate mass concentrations do cause bars to
weaken and to become more oval (\S\ref{sec.dissol}), but no author has
commented, as far as this reviewer is aware, that the weakened bar in
a simulation inhabits a lenslike structure.  Nevertheless, lenses are
established features of galaxies that seem most likely to have been
created through disk evolution.  The fact that we do not yet have a
satisfactory explanation for their origin is part of the reason why
galaxy evolution remains so fascinating.

\section{Conclusions}
This review has been rather narrowly focused on the internal evolution
of isolated disk galaxies.  The environment surely does play a
substantial role in galaxy evolution; it is probably responsible for
warps, lop-sidedness, tidal bridges and tails, and a whole host of
phenomena related to halo substructure, halo streams, galaxy
transformations, dry mergers, \etc, but broadening this review to
include all, or even some, of these topics would have necessitated
either a shallower treatment or a greatly increased length.

The internally driven evolution of galaxy disks would scarcely be of
any interest if the disk were composed of stars alone.  Spiral
activity would heat the disk on the time-scale of a few disk
rotations, causing later spiral episodes to be progressively weaker
and less distinct.  The extent to which the overall distribution of
angular momentum among the stars could be rearranged on large scales is
strongly limited, since redistributive changes necessarily increase
random motion.  The fractional change in angular momentum of a
distribution of stars (\S\ref{sec.Lchange}) with radial velocity
dispersion $\sigma_R$ and typical radial excursion $a \sim
\sigma_R/\kappa$ is bounded by
\begin{equation}
\left|{\Delta L_z \over L_z}\right| \ltsim {a \over Rm} {\sigma_R \over V_c},
\label{eq.Lbound}
\end{equation}
where $V_c$ is the circular orbit speed at radius $R$, and $m \gtsim
2$ is the angular periodicity of the spiral patterns.  Thus the small
value of both factors on the right-hand side provides a very tight
constraint on the extent to which the distribution of angular momentum
among the stars of a galaxy disk can have changed since their birth.

However, this constraint does not apply to individual stars, which can
migrate radially for large distances within the disk through
interactions near the corotation resonance of spirals
(\S\ref{sec.rm}).  Since gains by some stars are roughly matched by
losses by others in every diffusive step, these changes alter only the
distribution of metals in the disk with almost no change to its
dynamical structure.  In particular, they neither lead to increased
random motion, nor do they cause the disk to spread.

Note also that Eq.~(\ref{eq.Lbound}) does not limit the possible
angular momentum changes of the gas component.  The random motions of
gas clouds, which experience similar radial accelerations from
nonaxisymmetric disturbances as do the stars, are quickly damped
through dissipative collisions with other clouds.  Furthermore, the
low velocity dispersion of the clouds makes them highly responsive to
nonaxisymmetric disturbances, allowing them to exchange angular
momentum with the driving potential to a greater extent than for the
stars.  Thus secular evolution in galaxies is greatly accelerated by
the gas component.  Since gas is consumed by star formation, it
requires constant replenishment, as is expected in hierarchical
structure formation models \citep[\eg][]{Gunn82}.

The rising velocity dispersion of disk stars with age is now thought
(\S\ref{sec.scatt}.6) to be driven by the combined influence of
deflections away from circular orbits by scattering at the resonances
of spiral patterns, with the resulting in-plane peculiar motions being
efficiently redirected into the third dimension by encounters with
massive gas cloud complexes.  No other combination of heating and
scattering can account for both the high dispersion of the older disk
stars and the fact that the velocity ellipsoid maintains a roughly
constant shape as it grows in size.  This combination of factors has
not been tested in fully self-consistent simulations because particle
masses in most simulations are too large to mimic the two processes
separately.  The vertical heating that has been reported in some
simulations is probably due to collisional relaxation \citep{Sell13b}.

Bars are another important agent of secular evolution.  The formation
of a bar causes the largest change in the distribution of angular
momentum among the stars of a disk, and further evolution occurs only
through the influence of the outer disk, halo, and/or gas component.
Bars can continue to grow, losing angular momentum to the outer disk,
or to the halo, and the fact that bars are usually surrounded by an
extensive disk suggests that halos cannot be dense enough to cause
them to grow excessively (\S\ref{sec.bargrow}).\Ignore{ Gas in the
  disk beyond the bar can also be driven outwards to form outer rings,
  which must again remove some angular momentum from the bar.}

Bars slow, as well as grow, through dynamical friction from the halo
(\S\ref{sec.dynfr}).  The loss of angular momentum by this mechanism
also causes the disk mass to contract slightly, which actually deepens
the gravitational potential, overwhelming any tendency for halo
density to decrease as a result of its energy gain from the disk.
While the central density rises, bars also grow in length as they
slow, and the fact that corotation of most bars today appears to lie
just beyond the bar end requires that the inner DM halos have lower
densities than is predicted by $\Lambda$CDM models of galaxy formation
(\S\ref{sec.barfr}).

Bars also drive gas in towards their centers, causing the build up of
gas-rich nuclear rings (\S\ref{sec.flows}) where stars are seen to
form at a high rate (\S\ref{sec.rings}).  The integrated inflow over
the lifetime of a galaxy can lead to the build up of concentrations of
stars and gas in the center that may be able to destroy the bar and to
form a pseudobulge (\S\ref{sec.korm}).

Substantial evolutionary changes to the structure of disks could also
occur through outside intervention, although the degree to which minor
mergers could be important is again strongly constrained by data
(\S\ref{sec.thick}).  The infrequency of classical bulges
\citep{Korm10} places strong constraints on past merging activity, as
does both the thinness of the main disks, and the absence of young
stars in thick disks.

The realization that secular evolution is capable of rearranging the
structure of disk galaxies from their initially endowed properties has
been gradual.  The topic was perhaps begun by \citet{Korm79}, and it
has gradually gained credence, largely through his constant advocacy.
Despite the enormous progress described in this review, there are many
areas where more work, such as the shaping of rotation curves
(\S\ref{sec.smooth}) and the weakening of bars by spirals
(\S\ref{sec.dissol}), or even new ideas, such as to account for the
observed fraction of galaxies that host bars (\S\ref{sec.barfrac}) or
the formation of double bars (\S\ref{sec.twobar}) or of lens
components (\S\ref{sec.korm}), are needed.  Above all, we need better
algorithms, with low numerical viscosity (\S\ref{sec.flows}), to
capture the role of gas in more realistic manner -- a need that is
also recognized in galaxy formation.

\section*{Acknowledgment}
I thank Tad Pryor, Michael Solway, and Ortwin Gerhard for helpful
conversations, and the editor for his patience.  Comments by an
anonymous referee, Rok Ro\v skar, Victor Debattista, James Binney, and
especially by John Kormendy were extremely valuable.  This work was
supported in part by NSF Grants AST-1108977 and AST-1211793.


\begin{thebibliography}{}

\def\aap{A\&A}
\def\aapr{A\&A Rev.}
\def\aj{AJ}
\def\an{Astron.\ Nach.}
\def\anp{Ann.\ der Phys.}
\def\apj{ApJ}
\def\apjl{ApJL}
\def\apjs{ApJS}
\def\apss{Ap.\ Sp.\ Sci.}
\def\araa{ARAA}
\def\astl{Astron. Lett.}
\def\fcp{Fund.\ Cosmic Phys.}
\def\jcop{J. Comp.\ Phys.}
\def\mnras{MNRAS}
\def\newa{New. Astron.}
\def\pasj{PASJ}
\def\PhD{PhD.\ thesis}
\def\phr{Phys.\ Rep.}
\def\nat{Nature}
\def\raa{RAA}
\def\rpp{Rep.\ Prog.\ Phys.}
\def\sci{Science}

\def\skip#1{ \etal\ }

\bibitem[\protect\citeauthoryear{Abadi \etal}{2003}]{Abad03}
Abadi, M. G.,{ Navarro, J. F., Steinmetz, M. \& Eke, V. R.} 2003, \apj, {\bf 597}, 21

\bibitem[\protect\citeauthoryear{Agertz \etal}{2011}]{Ager11}
Agertz, O., Teyssier, R. \& Moore, B. 2011, \mnras, {\bf 410}, 1391

\bibitem[\protect\citeauthoryear{Agertz \etal}{2007}]{Ager07}
Agertz, O.,\skip{ Moore, B., Stadel, J., Potter, D., Miniati, F., Read, J., Mayer, L., Gawryszczak, A., Kravtsov, A., Nordlund, \o A, Pearce, F., Quilis, V., Rudd, D., Springel, V., Stone, J., Tasker, E., Teyssier, R., Wadsley, J. \& Walder, R.} 2007, \mnras, {\bf 380}. 963

\bibitem[\protect\citeauthoryear{Alam \etal}{2002}]{Alam02}
Alam, S. M. K., Bullock, J. S. \& Weinberg, D. H. 2002, \apj, {\bf 572}, 34

\bibitem[\protect\citeauthoryear{Allende Prieto \etal}{2008}]{Alle08}
Allende Prieto, C.,\skip{ Majewski, S. R., Schiavon, R., Cunha, K., Frinchaboy, P., Holtzman, J., Johnston, K., Shetrone, M., Skrutskie, M., Smith, V. \& Wilson, J.} 2008, \an, {\bf 329}, 1018

\bibitem[\protect\citeauthoryear{Andredakis \& Sanders}{1994}]{AS94}
Andredakis, Y. C. \& Sanders, R. H. 1994, \mnras, {\bf 267}, 283

\bibitem[\protect\citeauthoryear{Antoja \etal}{2009}]{Anto09}
Antoja, T.,{ Valenzuela, O., Pichardo, B., Moreno, E., Figueras, F. \& Fern\'andez, D.} 2009, \apjl, {\bf 700}, L78

\bibitem[\protect\citeauthoryear{Antoja \etal}{2011}]{Anto11}
Antoja, T.,\skip{ Figueras, F., Romero-G\'omez, M., Pichardo, B., Valenzuela, O. \& Moreno, E.} 2011, \mnras, {\bf 418}, 1423

\bibitem[\protect\citeauthoryear{Antoja \etal}{2012}]{Anto12}
Antoja, T.,\skip{ Helmi, A., Bienayme, O., Bland-Hawthorn, J., Famaey, B., Freeman, K., Gibson, B. K., Gilmore, G., Grebel, E. K., Minchev, I., Munari, U., Navarro, J., Parker, Q., Reid, W., Seabroke, G. M., Siebert, A., Siviero, A., Steinmetz, M., Williams, M., Wyse, R. \& Zwitter, T.} 2012, \mnras, {\bf 426}, L1

\bibitem[\protect\citeauthoryear{Araki}{1985}]{Arak85}
Araki, S. 1985, \PhD, MIT.

\bibitem[\protect\citeauthoryear{Arena \& Bertin}{2007}]{AB07}
Arena, S. E. \& Bertin, G. 2007, \aap, {\bf 463}, 921

\bibitem[\protect\citeauthoryear{Athanassoula}{1992}]{Atha92}
Athanassoula, E. 1992, \mnras, {\bf 259}, 345

\bibitem[\protect\citeauthoryear{Athanassoula}{2002}]{Atha02}
Athanassoula, E. 2002, \apjl, {\bf 569}, L83

\bibitem[\protect\citeauthoryear{Athanassoula}{2003}]{Atha03}
Athanassoula, E. 2003, \mnras, {\bf 341}, 1179

\bibitem[\protect\citeauthoryear{Athanassoula}{2005}]{Atha05}
Athanassoula, E. 2005, \mnras, {\bf 358}, 1477

\bibitem[\protect\citeauthoryear{Athanassoula}{2007}]{Atha07}
Athanassoula, E. 2007, \mnras, {\bf 377}, 1569

\bibitem[\protect\citeauthoryear{Athanassoula \etal}{1987}]{ABP87}
Athanassoula, E., Bosma, A. \& Papaioannou, S. 1987, \aap, {\bf 179}, 23

\bibitem[\protect\citeauthoryear{Athanassoula \etal}{2005}]{ALD05}
Athanassoula, E., Lambert, J. C. \& Dehnen, W. 2005, \mnras, {\bf 363}, 496

\bibitem[\protect\citeauthoryear{Athanassoula \& Misiriotis}{2002}]{AM02}
Athanassoula, E. \& Misiriotis, A. 2002, \mnras, {\bf 330}, 35

\bibitem[\protect\citeauthoryear{Athanassoula \etal}{1990}]{Atha90}
Athanassoula, E.,\skip{ Morin, S., Wozniak, H., Puy, D., Pierce, M. J., Lombard, J. \& Bosma, A.} 1990, \mnras, {\bf 245}, 130

\bibitem[\protect\citeauthoryear{Athanassoula \& Sellwood}{1986}]{AS86}
Athanassoula, E. \& Sellwood, J. A. 1986, \mnras, {\bf 221}, 213

\bibitem[\protect\citeauthoryear{Aumer \& Binney}{2009}]{AB09}
Aumer, M. \& Binney, J. J. 2009, \mnras, {\bf 397}, 1286

\bibitem[\protect\citeauthoryear{Baba \etal}{2013}]{Baba13}
Baba, J., Saitoh, T. R. \& Wada, K. 2013, \apj, {\bf 763}, 46

\bibitem[\protect\citeauthoryear{Bagley \etal}{2009}]{Bagl09}
Bagley, M., Minchev, I. \& Quillen, A. C. 2009, \mnras, {\bf 395}, 537

\bibitem[\protect\citeauthoryear{Bahcall \& Casertano}{1985}]{BC85}
Bahcall, J. N. \& Casertano, S. 1985, \apjl, {\bf 293}, L7

\bibitem[\protect\citeauthoryear{Balbus \& Hawley}{1998}]{BH98}
Balbus, S. A. \& Hawley, J. F. 1998, \rmp, {\bf 70}, 1

\bibitem[\protect\citeauthoryear{Baldry \etal}{2004}]{Bald04}
Baldry, I. K.,{ Glazebrook, K., Brinkmann, J., Ivezi\'c, \v Z., Lupton, R. H., Nichol, R. C. \& Szalay \& A. S.} 2004, \apj, {\bf 600}, 681

\bibitem[\protect\citeauthoryear{Barbanis \& Woltjer}{1967}]{BW67}
Barbanis, B. \& Woltjer, L. 1967, \apj, {\bf 150}, 461

\bibitem[\protect\citeauthoryear{Bardeen \etal}{1986}]{Bard86}
Bardeen, J. M.,{ Bond, J. R., Kaiser, N. \& Szalay, A. S.} 1986, \apj, {\bf 304}, 15

\bibitem[\protect\citeauthoryear{Barnes \& Hernquist}{1992}]{BH92}
Barnes, J. E. \& Hernquist, L. 1992, \araa, {\bf 30}, 705

\bibitem[\protect\citeauthoryear{Barazza \etal}{2008}]{Barr08}
Barazza, F. D., Jogee, S. \& Marinova, I. 2008, \apj, {\bf 675}, 194

\bibitem[\protect\citeauthoryear{Belokurov \etal}{2006}]{Belo06}
Belokurov, V.,\skip{ Zucker, D. B., Evans, N. W., Gilmore, G., Vidrih, S., Bramich, D. M., Newberg, H. J., Wyse, R. F. G., Irwin, M. J., Fellhauer, M., Hewett, P. C., Walton, N. A., Wilkinson, M. I., Cole, N., Yanny, B., Rockosi, C. M., Beers, T. C., Bell, E. F., Brinkmann, J., Ivezi\'c, \u Z. \& Lupton, R.} 2006, \apjl, {\bf 642}, L137

\bibitem[\protect\citeauthoryear{Benedict \etal}{2002}]{Bene02}
Benedict, G. F.,{ Howell, A., Jorgensen, I., Kenney, J. \& Smith, B. J.} 2002, \aj, {\bf 123}, 1411

\bibitem[\protect\citeauthoryear{Benjamin \etal}{2005}]{Benj05}
Benjamin, R. A.,\skip{ Churchwell, E., Babler, B. L., Indebetouw, R., Meade, M. R., Whitney, B. A., Watson, C., Wolfire, M. G., Wolff, M. J., Ignace, R., Bania, T. M., Bracker, S., Clemens, D. P., Chomiuk, L., Cohen, M., Dickey, J. M., Jackson, J. M., Kobulnicky, H. A., Mercer, E. P., Mathis, J. S., Stolovy, S. R., Uzpen, B.} 2005, \apj, {\bf 630}, L149

\bibitem[\protect\citeauthoryear{Bensby}{2013}]{Bens13}
Bensby, T. 2013, arXiv:1308.5191

\bibitem[\protect\citeauthoryear{Bensby \etal}{2005}]{BF05}
Bensby, T.,{ Feltzing, S., Lundstr\"om, I. \& Ilyin, I.} 2005, \aap, {\bf 433}, 185

\bibitem[\protect\citeauthoryear{Bensby \etal}{2007}]{Bens07}
Bensby, T.,{ Oey, M. S., Feltzing, S. \& Gustafsson, B.} 2007, \apjl, {\bf 655}, L89

\bibitem[\protect\citeauthoryear{Bensby \etal}{2011}]{Bens11}
Bensby, T.,{ Alves-Brito, A., Oey, M. S., Yong, D. \& Mel\'endez, J.} 2011, \apjl, {\bf 735} L46

\bibitem[\protect\citeauthoryear{Berentzen \etal}{2003}]{Bere03}
Berentzen, I.,{ Athanassoula, E., Heller, C. H. \& Fricke, K. J.} 2003, \mnras, {\bf 341}, 343

\bibitem[\protect\citeauthoryear{Berentzen \etal}{1998}]{Bere98}
Berentzen, I.,{ Heller, C. H., Shlosman, I. \& Fricke, K.} 1998, \mnras, {\bf 300}, 49

\bibitem[\protect\citeauthoryear{Berentzen \etal}{2007}]{Bere07}
Berentzen, I.,{ Shlosman, I., Martinez-Valpuesta, I. \& Heller, C. H.} 2007, \apj, {\bf 666}, 189

\bibitem[\protect\citeauthoryear{Bershady \etal}{2011}]{Bers11}
Bershady, M. A.,{ Martinsson, T. P. K., Verheijen, M. A. W., Westfall, K. B., Andersen, D. R. \& Swaters, R. A.} 2011, \apjl, {\bf 739}, L47

\bibitem[\protect\citeauthoryear{Bertin \& Lin}{1996}]{BL96}
Bertin, G. \& Lin, C. C. 1996, {\it Spiral Structure in Galaxies\/} (Cambridge, MA: The MIT Press)

\bibitem[\protect\citeauthoryear{Bertin \etal}{1989}]{Bert89}
Bertin, G.,{ Lin, C. C., Lowe, S. A. \& Thurstans, R. P.} 1989, \apj, {\bf 338}, 78

\bibitem[\protect\citeauthoryear{Binney}{2010}]{Binn10}
Binney, J. J. 2010, \mnras, {\bf 401}, 2318

\bibitem[\protect\citeauthoryear{Binney \& Lacey}{1988}]{BL88}
Binney, J. J. \& Lacey, C. G. 1988, \mnras, {\bf 230}, 597

\bibitem[\protect\citeauthoryear{Binney \& Tremaine}{2008}]{BT08}
Binney, J. \& Tremaine, S. 2008, Galactic Dynamics. Princeton Univ. Press, Princeton (\BTii)

\bibitem[\protect\citeauthoryear{Bird \etal}{2012}]{Bird11}
Bird, J. C., Kazantzidis, S. \& Weinberg, D. H. 2012, \mnras, {\bf 420}, 913

\bibitem[\protect\citeauthoryear{Bird \etal}{2013}]{Bird13}
Bird, J. C.,{ Kazantzidis, S., Weinberg, D. H., Guedes, J., Callegari, S., Mayer, L. \& Madau, P.} 2013, \apj, {\bf 773}, 43

\bibitem[\protect\citeauthoryear{Bland-Hawthorn \etal}{2010}]{Blan10}
Bland-Hawthorn, J., Krumholz, M. R. \& Freeman, K. 2010, \apj, {\bf 713}, 166

\bibitem[\protect\citeauthoryear{Bland-Hawthorn \etal}{2005}]{BH05}
Bland-Hawthorn, J.,\skip{ Vlaji\'c, M., Freeman, K. C. \& Draine, B. T.} 2005, \apj, {\bf 629}, 239

\bibitem[\protect\citeauthoryear{Blitz \& Spergel}{1991}]{BS91}
Blitz, L. \& Spergel, D. N. 1991, \apj, {\bf 379}, 631

\bibitem[\protect\citeauthoryear{Block \& Wainscoat}{1991}]{BW91}
Block, D. L. \& Wainscoat, R. J. 1991, \nat, {\bf 353}, 48

\bibitem[\protect\citeauthoryear{Bontekoe \& van Albada}{1987}]{BvA87}
Bontekoe, Tj. R. \& van Albada, T. S. 1987, \mnras, {\bf 224}, 349

\bibitem[\protect\citeauthoryear{Bosma}{1996}]{Bosm96}
Bosma, A. 1996, in IAU Colloq.\ {\bf 157}, {\it Barred Galaxies}, ed.\ R. Buta, D. A. Crocker \& B. G. Elmegreen (San Francisco: ASP Conf series {\bf 91}), 132

\bibitem[\protect\citeauthoryear{Bottema}{1993}]{Bott93}
Bottema, R. 1993, \aap, {\bf 275}, 16

\bibitem[\protect\citeauthoryear{Bournaud \& Combes}{2002}]{BC02}
Bournaud, F. \& Combes, F. 2002, \aap, {\bf 392}, 83

\bibitem[\protect\citeauthoryear{Bournaud \etal}{2005}]{BCS05}
Bournaud, F., Combes, F. \& Semelin, B. 2005, \mnras, {\bf 364}, L18

\bibitem[\protect\citeauthoryear{Bournaud \etal}{2009}]{Bour09}
Bournaud, F., Elmegreen, B. G. \& Martig, M. 2009, \apj, {\bf 707}, L1

\bibitem[\protect\citeauthoryear{Bovy \etal}{2009}]{Bovy09}
Bovy, J., Hogg, D. W. \& Roweis, S. T. 2009, \apj, {\bf 700}, 1794

\bibitem[\protect\citeauthoryear{Bovy \etal}{2012a}]{Bovy11a}
Bovy, J., Rix, H-W. \& Hogg, D. W. 2012a, \apj, {\bf 751}, 131

\bibitem[\protect\citeauthoryear{Bovy \etal}{2012b}]{Bovy11b}
Bovy, J.,{ Rix, H-W., Liu, C., Hogg, D. W., Beers, T. C. \& Lee, Y. S.} 2012b, \apj, {\bf 753}, 148

\bibitem[\protect\citeauthoryear{Boylan-Kolchin \etal}{2012}]{Boyl12}
Boylan-Kolchin, M., Bullock, J. S. \& Kaplinghat, M. 2012, \mnras, {\bf 422}, 1203

\bibitem[\protect\citeauthoryear{Brook \etal}{2004}]{Broo04}
Brook, C. B.,{ Kawata, D., Gibson, B. K. \& Freeman, K. C.} 2004, \apj, {\bf 612}, 894

\bibitem[\protect\citeauthoryear{Brook \etal}{2005}]{Broo05}
Brook, C. B.,{ Gibson, B. K., Martel, H. \& Kawata, D.} 2005, \apj, {\bf 630}, 298

\bibitem[\protect\citeauthoryear{Brunetti \etal}{2011}]{BC11}
Brunetti, M., Chiappini, C. \& Pfenniger, D. 2011, \aap, {\bf 534}, A75

\bibitem[\protect\citeauthoryear{Bullock \etal}{2001}]{Bull01}
Bullock, J. S.,{ Kolatt, T. S., Sigad, Y., Somerville, R. S., Kravtsov, A. V., Klypin, A. A., Primack, J. R. \& Dekel, A.} 2001, \mnras, {\bf 321}, 559

\bibitem[\protect\citeauthoryear{Bureau \& Athanassoula}{2005}]{BA05}
Bureau, M. \& Athanassoula, E. 2005, \apj, {\bf 626}, 159

\bibitem[\protect\citeauthoryear{Burstein}{1979}]{Burs79}
Burstein, D. 1979, \apj, {\bf 234}, 829

\bibitem[\protect\citeauthoryear{Buta}{1995}]{Buta95}
Buta, R. 1995, \apjs, {\bf 96}, 39

\bibitem[\protect\citeauthoryear{Buta \& Combes}{1996}]{BC96}
Buta, R. \& Combes, F. 1996, \fcp, {\bf 17}, 95 

\bibitem[\protect\citeauthoryear{Buta \etal}{2009}]{Buta09}
Buta, R. J.,\skip{ Knapen, J. H., Elmegreen, B. G., Salo, H., Laurikainen, E., Elmegreen, D. M., Puerari, I. \& Block, D. L.} 2009, \aj, {\bf 137}, 4487

\bibitem[\protect\citeauthoryear{Buta \etal}{2010a}]{Buta10a}
Buta, R.,\skip{ Laurikainen, E., Salo, H. \& Knapen, J. H.} 2010a, \apj, {\bf 721}, 259

\bibitem[\protect\citeauthoryear{Buta \etal}{2010b}]{Buta10b}
Buta, R. J.,\skip{ Sheth, K., Regan, M., Hinz, J. L., Gil de Paz, A., Men\'endez-Delmestre, K., Munoz-Mateos, J-C., Seibert, M., Laurikainen, E., Salo, H., Gadotti, D. A., Athanassoula, E., Bosma, A., Knapen, J. H., Ho, L. C., Madore, B. F., Elmegreen, D. M., Masters, K. L., Comer\'on, S., Aravena, M. \& Kim, T.} 2010b, \apjs, {\bf 190}, 147

\bibitem[\protect\citeauthoryear{Byrd \etal}{1986}]{Byrd86}
Byrd, G. G.,{ Valtonen, M. J., Valtaoja, L. \& Sundelius, B.} 1986, \aap, {\bf 166}, 75

\bibitem[\protect\citeauthoryear{Cameron \etal}{2010}]{Came10}
Cameron, E.,\skip{ Carollo, C. M., Oesch, P., Aller, M. C., Bschorr, T., Cerulo, P., Aussel, H., Capak, P., Le Floc'h, E., Ilbert, O., Kneib, J.-P., Koekemoer, A., Leauthaud, A., Lilly, S. J., Massey, R., McCracken, H. J., Rhodes, J., Salvato, M., Sanders, D. B., Scoville, N., Sheth, K., Taniguchi, Y. \& Thompson, D.} 2010, \mnras, {\bf 409}, 346

\bibitem[\protect\citeauthoryear{Carlberg}{1987}]{Carl87}
Carlberg, R. G. 1987, \apj, {\bf 322}, 59

\bibitem[\protect\citeauthoryear{Carlberg \& Freedman}{1985}]{CF85}
Carlberg, R. G. \& Freedman, W. L. 1985, \apj, {\bf 298}, 486

\bibitem[\protect\citeauthoryear{Carlberg \& Sellwood}{1985}]{CS85}
Carlberg, R. G. \& Sellwood, J. A. 1985, \apj, {\bf 292}, 79

\bibitem[\protect\citeauthoryear{Carollo \etal}{1998}]{Caro98}
Carollo, C. M., Stiavelli, M. \& Mack, J. 1998, \aj, {\bf 116}, 68

\bibitem[\protect\citeauthoryear{Casagrande \etal}{2011}]{Casa11}
Casagrande, L.,{ Sch\"onrich, R., Asplund, M., Cassisi, S., Ramirez, I., Mel\'endez, J., Bensby, T. \& Feltzing, S.} 2011, \aap, {\bf 530}, A138

\bibitem[\protect\citeauthoryear{Casetti-Dinescu \etal}{2011}]{Case11}
Casetti-Dinescu, D. I.,\skip{ Girard, T. M., Korchagin, V. I. \& van Altena, W. F.} 2011, \apj, {\bf 728}, 7

\bibitem[\protect\citeauthoryear{Ceverino \& Klypin}{2007}]{CK07}
Ceverino, D. \& Klypin, A. 2007, \mnras, {\bf 379}, 1155

\bibitem[\protect\citeauthoryear{Chakrabarty}{2007}]{Chak07}
Chakrabarty, D. 2007, \aap, {\bf 467}, 145

\bibitem[\protect\citeauthoryear{Chandrasekhar}{1943}]{Chan43}
Chandrasekhar, S. 1943, \apj, {\bf 97}, 255

\bibitem[\protect\citeauthoryear{Chandrasekhar}{1987}]{Chan87}
Chandrasekhar, S. 1987, {\it Ellipsoidal Figures of Equilibrium}, Dover: New York.

\bibitem[\protect\citeauthoryear{Chemin \etal}{2006}]{Chem06}
Chemin, L.,\skip{ Balkowski, C., Cayatte, V., Carignan, C., Amram, P., Garrido, O., Hernandez, O., Marcelin, M., Adami, C., Boselli, A. \& Boulesteix, J.} 2006, \mnras, {\bf 366}, 812

\bibitem[\protect\citeauthoryear{Chemin \& Hernandez}{2009}]{CH09}
Chemin, L. \& Hernandez, O. 2009, \aap, {\bf 499}, L25

\bibitem[\protect\citeauthoryear{Cheng \etal}{2012}]{Chen12}
Cheng, J. Y.,\skip{ Rockosi, C. M., Morrison, H. L., Lee, Y. S., Beers, T. C., Bizyaev, D., Harding, P., Malanushenko, E., Malanushenko, V., Oravetz, D., Pan, K., Schlesinger, K. J., Schneider, D. P., Simmons, A. \& Weaver, B. A.} 2012, \apj, {\bf 752}, 51 

\bibitem[\protect\citeauthoryear{Chiba \& Beers}{2000}]{CB00}
Chiba, M. \& Beers, T. C. 2000, \aj, {\bf 119}, 2843

\bibitem[\protect\citeauthoryear{Chirikov}{1979}]{Chir79}
Chirikov, B. V. 1979, \phr, {\bf 52}, 265-379

\bibitem[\protect\citeauthoryear{Christodoulou \etal}{1995}]{Chri95}
Christodoulou, D. M., Shlosman, I. \& Tohline, J. E. 1995, \apj, {\bf 443}, 551

\bibitem[\protect\citeauthoryear{Cole \etal}{2011}]{Cole11}
Cole, D. R., Dehnen, W. \& Wilkinson, M. I. 2011, \mnras, {\bf 416}, 1118

\bibitem[\protect\citeauthoryear{Col\'\i n \etal}{2006}]{CVK06}
Col\'\i n, P., Valenzuela, O. \& Klypin, A. 2006, \apj, {\bf 644}, 687

\bibitem[\protect\citeauthoryear{Combes}{2008}]{Comb08}
Combes, F. 2008, arXiv:0811.0153

\bibitem[\protect\citeauthoryear{Combes \etal}{1990}]{Comb90}
Combes, F.,{ Debbasch, F., Friedli, D. \& Pfenniger, D.} 1990, \aap, {\bf 233}, 82

\bibitem[\protect\citeauthoryear{Combes \& Sanders}{1981}]{CS81}
Combes, F. \& Sanders, R. H. 1981, \aap, {\bf 96}, 164

\bibitem[\protect\citeauthoryear{Comer\'on \etal}{2011}]{Come11}
Comer\'on, S.,\skip{ Elmegreen, B. G., Knapen, J. H., Salo, H., Laurikainen, E., Laine, J., Athanassoula, E., Bosma, A., Sheth, K., Regan, M. W., Hinz, J. L., Gil de Paz, A., Men\'endez-Delmestre, K., Mizusawa, T., Mu\u noz-Mateos, J-C., Seibert, M., Kim, T., Elmegreen, D. M., Gadotti, D. A., Ho, L. C., Holwerda, B. W., Lappalainen, J., Schinnerer, E., Skibba, R.} 2011, \apj, {\bf 741}, 28

\bibitem[\protect\citeauthoryear{Contopoulos}{1980}]{Cont80}
Contopoulos, G. 1980, \aap, {\bf 81}, 198

\bibitem[\protect\citeauthoryear{Contopoulos \& Papayannopoulos}{1980}]{CP80}
Contopoulos, G. \& Papayannopoulos, Th. 1980, \aap, {\bf 92}, 33

\bibitem[\protect\citeauthoryear{Corder \etal}{2008}]{Cord08}
Corder, S.,{ Sheth, K., Scoville, N. Z., Koda, J., Vogel, S. N. \& Ostriker, E.} 2008, \apj, {\bf 689}, 148

\bibitem[\protect\citeauthoryear{Corsini}{2008}]{Cors08}
Corsini, E. M. 2008, in {\it Formation and Evolution of Galaxy Bulges}, IAU Symp.~{\bf 245} (Dordrecht: Kluwer) p.~125

\bibitem[\protect\citeauthoryear{Corsini \etal}{2003}]{CDA03}
Corsini, E. M., Debattista, V. P. \& Aguerri, J. A. L. 2003, \apjl, {\bf 599}, L29

\bibitem[\protect\citeauthoryear{Courteau \etal}{2003}]{Cour03}
Courteau, S.,\skip{ Andersen, D. R., Bershady, M. A., MacArthur, L. A. \& Rix, H-W.} 2003, \apj, {\bf 594}, 208

\bibitem[\protect\citeauthoryear{Curir \etal}{2006}]{Curi06}
Curir, A., Mazzei, P. \& Murante, G. 2006, \aap, {\bf 447}, 453

\bibitem[\protect\citeauthoryear{Debattista \etal}{2004}]{Deba04}
Debattista, V. P.,\skip{ Carollo, C. M., Mayer, L. \& Moore, B.} 2004, \apjl, {\bf 604}, L93

\bibitem[\protect\citeauthoryear{Debattista \etal}{2006}]{Deba06}
Debattista, V. P.,{ Mayer, L., Carollo, C. M., Moore, B., Wadsley, J. \& Quinn, T.} 2006, \apj, {\bf 645}, 209

\bibitem[\protect\citeauthoryear{Debattista \& Sellwood}{1998}]{DS98}
Debattista, V. P. \& Sellwood, J. A. 1998, \apjl, {\bf 493}, L5

\bibitem[\protect\citeauthoryear{Debattista \& Sellwood}{1999}]{DS99}
Debattista, V. P. \& Sellwood, J. A. 1999, \apjl, {\bf 513}, L107

\bibitem[\protect\citeauthoryear{Debattista \& Sellwood}{2000}]{DS00}
Debattista, V. P. \& Sellwood, J. A. 2000, \apj, {\bf 543}, 704

\bibitem[\protect\citeauthoryear{Debattista \& Shen}{2007}]{DS07}
Debattista, V. P. \& Shen, J. A. 2007, \apjl, {\bf 654}, L127

\bibitem[\protect\citeauthoryear{Dehnen}{1998}]{Dehn98}
Dehnen, W. 1998, \aj, {\bf 115}, 2384

\bibitem[\protect\citeauthoryear{Dehnen}{2000}]{Dehn00}
Dehnen, W. 2000, \aj, {\bf 119}, 800

\bibitem[\protect\citeauthoryear{Dekker}{1976}]{Dekk76}
Dekker, E. 1976, Phys.\ Rep., {\bf 24}, 315

\bibitem[\protect\citeauthoryear{Deng \etal}{2012}]{Deng12}
Deng, L-C.,\skip{ Newberg, H. J., Liu, C., Carlin, J. L., Beers, T. C., Chen, L., Chen, Y-Q., Christlieb, N., Grillmair, C. J., Guhathakurta, P., Han, Z-W., Hou, J-L., Lee, H.-T., L\'epine, S., Li, J., Liu, X-W., Pan, K-K., Sellwood, J. A., Wang, B., Wang, H-C., Yang, F., Yanny, B., Zhang, H-T., Zhang, Y-Y., Zheng, Z. \& Zhu, Z.} 2012, \raa, {\bf 12}, 735

\bibitem[\protect\citeauthoryear{De Simone \etal}{2004}]{dSWT}
De Simone, R. S., Wu, X. \& Tremaine, S. 2004, \mnras, {\bf 350}, 627

\bibitem[\protect\citeauthoryear{Dickey \etal}{1990}]{Dick90}
Dickey, J. M., Hanson, M. M. \& Helou, G. 1990, \apj, {\bf 352} 522

\bibitem[\protect\citeauthoryear{Dierickx \etal}{2010}]{Dier10}
Dierickx, M.,{ Klement, R., Rix, H-W. \& Liu, C.} 2010, \apjl, {\bf 725}, L186

\bibitem[\protect\citeauthoryear{D'Onghia \etal}{2013}]{DVH12}
D'Onghia, E., Vogelsberger, M. \& Hernquist, L. 2013, \apj, {\bf 766}, 34

\bibitem[\protect\citeauthoryear{Drury}{1980}]{Drur80}
Drury, L. O'C. 1980, \mnras, {\bf 193}, 337

\bibitem[\protect\citeauthoryear{Dubinski \etal}{2009}]{DBS09}
Dubinski, J., Berentzen, I. \& Shlosman, I. 2009, \apj, {\bf 697}, 293

\bibitem[\protect\citeauthoryear{Edvardsson \etal}{1993}]{Edva93}
Edvardsson, B.,{  Andersen, B., Gustafsson, B., Lambert, D. L., Nissen, P. E. \& Tomkin, J.} 1993, \aap, {\bf 275}, 101

\bibitem[\protect\citeauthoryear{Efremov}{2010}]{Efre10}
Efremov, Yu. N. 2010, \mnras, {\bf 405}, 1531

\bibitem[\protect\citeauthoryear{Efstathiou \etal}{1982}]{ELN82}
Efstathiou, G., Lake, G. \& Negroponte, J. 1982, \mnras, {\bf 199}, 1069

\bibitem[\protect\citeauthoryear{Eliche-Moral \etal}{2011}]{Elic11}
Eliche-Moral, M. C.,{ Gonz\'alez-García, A. C., Balcells, M., Aguerri, J. A. L., Gallego, J., Zamorano, J. \& Prieto, M.} 2011, \aap, {\bf 533}, A104

\bibitem[\protect\citeauthoryear{Elmegreen \& Elmegreen}{1982}]{EE82}
Elmegreen, D. M. \& Elmegreen, B. G. 1982, \mnras, {\bf 201}, 1021

\bibitem[\protect\citeauthoryear{Elmegreen}{1996}]{Elme96}
Elmegreen, B. 1996, in IAU Colloq.\ {\bf 157}, {\it Barred Galaxies}, ed.\ R. Buta, D. A. Crocker \& B. G. Elmegreen (San Francisco: ASP Conf series {\bf 91}), 197

\bibitem[\protect\citeauthoryear{El-Zant \etal}{2001}]{ElZa01}
El-Zant, A., Shlosman, I. \& Hoffman, Y. 2001, \apj, {\bf 560}, 636

\bibitem[\protect\citeauthoryear{Englmaier \& Shlosman}{2004}]{ES04}
Englmaier, P. \& Shlosman, I. 2004, \apjl, {\bf 617}, L115

\bibitem[\protect\citeauthoryear{Erwin}{2005}]{Erwi05}
Erwin, P. 2005, \mnras, {\bf 364}, 283

\bibitem[\protect\citeauthoryear{Erwin \etal}{2005}]{EBP05}
Erwin, P., Beckman, J. E. \& Pohlen, M. 2005, \apjl, {\bf 626}, L81

\bibitem[\protect\citeauthoryear{Erwin \& Debattista}{2013}]{ED13}
Erwin, P. \& Debattista, V. P. 2013, \mnras, {\bf 431}, 3060

\bibitem[\protect\citeauthoryear{Erwin \etal}{2008a}]{Erwi08}
Erwin, P., Pohlen, M. \& Beckman, J. E.	2008a, \aj, {\bf 135}, 20

\bibitem[\protect\citeauthoryear{Erwin \etal}{2008b}]{EPGB}
Erwin, P.,{ Pohlen, M., Guti\'errez, L. \& Beckman, J. E.} 2008b, in {\it Formation and Evolution of Galaxy Disks}, eds.\ J. G. Funes SJ \&  E. M. Corsini (San Francisco: ASP) {\bf 396}, p.~207

\bibitem[\protect\citeauthoryear{Erwin \& Sparke}{2002}]{ES02}
Erwin, P. \& Sparke, L. S. 2002, \aj, {\bf 124}, 65

\bibitem[\protect\citeauthoryear{Eskridge \etal}{2000}]{Eskr00}
Eskridge, P. B.,\skip{ Frogel, J. A., Pogge, R. W., Quillen, A. C., Davies, R. L., DePoy, D. L., Houdashelt, M. L., Kuchinski, L. E., Ram\'\i rez, S. V., Sellgren, K., Terndrup, D. M. \& Tiede, G. P.} 2000, \aj, {\bf 119}, 536.

\bibitem[\protect\citeauthoryear{Evans \& Read}{1998}]{ER98}
Evans, N. W. \& Read, J. C. A. 1998, \mnras, {\bf 300}, 106

\bibitem[\protect\citeauthoryear{Famaey \etal}{2007}]{Fama07}
Famaey, B.,{ Pont, F., Luri, X., Udry, S., Mayor, M. \& Jorrissen, A.} 2007, \aap, {\bf 461}, 957

\bibitem[\protect\citeauthoryear{Fathi \etal}{2007}]{Fath07}
Fathi, K.,{ Toonen, S., Falc\'on-Barroso, J., Beckman, J. E., Hernandez, O., Daigle, O., Carignan, C. \& de Zeeuw, T.} 2007. \apjl, {\bf 667}, L137

\bibitem[\protect\citeauthoryear{Fathi \etal}{2009}]{Fath09}
Fathi, K.,{ Beckman, J. E., Pi\~nol-Ferrer, N., Hernandez, O., Mart\'\i nez-Valpuesta, I. \& Carignan, C.} 2009, \apj, {\bf 704}, 1657

\bibitem[\protect\citeauthoryear{Fisher \& Drory}{2008}]{FD08}
Fisher, D. B. \& Drory, N. 2008, \aj, {\bf 136}, 773

\bibitem[\protect\citeauthoryear{Fisher \& Drory}{2011}]{FD11}
Fisher, D. B. \& Drory, N. 2011, \apjl, {\bf 733}, L47

\bibitem[\protect\citeauthoryear{Fisher \etal}{2009}]{Fish09}
Fisher, D. B., Drory, N. \& Fabricius, M. H. 2009, \apj, {\bf 697}, 630

\bibitem[\protect\citeauthoryear{Folye \etal}{2008}]{Foyl08}
Foyle, K., Courteau, S. \& Thacker, R. J. 2008, \mnras, {\bf 386}, 1821

\bibitem[\protect\citeauthoryear{Fraternali \& Tomassetti}{2012}]{FT12}
Fraternali, F. \& Tomassetti, M. 2012, \mnras, {\bf 426}, 2166

\bibitem[\protect\citeauthoryear{Freeman}{1970}]{Free70}
Freeman, K. C. 1970, \apj, {\bf 160}, 811

\bibitem[\protect\citeauthoryear{Freeman \etal}{2013}]{Free13}
Freeman, K.,\skip{ Ness, M., Wylie-de-Boer, E., Athanassoula, E., Bland-Hawthorn, J., Asplund, M., Lewis, G., Yong, D., Lane, R., Kiss, L. \& Ibata, R.} 2013, \mnras, {\bf 428}, 3660

\bibitem[\protect\citeauthoryear{Frenk \& White}{2012}]{FW12}
Frenk, C. S. \& White, S. D. M. 2012, \anp, {\bf 524}, 507

\bibitem[\protect\citeauthoryear{Fridman \& Polyachenko}{1984}]{FP84}
Fridman, A. M. \& Polyachenko, V. L. 1984. {\it Physics of Gravitating Systems} (New York: Springer-Verlag)

\bibitem[\protect\citeauthoryear{Friedli \etal}{1994}]{FBK94}
Friedli, D., Benz, W. \& Kennicutt, R. 1994, \apjl, {\bf 430}, L105

\bibitem[\protect\citeauthoryear{Friedli \& Martinet}{1993}]{FM93}
Friedli, D. \& Martinet, L. 1993, \aap, {\bf 277}, 27

\bibitem[\protect\citeauthoryear{Friedli \& Pfenniger}{1994}]{FP90}
Friedli, D. \& Pfenniger, D. 1990, in ``Bulges of Galaxies'', eds. B. J. Jarvis \& D. M. Terndrup (Garching: ESO workshop {\bf 35}) p.~265

\bibitem[\protect\citeauthoryear{Fuchs \etal}{2005}]{Fuch05}
Fuchs, B., Dettbarn, C. \& Tsuchiya, T. 2005, \aap, {\bf 444}, 1

\bibitem[\protect\citeauthoryear{Fuhrmann}{2008}]{Fuhr08}
Fuhrmann, K. 2008, \mnras, {\bf 384}, 173

\bibitem[\protect\citeauthoryear{Fujii \etal}{2011}]{Fuji11}
Fujii, M. S.,{ Baba, J., Saitoh, T. R., Makino, J., Kokubo, E. \& Wada, K.} 2011, \apj, {\bf 730}, 109

\bibitem[\protect\citeauthoryear{Fux}{1999}]{Fux99}
Fux, R. 1999, \aap, {\bf 345}, 787

\bibitem[\protect\citeauthoryear{Gadotti}{2011}]{Gado11}
Gadotti, D. A. 2011, \mnras, {\bf 415}, 3308

\bibitem[\protect\citeauthoryear{Gadotti \etal}{2007}]{Gado07}
Gadotti, D. A.,{ Athanassoula, E., Carrasco, L., Bosma, A., de Souza, R. E. \& Recillas, E.} 2007, \mnras, {\bf 381}, 943

\bibitem[\protect\citeauthoryear{Gao \etal}{2011}]{Gao11}
Gao, L.,{ Frenk, C. S., Boylan-Kolchin, M., Jenkins, A., Springel, V. \& White, S. D. M.} 2011, \mnras, {\bf 410}, 2309

\bibitem[\protect\citeauthoryear{Garcia-Barreto \etal}{1991b}]{GB91b}
Garcia-Barreto, J. A.,{ Downes, D., Combes, F., Gerin, M., Carrasco, L. \& Cruz-Gonzales, I.} 1991b, \aap, {\bf 252}, 19

\bibitem[\protect\citeauthoryear{Garcia-Barreto \etal}{1991a}]{GB91a}
Garcia-Barreto, J. A.,{ Downes, D., Combes, F., Gerin, M., Magri, C., Carrasco, L. \& Cruz-Gonzalez, I.} 1991a, \aap, {\bf 244}, 257

\bibitem[\protect\citeauthoryear{Gerhard \& Binney}{1985}]{GB85}
Gerhard, O. E. \& Binney, J. 1985, \mnras, {\bf 216}, 467

\bibitem[\protect\citeauthoryear{Gerhard \& Martinez-Valpuesta}{2012}]{GM12}
Gerhard, O. E. \& Martinez-Valpuesta, I. 2012, \apjl, {\bf 744}, L8

\bibitem[\protect\citeauthoryear{Gerin \etal}{1990}]{Geri90}
Gerin, M., Combes, F. \& Athanassoula, E. 1990, \aap, {\bf 230}, 37

\bibitem[\protect\citeauthoryear{Gerin \etal}{1988}]{Geri88}
Gerin, M., Combes, F. \& Nakai, N. 1988, \aap, {\bf 203}, 44-50

\bibitem[\protect\citeauthoryear{Gerssen \etal}{2000}]{Gers00}
Gerssen, J., Kuijken, K. \& Merrifield, G. K. 2000, \mnras, {\bf 317}, 545

\bibitem[\protect\citeauthoryear{Gerssen \& Shapiro}{2012}]{GS12}
Gerssen, J. \& Shapiro, M. R. 2012, \mnras, {\bf 423}, 2726

\bibitem[\protect\citeauthoryear{Gilmore \& Reid}{1983}]{GR83}
Gilmore, G. \& Reid, N. 1983, \mnras, {\bf 202}, 1025

\bibitem[\protect\citeauthoryear{Gnedin \etal}{1996}]{GGF95}
Gnedin, O. Y., Goodman, J. \& Frei, Z. 1995, \aj, {\bf 110}, 1105

\bibitem[\protect\citeauthoryear{Goerdt \etal}{2010}]{Goer10}
Goerdt, T.,{ Moore, B., Read, J. I. \& Stadel, J.} 2010, \apj, {\bf 725}, 1707

\bibitem[\protect\citeauthoryear{Goldreich \& Lynden-Bell}{1965}]{GLB65}
Goldreich, P. \& Lynden-Bell, D. 1965, \mnras, {\bf 130}, 125

\bibitem[\protect\citeauthoryear{Gonz\'alez-Fern\'andez \etal}{2012}]{Gonz12}
Gonz\'alez-Fern\'andez, C.,{ L\'opez-Corredoira, M., Am\^ores, E. B., Minniti, D., Lucas, P. \& Toledo, I.} 2012, \aap, {\bf 546}, A107

\bibitem[\protect\citeauthoryear{Governato \etal}{2009}]{Gove09}
Governato, F.,\skip{ Brook, C. B., Brooks, A. M., Mayer, L., Willman, B., Jonsson, P., Stilp, A. M., Pope, L., Christensen, C., Wadsley, J. \& Quinn, T.} 2009, \mnras, {\bf 398}, 312

\bibitem[\protect\citeauthoryear{Governato \etal}{2010}]{Gove10}
Governato, F.\skip{ Brook, C., Mayer, L., Brooks, A., Rhee, G., Wadsley, J., Jonsson, P., Willman, B., Stinson, G., Quinn, T. \& Madau, P.} 2010, \nat, {\bf 463}, 203

\bibitem[\protect\citeauthoryear{Grand \etal}{2012}]{Gran12}
Grand, R. J. J., Kawata, D. \& Cropper, M.  2012, \mnras, {\bf 421}, 1529

\bibitem[\protect\citeauthoryear{Gratier \etal}{2010}]{Grat10}
Gratier, P.,\skip{ Braine, J., Rodriguez-Fernandez, N. J., Schuster, K. F., Kramer, C., Xilouris, E. M., Tabatabaei, F. S., Henkel, C., Corbelli, E., Israel, F., van der Werf, P. P., Calzetti, D., Garcia-Burillo, S., Sievers, A., Combes, F., Wiklind, T., Brouillet, N., Herpin, F., Bontemps, S., Aalto, S., Koribalski, B., van der Tak, F., Wiedner, M. C., R\"ollig, M. \& Mookerjea, B.} 2010, \aap. {\bf 522}, A3

\bibitem[\protect\citeauthoryear{Grosb\o l \etal}{2004}]{GPP04}
Grosb\o l, P., Patsis, P. A. \& Pompei, E. 2004, \aap, {\bf 423}, 849

\bibitem[\protect\citeauthoryear{Guedes \etal}{2013}]{Gued12}
Guedes, J.,{ Mayer, L., Carollo, M. \& Madau, P.} 2013, \apj, {\bf 772}, 36

\bibitem[\protect\citeauthoryear{G\"ultekin \etal}{2009}]{Gult09}
G\"ultekin, K.,\skip{ Richstone, D. O., Gebhardt, K., Lauer, T. R., Tremaine, S., Aller, M. C., Bender, R., Dressler, A., Faber, S. M., Filippenko, A. V., Green, R., Ho, L. C., Kormendy, J., Magorrian, J., Pinkney, J. \& Siopis, C.} 2009, \apj, {\bf 698}, 198

\bibitem[\protect\citeauthoryear{Gunn}{1982}]{Gunn82}
Gunn, J. E. 1982, in {\it Astrophysical Cosmology}, eds.\ H. A. Br\"uck, G. V. Coyne \& M. S. Longair (Vatican City: Pontificia Academia Scientiarum) p.~233

\bibitem[\protect\citeauthoryear{Haan \etal}{2009}]{Haan09}
Haan, S.,{ Schinnerer, E., Emsellem, E., García-Burillo, S., Combes, F., Mundell, C. G. \& Rix, H.-W.} 2009, \apj, {\bf 692}, 1623

\bibitem[\protect\citeauthoryear{Hahn \etal}{2011}]{Hahn11}
Hahn, C. H., Sellwood, J. A. \& Pryor, C. 2011, \mnras, {\bf 418}, 2459

\bibitem[\protect\citeauthoryear{H\"anninen \& Flynn}{2002}]{HF02}
H\"anninen, J. \& Flynn, C. 2002, \mnras, {\bf 337}, 731

\bibitem[\protect\citeauthoryear{Hao \etal}{2009}]{Hao09}
Hao, L.,{ Jogee, S., Barazza, F. D., Marinova, I. \& Shen, J.} 2009, in ASP Conf.\ Ser.\ {\bf 419}, Galaxy Evolution: Emerging Insights and Future Challenges, ed.\ S. Jogee,\skip{ I. Marinova, L. Hao, \& G. A. Blanc} (San Francisco, CA: ASP), 402

\bibitem[\protect\citeauthoryear{Hawarden \etal}{1986}]{Hawa86}
Hawarden, T. G.,{ Mountain, C. M., Leggett, S. K. \& Puxley, P. J.} 1986, \mnras, {\bf 221}, 41P

\bibitem[\protect\citeauthoryear{Haywood}{2008}]{Hayw08}
Haywood, M. 2008, \mnras, {\bf 388}, 1175

\bibitem[\protect\citeauthoryear{Haywood}{2012}]{Hayw12}
Haywood, M. 2012, in {\it Assembling the Puzzle of the Milky Way} Eds.\ C. Reyl\'e, A. Robin \& M. Schultheis (EPJ Web of Conferences, {\bf 19}) id.05001

\bibitem[\protect\citeauthoryear{Heller \etal}{2007}]{Hell07}
Heller, C. H., Shlosman, I. \& Athanassoula, E. 2007, \apj, {\bf 657}, L65

\bibitem[\protect\citeauthoryear{Heller \etal}{2001}]{HSE01}
Heller, C., Shlosman, I. \& Englmaier, P. 2001, \apj, {\bf 553}, 661

\bibitem[\protect\citeauthoryear{H\'enon}{1973}]{Heno73}
H\'enon, M. 1973, in {\it Dynamical Structure and Evolution of Stellar Systems}, ed.\ L. Martinet \& M. Mayor (Sauverny: Geneva Observatory) p.~182

\bibitem[\protect\citeauthoryear{Hernandez \etal}{2005}]{Hern05}
Hernandez, O.,{ Carignan, C., Amram, P., Chemin, L. \& Daigle, O.} 2005, \mnras, {\bf 360}, 1201

\bibitem[\protect\citeauthoryear{Hernquist \& Weinberg}{1992}]{HW92}
Hernquist, L. \& Weinberg, M. D. 1992, \apj, {\bf 400}, 80

\bibitem[\protect\citeauthoryear{Herrmann \& Ciardullo}{2009}]{HC09}
Herrmann, K. A. \& Ciardullo, R. 2009, \apj, {\bf 705}, 1686
	
\bibitem[\protect\citeauthoryear{Hill \etal}{2012}]{Hill12}
Hill, V.,{ Babusiaux, C., Gómez, A., Haywood, M., Katz, D. \& Royer, F.} 2012, in {\it Assembling the Puzzle of the Milky Way} Eds.\ C. Reyl\'e, A. Robin \& M. Schultheis (EPJ Web of Conferences, {\bf 19}) id.06001

\bibitem[\protect\citeauthoryear{Hinshaw \etal}{2013}]{Hins12}
Hinshaw, G.,\skip{ Larson, D., Komatsu, E., Spergel, D. N., Bennett, C. L., Dunkley, J., Nolta, M. R., Halpern, M., Hill, R. S., Odegard, N., Page, L., Smith, K. M., Weiland, J. L., Gold, B., Jarosik, N., Kogut, A., Limon, M., Meyer, S. S., Tucker, G. S., Wollack, E. \& Wright, E. L.} 2013, \apjs, {\bf 208}, 19

\bibitem[\protect\citeauthoryear{Hohl}{1971}]{Hohl71}
Hohl, F. 1971, \apj, {\bf 168}, 343

\bibitem[\protect\citeauthoryear{Hohl}{1973}]{Hohl73}
Hohl, F. 1973, \apj, {\bf 184}, 353

\bibitem[\protect\citeauthoryear{Holley-Bockelmann \etal}{2005}]{HWK05}
Holley-Bockelmann, K., Weinberg, M. \& Katz, N. 2005, \mnras, {\bf 363}, 991

\bibitem[\protect\citeauthoryear{Holmberg \etal}{2007}]{HNA07}
Holmberg, J., Nordstr\"om, B. \& Andersen, J. 2007, \aap, {\bf 475}, 519

\bibitem[\protect\citeauthoryear{Holmberg \etal}{2009}]{HNA09}
Holmberg, J., Nordstr\"om, B. \& Andersen, J. 2009, \aap, {\bf 501}, 941

\bibitem[\protect\citeauthoryear{Hopkins}{2013}]{Hopk13}
Hopkins, P. F. 2013 \mnras, {\bf 428}, 2840

\bibitem[\protect\citeauthoryear{House \etal}{2011}]{Hous11}
House, E. L.,\skip{ Brook, C. B., Gibson, B. K., S\'anchez-Bl\'azquez, P., Courty, S., Few, C. G., Governato, F., Kawata, D., Ro\v skar, R., Steinmetz, M., Stinson, G. S. \& Teyssier, R.} 2011, \mnras, {\bf 415}, 2652

\bibitem[\protect\citeauthoryear{Hoyle \etal}{2011}]{Hoyl11}
Hoyle, B.,\skip{ Masters, K.. L., Nichol, R. C., Edmondson, E. M., Smith, A. M., Lintott, C., Scranton, R., Bamford, S., Schawinski, K. \& Thomas, D.} 2011, \mnras, {\bf 415}, 3627 

\bibitem[\protect\citeauthoryear{Huang \& Carlberg}{1997}]{HC97}
Huang, S. \& Carlberg, R. G. 1997, \apj, 480, 503

\bibitem[\protect\citeauthoryear{Hunter \& Elmegreen}{2006}]{HE06}
Hunter, D. A. \& Elmegreen, B. G. 2006, \apjs, {\bf 162}, 49

\bibitem[\protect\citeauthoryear{Ida \etal}{1993}]{Ida93}
Ida, S., Kokuba, E. \& Makino, J. 1993, \mnras, {\bf 263}, 875

\bibitem[\protect\citeauthoryear{Ivezi\'c \etal}{2008}]{Ivez08}
Ivezi\'c, \v Z. 2008, \aj, {\bf 684}, 287

\bibitem[\protect\citeauthoryear{Jalali}{2007}]{Jala07}
Jalali, M. A. 2007, \apj, {\bf 669}, 218

\bibitem[\protect\citeauthoryear{James \& Sellwood}{1978}]{JS78}
James, R. A. \& Sellwood, J. A. 1978, \mnras, {\bf 182}, 331

\bibitem[\protect\citeauthoryear{Jardel \& Sellwood}{2009}]{JS09}
Jardel, J. \& Sellwood, J. A. 2009, \apj, {\bf 691}, 1300

\bibitem[\protect\citeauthoryear{Jenkins \& Binney}{1990}]{JB90}
Jenkins, A. \& Binney, J. J. 1990, \mnras, {\bf 245}, 305

\bibitem[\protect\citeauthoryear{Julian \& Toomre}{1966}]{JT66}
Julian, W. H. \& Toomre, A. 1966, \apj, {\bf 146}, 810

\bibitem[\protect\citeauthoryear{Juri\'c \etal}{2008}]{Juri08}
Juri\'c, M. 2008, \aj, {\bf 673}, 864

\bibitem[\protect\citeauthoryear{Kalnajs}{1972}]{Kaln72}
Kalnajs, A. J. 1972, \apj, {\bf 175}, 63

\bibitem[\protect\citeauthoryear{Kalnajs}{1973}]{Kaln73}
Kalnajs, A. J. 1973, {\it Proc. Astron. Soc. Australia}, {\bf 2}, 174

\bibitem[\protect\citeauthoryear{Kalnajs}{1976}]{Kaln76}
Kalnajs, A. J. 1976, \apj, {\bf 205}, 751

\bibitem[\protect\citeauthoryear{Kalnajs}{1977}]{Kaln77}
Kalnajs, A. J. 1977, \apj, {\bf 212}, 637

\bibitem[\protect\citeauthoryear{Kalnajs}{1978}]{Kaln78}
Kalnajs, A. J. 1978, in IAU Symposium {\bf 77} {\it Structure and Properties of Nearby Galaxies} eds.\ E. M. Berkhuisjen \& R. Wielebinski (Dordrecht:Reidel) p.~113

\bibitem[\protect\citeauthoryear{Kalnajs}{1983}]{Kaln83}
Kalnajs, A. J. 1983, in {\it Internal Kinematics and Dynamics of Galaxies}, IAU Symp.\ {\bf 100}, ed.\ E. Athanassoula (Dordrecht: Reidel) p~87

\bibitem[\protect\citeauthoryear{Kalnajs}{1991}]{Kaln91}
Kalnajs, A. J. 1991, in {\it Dynamics of Disc Galaxies}, ed.\ B. Sundelius ( Gothenburg: G\"oteborgs University) p.~323

\bibitem[\protect\citeauthoryear{Kaufmann \etal}{2006}]{Kauf06}
Kaufmann, T., Mayer, L., Wadsley, J., Stadel, J. \& Moore, B. 2006, \mnras, {\bf 370}, 1612

\bibitem[\protect\citeauthoryear{Kazantzidis \etal}{2009}]{Kaza09}
Kazantzidis, S.,{ Zentner, A. R., Kravtsov, A. V., Bullock, J. S., \& Debattista, V. P.} 2009, \apj, {\bf 700}, 1896

\bibitem[\protect\citeauthoryear{Kendall \etal}{2011}]{KKC11}
Kendall, S., Kennicutt, R. C. \& Clarke, C. 2011, \mnras, {\bf 414}, 538

\bibitem[\protect\citeauthoryear{Kent}{1986}]{Kent86}
Kent, S. M. 1986, \aj, {\bf 91}, 1301

\bibitem[\protect\citeauthoryear{Kim \etal}{2012}]{Kim12}
Kim, W-T.,{ Seo, W-Y., Stone, J. M., Yoon, D. \& Teuben, P. J.} 2012, \apj, {\bf 747}, 60

\bibitem[\protect\citeauthoryear{Kim \& Stone}{2012}]{KS12}
Kim, W-T. \& Stone, J. M. 2012, \apj, {\bf 751}, 124

\bibitem[\protect\citeauthoryear{Klypin \etal}{2009}]{Klyp09}
Klypin, A.,{ Valenzuela, O., Col\'\i n, P. \& Quinn, T.} 2009, \mnras, {\bf 398}, 1027

\bibitem[\protect\citeauthoryear{Knapen \etal}{2000}]{Knap00}
Knapen, J. H., Shlosman, I. \& Peletier, R. F. 2000, \apj, {\bf 529}, 93

\bibitem[\protect\citeauthoryear{Koda}{2009}]{Koda09}
Koda, J.,\skip{ Scoville, N., Sawada, T., La Vigne, M. A., Vogel, S. N., Potts, A. E., Carpenter, J. M., Corder, S. A., Wright, M. C. H., White, S. M., Zauderer, B. A., Patience, J., Sargent, A. I., Bock, D. C. J., Hawkins, D., Hodges, M., Kemball, A., Lamb, J. W., Plambeck, R. L., Pound, M. W., Scott, S. L., Teuben, P. \& Woody, D. P.} 2009, \apjl, {\bf 700}, L132

\bibitem[\protect\citeauthoryear{Kormendy}{1979}]{Korm79}
Kormendy, J. 1979, \apj, {\bf 227}, 714

\bibitem[\protect\citeauthoryear{Kormendy}{1983}]{Korm83}
Kormendy, J. 1983, \apj, {\bf 275}, 529

\bibitem[\protect\citeauthoryear{Kormendy}{2012}]{Korm12}
Kormendy, J. 2012, In XXIII Canary Islands Winter School of Astrophysics, ``Secular Evolution of Galaxies'' eds. J. Falc\'on-Barroso \& J. H. Knapen (Cambridge: Cambridge University Press) (to appear)

\bibitem[\protect\citeauthoryear{Kormendy \etal}{2010}]{Korm10}
Kormendy, J.,{ Drory, N., Bender, R. \& Cornell, M. E.} 2010, \apj, {\bf 723}, 54

\bibitem[\protect\citeauthoryear{Kormendy \& Illingworth}{1982}]{KI82}
Kormendy, J. \& Illingworth, G. 1982, \apj, {\bf 256}, 460

\bibitem[\protect\citeauthoryear{Kormendy \& Kennicutt}{2004}]{KK04}
Kormendy, J. \& Kennicutt, R. C. 2004, \araa, {\bf 42}, 603

\bibitem[\protect\citeauthoryear{Kormendy \& Norman}{1979}]{KN79}
Kormendy, J. \& Norman, C. A. 1979, \apj, {\bf 233}, 539

\bibitem[\protect\citeauthoryear{Kraljic \etal}{2012}]{KBM12}
Kraljic, K., Bournaud, F. \& Martig, M. 2012, \apj, {\bf 757}, 60

\bibitem[\protect\citeauthoryear{Kravtsov \etal}{2002}]{Krav02}
Kravtsov, A. V., Klypin, A, \& Hoffman, Y. 2002, \apj, {\bf 571}, 563

\bibitem[\protect\citeauthoryear{Kregel \etal}{2002}]{Kreg02}
Kregel, M., van der Kruit, P. C. \& de Grijs, R. 2002, \mnras, {\bf 334}, 646

\bibitem[\protect\citeauthoryear{Krolik}{1999}]{Krol99}
Krolik, J. 1999, Active Galactic Nuclei. Princeton Univ. Press, Princeton (\BTii)

\bibitem[\protect\citeauthoryear{Kroupa}{2002}]{Krou02}
Kroupa, P. 2002, \mnras, {\bf 330}, 707

\bibitem[\protect\citeauthoryear{Kulsrud \etal}{1971}]{Kuls71}
Kulsrud, R. M., Mark, J. W-K. \& Caruso, A. 1971, \apss, {\bf 14}, 52

\bibitem[\protect\citeauthoryear{Kuzio de Naray \& Spekkens}{2011}]{KS11}
Kuzio de Naray, R. \& Spekkens, K. 2011, \apjl, {\bf 741}, L29

\bibitem[\protect\citeauthoryear{Lacey}{1984}]{Lace84}
Lacey, C. G. 1984, \mnras, {\bf 208}, 687

\bibitem[\protect\citeauthoryear{Lacey}{1991}]{Lace91}
Lacey, C. G. 1991, in {\it Dynamics of Disc Galaxies}, ed.\ B. Sundelius (Gothenburg: G\"oteborgs University) p.~257

\bibitem[\protect\citeauthoryear{Laine \etal}{2002}]{Lain02}
Laine, S.,{ Shlosman, I., Knapen, J. H. \& Peletier, R. F.} 2002, \apj, {\bf 567}, 97

\bibitem[\protect\citeauthoryear{Larsen \etal}{2008}]{Lars08}
Larsen, J. A., Humphreys, R. M. \& Cabanela, J, E. 2008, \apjl, {\bf 687}, L17

\bibitem[\protect\citeauthoryear{Larson \etal}{2011}]{Lars11}
Larson, D.,\skip{ Dunkley, J., Hinshaw, G., Komatsu, E., Nolta, M. R., Bennett, C. L., Gold, B., Halpern, M., Hill, R. S., Jarosik, N., Kogut, A., Limon, M., Meyer, S. S., Odegard, N., Page, L., Smith, K. M., Spergel, D. N., Tucker, G. S., Weiland, J. L., Wollack, E. \& Wright, E. L.} 2011, \apjs, {\bf 192}, 16

\bibitem[\protect\citeauthoryear{Laurikainen \etal}{2004}]{Laur04}
Laurikainen, E., Salo, H. \& Buta, R. 2004, \apj, {\bf 607}, 103

\bibitem[\protect\citeauthoryear{Lee \etal}{2012}]{Lee12}
Lee, G-H.,{ Woo, J-H., Lee, M. G., Hwang, H. S., Lee, J. C., Sohn, J. \& Lee, J. H.} 2012, \apj, {\bf 750}, 141

\bibitem[\protect\citeauthoryear{Lee \etal}{2011}]{Lee11}
Lee, Y. S. 2011, \apj, {\bf 738}, 187

\bibitem[\protect\citeauthoryear{Lewis \& Freeman}{1989}]{LF89}
Lewis, J. R. \& Freeman, K. C. 1989, \aj, {\bf 97}, 139

\bibitem[\protect\citeauthoryear{Li \& Shen}{2012}]{LS12}
Li, Z-Y. \& Shen, J. 2012, \apjl, {\bf 757}, L7

\bibitem[\protect\citeauthoryear{Lin \& Tremaine}{1983}]{LT83}
Lin, D. N. C. \& Tremaine, S. 1983, \apj, {\bf 264}, 364

\bibitem[\protect\citeauthoryear{Lindblad \etal}{1996}]{Lind96}
Lindblad, P. A. B., Lindblad, P. O. \& Athanassoula, E. 1996, \aap, {\bf 313}, 65

\bibitem[\protect\citeauthoryear{Little \& Carlberg}{1991}]{LC91}
Little, B. \& Carlberg, R. G. 1991, \mnras, {\bf 251}, 227

\bibitem[\protect\citeauthoryear{Liu \& van de Ven}{2012}]{LvdV12}
Liu, C. \& van de Ven, G. 2012, \mnras, {\bf 425}, 2144

\bibitem[\protect\citeauthoryear{Liu \& Chaboyer}{2000}]{LC00}
Liu, W. M. \& Chaboyer, B. 2000, \apj, {\bf 544}, 818

\bibitem[\protect\citeauthoryear{Loebman \etal}{2011}]{Loeb11}
Loebman, S. R.,{ Ro\v skar, R., Debattista, V. P., Ivezi\'c, \u Z., Quinn, T. R. \& Wadsley, J.} 2011, \apj, {\bf 737}, 8

\bibitem[\protect\citeauthoryear{Louis \& Gerhard}{1988}]{LG88}
Louis, P. D. \& Gerhard, O. E. 1988, \mnras, {\bf 233}, 337

\bibitem[\protect\citeauthoryear{Lovelace \& Hohlfeld}{1978}]{LH78}
Lovelace, R. V. E. \& Hohlfeld, R. G. 1978, \apj, {\bf 221}, 51

\bibitem[\protect\citeauthoryear{L\"utticke \etal}{2000}]{Lutt00}
L\"utticke, R., Dettmar, R.-J. \& Pohlen, M. 2000, \aap, {\bf 362}, 435

\bibitem[\protect\citeauthoryear{Lynden-Bell}{1962}]{LB62}
Lynden-Bell, D. 1962, \mnras, {\bf 124}, 1 

\bibitem[\protect\citeauthoryear{Lynden-Bell}{1963}]{LB63}
Lynden-Bell, D. 1963, {\it The Observatory}, {\bf 83}, 23

\bibitem[\protect\citeauthoryear{Lynden-Bell}{1979}]{LB79}
Lynden-Bell, D. 1979, \mnras, {\bf 187}, 101

\bibitem[\protect\citeauthoryear{Lynden-Bell \& Kalnajs}{1972}]{LBK72}
Lynden-Bell, D. \& Kalnajs, A. J. 1972, \mnras, {\bf 157}, 1

\bibitem[\protect\citeauthoryear{Lynds \& Toomre}{1976}]{LT76}
Lynds, R. \& Toomre, A. 1976, \apj, 209, 382

\bibitem[\protect\citeauthoryear{Ma \& Boylan-Kolchin}{2004}]{MBK04}
Ma, C-P. \& Boylan-Kolchin, M. 2004, \prl, {\bf 93}, 21301

\bibitem[\protect\citeauthoryear{Maciejewski}{2006}]{Maci06}
Maciejewski, W. 2006, \mnras, {\bf 371}, 451

\bibitem[\protect\citeauthoryear{Maciejewski \& Sparke}{2000}]{MS00}
Maciejewski, W. \& Sparke, L. S. 2000, \mnras, {\bf 313}, 745

\bibitem[\protect\citeauthoryear{Maciejewski \etal}{2002}]{Maci02}
Maciejewski, W.,{ Teuben, P. J., Sparke, L. S. \&  Stone, J. M.} 2002, \mnras, {\bf 329}, 502

\bibitem[\protect\citeauthoryear{MacLow}{1999}]{MacL99}
MacLow, M.-M. 1999, \apj, {\bf 524}, 169

\bibitem[\protect\citeauthoryear{MacLow \& Ferrara}{1999}]{MF99}
MacLow, M.-M. \& Ferrara, A. 1999, \apj, {\bf 513}, 142

\bibitem[\protect\citeauthoryear{Majewski}{1993}]{Maje93}
Majewski, S. R. 1993, \araa, {\bf 31}, 575

\bibitem[\protect\citeauthoryear{Mark}{1974}]{Mark74}
Mark, J. W-K. 1974, \apj, {\bf 193}, 539

\bibitem[\protect\citeauthoryear{Maoz \etal}{2001}]{Maoz01}
Maoz, D.,{ Barth, A. J., Ho, L. C., Sternberg, A. \& Filippenko, A. V.} 2001, \aj, {\bf 121}, 3048

\bibitem[\protect\citeauthoryear{Marinova \& Jogee}{2007}]{MJ07}
Marinova, I. \& Jogee, S. 2007, \apj, {\bf 659}, 1176

\bibitem[\protect\citeauthoryear{Mart\'\i n-Navarro \etal}{2012}]{Mart12}
Mart\'\i n-Navarro, I.,\skip{ Bakos, J., Trujillo, I., Knapen, J. H., Athanassoula, E., Bosma, A., Comer\'on, S., Elmegreen, B. G., Erroz-Ferrer, S., Gadotti, D. A., Gil de Paz, A., Hinz, J. L., Ho, L. C., Holwerda, B. W., Kim, T., Laine, J., Laurikainen, E., Men\'endez-Delmestre, K.., Mizusawa, T., Mu\~noz-Mateos, J-C., Regan, M. W., Salo, H., Seibert, M. \& Sheth, K.} 2012, \mnras, {\bf 427}, 1102

\bibitem[\protect\citeauthoryear{Mart\'\i nez-Serrano \etal}{2009}]{Mart09}
Mart\'\i nez-Serrano, F. J.,{ Serna, A., Dom{\'e}nech-Moral, M. \& Dom{\'i}nguez-Tenreiro, R.} 2009, \apjl, {\bf 705}, L133

\bibitem[\protect\citeauthoryear{Martinez-Valpuesta \& Gerhard}{2011}]{MVG11}
Martinez-Valpuesta, I. \& Gerhard, 0. 2011, \apjl, {\bf 734}, L20

\bibitem[\protect\citeauthoryear{Martinez-Valpuesta \& Shlosman}{2004}]{MS04}
Martinez-Valpuesta, I. \& Shlosman, I. 2004, \apjl, {\bf 613}, L29

\bibitem[\protect\citeauthoryear{Martinez-Valpuesta \etal}{2006}]{Mart06}
Martinez-Valpuesta, I., Shlosman, I. \& Heller, C. 2006, \apj, {\bf 637}, 214

\bibitem[\protect\citeauthoryear{Martini \etal}{2003}]{Mart03}
Martini, P.,{ Regan, M. W., Mulchaey, J. S. \& Pogge, R. W.} 2003, \apj, {\bf 589}, 774

\bibitem[\protect\citeauthoryear{Mashchenko \etal}{2006}]{MCW06}
Mashchenko, S., Couchman, H. M. P. \& Wadsley, J. 2006, \nat, {\bf 442}, 539

\bibitem[\protect\citeauthoryear{Mashchenko \etal}{2008}]{MCW07}
Mashchenko, S., Wadsley, J. \& Couchman, H. M. P. 2008, \sci, {\bf 319}, 174

\bibitem[\protect\citeauthoryear{Masset \& Tagger}{1997}]{MT97}
Masset, F. \& Tagger, M. 1997, \aap, {\bf 322}, 442

\bibitem[\protect\citeauthoryear{Masters \etal}{2011}]{Mast11}
Masters, K. L.,\skip{ R. C., Hoyle, B., Lintott, C., Bamford, S., Edmondson, E. M., Fortson, L., Keel, W. C., Schawinski, K., Smith, A. \& Thomas, D.} 2011, \mnras, {\bf 411}, 2026

\bibitem[\protect\citeauthoryear{Matsuda \& Isaka}{1980}]{MT80}
Matsuda, T. \& Isaka, H. 1980, {\it Prog. Theor. Phys.}, {\bf 64}, 1265

\bibitem[\protect\citeauthoryear{Matteucci \& Francois}{1989}]{MF89}
Matteucci, F. \& Francois, P. 1989, \mnras, {\bf 239}, 885

\bibitem[\protect\citeauthoryear{Matthews}{2000}]{Matt00}
Matthews, L. D. 2000, \aj, {\bf 120}, 1764

\bibitem[\protect\citeauthoryear{Mayer \& Wadsley}{2004}]{MW04}
Mayer, L. \&  Wadsley, J. 2004, \mnras, {\bf 347}, 277

\bibitem[\protect\citeauthoryear{Mazzuca \etal}{2008}]{Mazz08}
Mazzuca, L. M.,{ Knapen, J. H., Veilleux, S. \& Regan, M. W.} 2008, \apjs, {\bf 174}, 337

\bibitem[\protect\citeauthoryear{Mazzuca \etal}{2011}]{Mazz11}
Mazzuca, L. M.,{ Swaters, R. A., Knapen, J. H. \& Veilleux, S.} 2011, \apj, {\bf 739}, 104

\bibitem[\protect\citeauthoryear{McMillan}{2011}]{McM11}
McMillan, P. J. 2011, \mnras, {\bf 418}, 1565

\bibitem[\protect\citeauthoryear{McMillan}{2013}]{McM13}
McMillan, P. J. 2013, \mnras, {\bf 430}, 3276

\bibitem[\protect\citeauthoryear{McMillan \& Binney}{2008}]{MB08}
McMillan, P. J. \& Binney, J. J. 2008, \mnras, {\bf 390}, 429

\bibitem[\protect\citeauthoryear{McMillan \& Dehnen}{2005}]{MD05}
McMillan, P. J. \& Dehnen, W. 2005, \mnras, {\bf 363}, 1205

\bibitem[\protect\citeauthoryear{McMillan \& Dehnen}{2007}]{MD07}
McMillan, P. J. \& Dehnen, W. 2007, \mnras, {\bf 378}, 541

\bibitem[\protect\citeauthoryear{McWilliam \& Zoccali}{2010}]{MZ10}
McWilliam, A. \& Zoccali, M. 2010, \apj, {\bf 724}, 1491

\bibitem[\protect\citeauthoryear{Meidt \etal}{2008}]{Meid08}
Meidt, S. E.,{ Rand, R. J., Merrifield, M. R., Debattista, V. P. \& Shen, J.} 2008, \apj, {\bf 676}, 899

\bibitem[\protect\citeauthoryear{Meidt \etal}{2009}]{MRM09}
Meidt, S. E., Rand, R. J. \& Merrifield, M. R. 2009, \apj, {\bf 702}, 277

\bibitem[\protect\citeauthoryear{M\'endez-Abreu \etal}{2012}]{Mend12}
M\'endez-Abreu, J.,{ S\'anchez-Janssen, R., Aguerri, J. A. L., Corsini, E. M. \& Zarattini, S.} 2012, \apjl, {\bf 761}, L6

\bibitem[\protect\citeauthoryear{Men\'endez-Delmestre \etal}{2007}]{Mene07}
Men\'endez-Delmestre, K.,{ Sheth, K., Schinnerer, E., Jarrett, T. H. \& Scoville, N. Z.} 2007, \apj, {\bf 657}, 790

\bibitem[\protect\citeauthoryear{Merritt \& Hernquist}{1991}]{MH91}
Merritt, D. \& Hernquist, L. 1991, \apj, {\bf 376}, 439

\bibitem[\protect\citeauthoryear{Merritt \etal}{2004}]{Merr04}
Merritt, D.,{ Piatek, S., Portegies Zwart, S. \& Hemsendorf, M.} 2004, \apjl, {\bf 608}, L25

\bibitem[\protect\citeauthoryear{Miller \& Smith}{1979}]{MS79}
Miller, R. H. \& Smith, B. F. 1979, \apj, {\bf 227}, 785

\bibitem[\protect\citeauthoryear{Minchev \etal}{2010}]{MBSB10}
Minchev, I.,{ Boily, C., Siebert, A. \& Bienayme, O.} 2010, \mnras, {\bf 407}, 2122

\bibitem[\protect\citeauthoryear{Minchev \etal}{2013}]{MCM12}
Minchev, I., Chiappini, C. \& Martig, M. 2013, \aap, to appear (arXiv:1208.1506)

\bibitem[\protect\citeauthoryear{Minchev \& Famaey}{2010}]{MF10}
Minchev, I. \& Famaey, B. 2010, \apj, {\bf 722}, 112

\bibitem[\protect\citeauthoryear{Minchev \etal}{2011}]{Minc11}
Minchev, I.,{ Famaey, B., Combes, F., Di Matteo, P., Mouhcine, M. \& Wozniak, H.} 2011, \aap, {\bf 527}, A147

\bibitem[\protect\citeauthoryear{Minchev \etal}{2012a}]{Minc12}
Minchev, I.,{ Famaey, B., Quillen, A. C., Di Matteo, P., Combes, F., Vlajić, M., Erwin, P. \& Bland-Hawthorn, J.} \etal\	2012a, \aap, {\bf 548}, A126

\bibitem[\protect\citeauthoryear{Minchev \etal}{2012b}]{MFQ12}
Minchev, I.,{ Famaey, B., Quillen, A. C., Dehnen, W., Martig, M. \& Siebert, A.} 2012b, \aap, {\bf 548}, A127

\bibitem[\protect\citeauthoryear{Minchev \& Quillen}{2006}]{MQ06}
Minchev, I. \& Quillen, A. C. 2006, \mnras, {\bf 368}, 623

\bibitem[\protect\citeauthoryear{Miwa \& Noguchi}{1998}]{MN98}
Miwa, T. \& Noguchi, M. 1998, \apj, {\bf 499}, 149

\bibitem[\protect\citeauthoryear{Moster}{2010}]{Most10}
Moster, B. P.,{ Macci\`o, A. V., Somerville, R. S., Johansson, P. H. \& Naab, T.} 2010, \mnras, {\bf 403}, 1009

\bibitem[\protect\citeauthoryear{Mould}{2005}]{Moul05}
Mould, J. 2005, \aj, {\bf 129}, 698

\bibitem[\protect\citeauthoryear{Munn \etal}{2004}]{Munn04}
Munn, J. A. 2004, \aj, {\bf 127}, 3034

\bibitem[\protect\citeauthoryear{Mu\~noz-Mateos \etal}{2013}]{Muno13}
Mu\~noz-Mateos, J. C.,\skip{ Sheth, K., Gil de Paz, A., Meidt, S., Athanassoula, E., Bosma, A., Comer\'on, S., Elmegreen, D. M., Elmegreen, B. G., Erroz-Ferrer, S., Gadotti, D. A., Hinz, J. L., Ho, L. C., Holwerda, B., Jarrett, T. H., Kim, T., Knapen, J. H., Laine, J., Laurikainen, E., Madore, B. F., Menendez-Delmestre, K., Mizusawa, T., Regan, M., Salo, H., Schinnerer, E., Seibert, M., Skibba, R. \& Zaritsky, D.} 2013, \apj, {\bf 771}, 59

\bibitem[\protect\citeauthoryear{Nataf \etal}{2010}]{Nata10}
Nataf, D. M.,{ Udalski, A., Gould, A., Fouqu\'e, P. \& Stanek, K. Z.} 2010, \apjl, {\bf 721}, L28

\bibitem[\protect\citeauthoryear{Ness \etal}{2012}]{Ness12}
Ness, M.,\skip{ Freeman, K., Athanassoula, E., Wylie-De-Boer, E., Bland-Hawthorn, J., Lewis, G. F., Yong, D., Asplund, M., Lane, R. R., Kiss, L. L. \& Ibata, R.} 2012, \apj, {\bf 756}, 22

\bibitem[\protect\citeauthoryear{Nieten \etal}{2006}]{Niet06}
Nieten, Ch.,{ Neininger, N., Gu\'elin, M., Ungerechts, H., Lucas, R., Berkhuijsen, E. M., Beck, R. \& Wielebinski, R.} 2006, \aap, {\bf 453}, 459

\bibitem[\protect\citeauthoryear{Nordstr\"om \etal}{2004}]{Nord04}
Nordstr\"om, B.,{ Mayor, M., Andersen, J., Holmberg, J., Pont, F., J\o rgensen, B. R., Olsen, E. H., Udry, S. \& Mowlavi, N.} 2004, \aap, {\bf 418}, 989

\bibitem[\protect\citeauthoryear{Noguchi}{1987}]{Nogu87}
Noguchi, M. 1987, \mnras, {\bf 228}, 635

\bibitem[\protect\citeauthoryear{Norman \etal}{1996}]{Norm96}
Norman, C. A., Sellwood, J. A. \& Hasan, H. 1996, \apj, {\bf 462}, 114

\bibitem[\protect\citeauthoryear{Okamoto}{2013}]{Okam13}
Okamoto, T. 2013, \mnras, {\bf 428}, 718

\bibitem[\protect\citeauthoryear{O'Neill \& Dubinski}{2003}]{OD03}
O'Neill, J. K. \& Dubinski, J. 2003, \mnras, {\bf 346}, 251

\bibitem[\protect\citeauthoryear{Oort}{1962}]{Oort62}
Oort, J. H. 1962, in {\it Interstellar Matter in Galaxies}, ed.\ L. Woltjer (New York: Benjamin), p.~234

\bibitem[\protect\citeauthoryear{Ostriker \& Peebles}{1973}]{OP73}
Ostriker, J. P. \& Peebles, P. J. E. 1973, \apj, {\bf 186}, 467

\bibitem[\protect\citeauthoryear{Palmer \etal}{1990}]{Palm90}
Palmer, P. L., Papaloizou, J. \& Allen, A. J. 1990, \mnras, {\bf 243}, 282

\bibitem[\protect\citeauthoryear{Palunas \& Williams}{2000}]{PW00}
Palunas, P. \& Williams, T. B. 2000, \aj, {\bf 120}, 2884

\bibitem[\protect\citeauthoryear{Parker \etal}{2004}]{Park04}
Parker, J. E., Humphreys, R. M. \& Beers, T. C. 2004, \aj, {\bf 127}, 1567

\bibitem[\protect\citeauthoryear{Patsis \etal}{2002}]{Pats02}
Patsis, P. A., Skokos, Ch. \& Athanassoula, E. 2002, \mnras, {\bf 337}, 578

\bibitem[\protect\citeauthoryear{P\'erez}{2008}]{Pere08}
P\'erez, I. 2008, \aap, {\bf 478}, 717

\bibitem[\protect\citeauthoryear{P\'erez \etal}{2004}]{Pere04}
P\'erez, I., Fux, R. \& Freeman, K. 2004, \aap, {\bf 424}, 799

\bibitem[\protect\citeauthoryear{Perlmutter \etal}{1999}]{Perl99}
Perlmutter, S.,\skip{ Aldering, G., Goldhaber, G., Knop, R. A., Nugent, P., Castro, P. G., Deustua, S., Fabbro, S., Goobar, A., Groom, D. E., Hook, I. M., Kim, A. G., Kim, M. Y., Lee, J. C., Nunes, N. J., Pain, R., Pennypacker, C. R., Quimby, R., Lidman, C., Ellis, R. S., Irwin, M., McMahon, R. G., Ruiz-Lapuente, P., Walton, N., Schaefer, B., Boyle, B. J., Filippenko, A. V., Matheson, T., Fruchter, A. S., Panagia, N., Newberg, H. J. M. \& Couch, W. J.} 1999, \apj, {\bf 517}, 565

\bibitem[\protect\citeauthoryear{Perryman \etal}{2001}]{Perr01}
Perryman, M. A. C.,\skip{ de Boer, K. S., Gilmore, G., H\o g, E., Lattanzi, M. G., Lindegren, L., Luri, X., Mignard, F., Pace, O. \& de Zeeuw, P. T.} 2001, \aap, {\bf 369}, 339

\bibitem[\protect\citeauthoryear{Pfenniger \& Friedli}{1991}]{PF91}
Pfenniger, D. \& Friedli, D. 1991, \aap, {\bf 252}, 75

\bibitem[\protect\citeauthoryear{Pfenniger \& Norman}{1990}]{PN90}
Pfenniger, D. \& Norman, C. 1990, \apj, {\bf 363}, 391

\bibitem[\protect\citeauthoryear{Piner \etal}{1995}]{Pine95}
Piner, B. G., Stone, J. M. \& Teuben, P. J. 1995, \apj, {\bf 449}, 508

\bibitem[\protect\citeauthoryear{Pohlen \etal}{2002}]{Pohl02}
Pohlen, M.,{ Dettmar, R.-J., L\"utticke, R. \& Aronica, G.} 2002, \aap, {\bf 392}, 807

\bibitem[\protect\citeauthoryear{Pohlen \& Trujillo}{2006}]{PT06}
Pohlen, M. \& Trujillo, I. 2006, \aap, {\bf 454}, 759

\bibitem[\protect\citeauthoryear{Polyachenko}{2004}]{Poly04}
Polyachenko, E. V. 2004, \mnras, {\bf 348}, 345

\bibitem[\protect\citeauthoryear{Polyachenko}{2013}]{Poly13}
Polyachenko, E. V. 2013, \astl, {\bf 39}, 72

\bibitem[\protect\citeauthoryear{Pomp\'eia \etal}{2011}]{Pomp11}
Pomp\'eia 2011, \mnras, {\bf 415}, 1138 

\bibitem[\protect\citeauthoryear{Pontzen \& Governato}{2012}]{PG12}
Pontzen, A. \& Governato, F. 2012, \mnras, {\bf 421}, 3464

\bibitem[\protect\citeauthoryear{Prendergast}{1962}]{Pren62}
Prendergast, K. H. 1962, in {\it Interstellar Matter in Galaxies}, ed.\ L. Woltjer (New York: Benjamin), p.~217

\bibitem[\protect\citeauthoryear{Prendergast}{1983}]{Pren83}
Prendergast, K. H. 1983, in {\it Internal Kinematics and Dynamics of Galaxies}, IAU Symp.\ {\bf 100}, ed.\ E. Athanassoula (Dordrecht: Reidel) p~215

\bibitem[\protect\citeauthoryear{Puech \etal}{2012}]{Puec12}
Puech, M.,{ Hammer, F., Hopkins, P. F., Athanassoula, E., Flores, H., Rodrigues, M., Wang, J. L. \& Yang, Y. B.} 2012, \apj, {\bf 753}, 128

\bibitem[\protect\citeauthoryear{Purcell \etal}{2009}]{Purc09}
Purcell, C. W., Kazantzidis, S. \& Bullock, J. S. 2009, \apjl, {\bf 694}, L98

\bibitem[\protect\citeauthoryear{Quillen}{2003}]{Quil03}
Quillen, A. C. 2003, \aj, {\bf 125}, 785

\bibitem[\protect\citeauthoryear{Quillen \etal}{1995}]{Quil95}
Quillen, A. C.,{ Frogel, J. A., Kenney, J. D. P., Pogge, R. W. \& Depoy, D. L.} 1995, \apj, {\bf 441}, 549

\bibitem[\protect\citeauthoryear{Quillen \& Garnett}{2001}]{QG01}
Quillen, A. C. \& Garnett, D. R. 2001, in {\it Galaxy Disks and Disk Galaxies\/}.eds.\ J. G. Funes SJ \&  E. M. Corsini (San Francisco: ASP) {\bf 230}, p.~87

\bibitem[\protect\citeauthoryear{Quillen \& Minchev}{2005}]{QM05}
Quillen, A. C. \& Minchev, I. 2005, \aj, {\bf 130}, 576

\bibitem[\protect\citeauthoryear{Quillen \etal}{2009}]{Quil09}
Quillen, A. C., Minchev, I., Bland-Hawthorn, J. \&  Haywood, M. 2009, \mnras, {\bf 397}, 1599

\bibitem[\protect\citeauthoryear{Quinn \& Goodman}{1986}]{QG86}
Quinn, P. J. \& Goodman, J. 1986, \apj, {\bf 309}, 472

\bibitem[\protect\citeauthoryear{Quinn \etal}{1993}]{Quin93}
Quinn, P. J., Hernquist, L. \& Fullagar, D. P. 1993, \apj, {\bf 403}, 74

\bibitem[\protect\citeauthoryear{Rafikov}{2001}]{Rafi01}
Rafikov, R. R. 2001, \mnras, {\bf 323}, 445

\bibitem[\protect\citeauthoryear{Raha \etal}{1991}]{Raha91}
Raha, N.,{ Sellwood, J. A., James, R. A. \& Kahn, F. D.} 1991, \nat, {\bf 352}, 411

\bibitem[\protect\citeauthoryear{Rautiainen \& Salo}{1999}]{RS99}
Rautiainen, P. \& Salo, H. 1999, \aap, {\bf 348}, 737

\bibitem[\protect\citeauthoryear{Rautiainen \etal}{2002}]{RSL02}
Rautiainen, P., Salo, H. \& Laurikainen, E. 2002, \mnras, {\bf 337}, 1233

\bibitem[\protect\citeauthoryear{Rautiainen \etal}{2008}]{RSL08}
Rautiainen, P., Salo, H. \& Laurikainen, E. 2008, \mnras, {\bf 388}, 1803

\bibitem[\protect\citeauthoryear{Read \& Gilmore}{2005}]{RG05}
Read, J. I. \& Gilmore, G. 2005, \mnras, {\bf 356}, 107

\bibitem[\protect\citeauthoryear{Read \& Hayfield}{2012}]{RH12}
Read, J. I. \& Hayfield, T. 2012, \mnras, {\bf 422}, 3037

\bibitem[\protect\citeauthoryear{Read \etal}{2008}]{Read08}
Read, J. I.,{ Lake, G., Agertz, O. \& Debattista, V. P.} 2008, \mnras, {\bf 389}, 1041

\bibitem[\protect\citeauthoryear{Reddy \etal}{2006}]{Redd06}
Reddy, B. E., Lambert, D. L. \& Allende Prieto, C. 2006, \mnras, {\bf 367}, 1329

\bibitem[\protect\citeauthoryear{Reese \etal}{2007}]{Rees07}
Reese, A.,{ Williams, T. B., Sellwood, J. A., Barnes, E. I. \& Powell, B. A.} 2007, \aj, {\bf 133}, 2846

\bibitem[\protect\citeauthoryear{Regan \etal}{2006}]{Rega06}
Regan, M. W.,\skip{ Thornley, M. D., Vogel, S. N., Sheth, K., Draine, B. T., Hollenbach, D. J., Meyer, M., Dale, D. A., Engelbracht, C. W., Kennicutt, R. C., Armus, L., Buckalew, B., Calzetti, D., Gordon, K. D., Helou, G., Leitherer, C., Malhotra, S., Murphy, E., Rieke, G. H., Rieke, M. J. \& Smith, J. D.} 2006, \apj, {\bf 652}, 1112

\bibitem[\protect\citeauthoryear{Reid \etal}{2007}]{Reid07}
Reid, I. N.,{ Turner, E. L., Turnbull, M. C., Mountain, M. \& Valenti, J. A.} 2007, \apj, {\bf 665}, 767

\bibitem[\protect\citeauthoryear{Riess \etal}{1998}]{Ries98}
Riess, A. G.,\skip{ Filippenko, A. V., Challis, P., Clocchiatti, A., Diercks, A., Garnavich, P. M., Gilliland, R. L., Hogan, C. J., Jha, S., Kirshner, R. P., Leibundgut, B., Phillips, M. M., Reiss, D., Schmidt, B. P., Schommer, R. A., Smith, R. C., Spyromilio, J., Stubbs, C., Suntzeff, N. B. \& Tonry, J.} 1998, \aj, {\bf 116}, 1009

\bibitem[\protect\citeauthoryear{Rix \& Bovy}{2013}]{RB13}
Rix, H-W. \& Bovy, J. 2013, \aapr, {\bf 21}, 61

\bibitem[\protect\citeauthoryear{Robertson \etal}{2006}]{Robe06}
Robertson, B.,{ Bullock, J. S., Cox, T. J., Di Matteo, T., Hernquist, L., Springel, V. \& Yoshida, N.} \etal, 2006, \apj, {\bf 645}, 986

\bibitem[\protect\citeauthoryear{Roca-F\`abrega \etal}{2013}]{Roca13}
Roca-F\`abrega, S.,{ Valenzuela, O., Figueras, F., Romero-G\'omez, M., Vel\'azquez, H., Antoja, T. \& Pichardo, B.} 2013, \mnras, {\bf 432}, 2878

\bibitem[\protect\citeauthoryear{Rodionov \& Sotnikova}{2013}]{RS13}
Rodionov, S. A. \& Sotnikova, N. Ya. 2013, \mnras, {\bf 434}, 2373

\bibitem[\protect\citeauthoryear{Romano-D\'\i az \etal}{2008a}]{Roma08a}
Romano-D\'\i az, E., Shlosman, I., Hoffman, Y. \& Heller C. 2008a, \apjl, {\bf 685}, L105

\bibitem[\protect\citeauthoryear{Romano-D\'\i az \etal}{2008b}]{Roma08b}
Romano-D\'\i az, E., Shlosman, I., Heller C. \& Hoffman, Y. 2008b, \apjl, {\bf 687}, L13

\bibitem[\protect\citeauthoryear{Romeo}{1992}]{Rome92}
Romeo, A. B. 1992, \mnras, {\bf 256}, 307

\bibitem[\protect\citeauthoryear{Romeo}{1998}]{Rome98}
Romeo, A. B. 1998, \aap, {\bf 335}, 922

\bibitem[\protect\citeauthoryear{Romeo \& Wiegert}{2011}]{RW11}
Romeo, A. B. \& Wiegert, J. 2011, \mnras, {\bf 416}, 1191

\bibitem[\protect\citeauthoryear{Ro\v skar \etal}{2008a}]{Rosk08a}
Ro\v skar, R.,{ Debattista, V. P., Quinn, T. R., Stinson, G. S. \& Wadsley, J.} 2008a, \apjl, {\bf 684}, L79

\bibitem[\protect\citeauthoryear{Ro\v skar \etal}{2008b}]{Rosk08b}
Ro\v skar, R.,{ Debattista, V. P., Stinson, G. S., Quinn, T. R., Kaufmann, T. \& Wadsley, J.} 2008b, \apjl, {\bf 675}, L65

\bibitem[\protect\citeauthoryear{Ro\v skar \etal}{2013}]{RDL12}
Ro\v skar, R., Debattista, V. P. \& Loebman, S. R. 2013, \mnras, {\bf 433}, 976

\bibitem[\protect\citeauthoryear{Ro\v skar \etal}{2012}]{Rosk12}
Ro\v skar, R.,{ Debattista, V. P., Quinn, T. R. \& Wadsley, J.} 2012, \mnras, {\bf 426}, 2089

\bibitem[\protect\citeauthoryear{Ruchti \etal}{2010}]{Ruch10}
Ruchti, G. R.,\skip{ Fulbright, J. P., Wyse, R. F. G., Gilmore, G. F., Bienaym\'e, O., Binney, J., Bland-Hawthorn, J., Campbell, R., Freeman, K. C., Gibson, B. K., Grebel, E. K., Helmi, A., Munari, U., Navarro, J. F., Parker, Q. A., Reid, W., Seabroke, G. M., Siebert, A., Siviero, A., Steinmetz, M., Watson, F. G., Williams, M., Zwitter, T.} 2010, \apjl, {\bf 721}, L92

\bibitem[\protect\citeauthoryear{Ruchti \etal}{2011}]{Ruch11}
Ruchti, G. R.,\skip{ Fulbright, J. P., Wyse, R. F. G., Gilmore, G. F., Bienaym\'e, O., Bland-Hawthorn, J., Gibson, B. K., Grebel, E. K., Helmi, A., Munari, U., Navarro, J. F., Parker, Q. A., Reid, W., Seabroke, G. M., Siebert, A., Siviero, A., Steinmetz, M., Watson, F. G., Williams, M. \& Zwitter, T.} 2011, \apj, {\bf 737}, 9

\bibitem[\protect\citeauthoryear{Rybicki}{1972}]{Rybi72}
Rybicki, G. B. 1972, in IAU Colloq.\ {\bf 10}, {\it Gravitational $N$-body Problem}, ed.\ M. Lecar (Dordrecht: Reidel), 22

\bibitem[\protect\citeauthoryear{Saha \etal}{2012}]{Saha12}
Saha, K., Martinez-Valpuesta, I. \& Gerhard, O. 2012, \mnras, {\bf 421}, 333

\bibitem[\protect\citeauthoryear{Sales \etal}{2009}]{Sale09}
Sales, L. V.,{ Helmi, A., Abadi, M. G., Brook, C. B., G\'omez, F. A., Ro\v skar, R., Debattista, V. P., House, E., Steinmetz, M. \& Villalobos, \'A.} 2009, \mnras, {\bf 400}, L61

\bibitem[\protect\citeauthoryear{Salo}{1991}]{Salo91}
Salo, H. 1991, \aap, {\bf 243}, 118

\bibitem[\protect\citeauthoryear{Sakamoto \etal}{1999}]{Saka99}
Sakamoto, K.,{ Okamura, S. K., Ishizuki, S. \& Scoville, N. Z.} 1999, \apj, {\bf 525}, 691

\bibitem[\protect\citeauthoryear{Samland \& Gerhard}{2003}]{SG03}
Samland, M. \& Gerhard, O. E. 2003, \aap, {\bf 399}, 961

\bibitem[\protect\citeauthoryear{S\'anchez-Bl\'azquez \etal}{2009}]{Sanc09}
S\'anchez-Bl\'azquez, P.,{ Courty, S., Gibson, B. K. \& Brook, C. B.} 2009, \mnras, {\bf 398}, 591

\bibitem[\protect\citeauthoryear{S\'anchez-Janssen \& Gadotti}{2013}]{SG12}
S\'anchez-Janssen, R. \& Gadotti, D. 2013, \mnras, {\bf 432}, L56

\bibitem[\protect\citeauthoryear{Sandage}{1961}]{Sand61}
Sandage, A. 1961, {\it The Hubble Atlas of Galaxies}, Carnegie Inst. of Washington. 

\bibitem[\protect\citeauthoryear{Sanders \& Huntley}{1976}]{SH76}
Sanders, R. H. \& Huntley, J. M. 1976, \apj, {\bf 209}, 53

\bibitem[\protect\citeauthoryear{Sanders \& Tubbs}{1980}]{ST80}
Sanders, R. H. \& Tubbs, A. D. 1980, \apj, {\bf 235}, 803

\bibitem[\protect\citeauthoryear{Scalo \& Elmegreen}{2004}]{SE04}
Scalo, J. \& Elmegreen, B. G. 2004, \araa, {\bf 42}, 275

\bibitem[\protect\citeauthoryear{Scannapieco \etal}{2011}]{Scan11}
Scannapieco, C.,\skip{ White, S. D. M, Springel, V. \& Tissera, P. B.} 2011, \mnras, {\bf 417}, 154

\bibitem[\protect\citeauthoryear{Schinnerer \etal}{2013}]{Schi13}
Schinnerer, E.,\skip{ Meidt, S. E., Pety, J., Hughes, A., Colombo, D., Garc\'\i a-Burillo, S., Schuster, K. F., Dumas, G., Dobbs, C. L., Leroy, A. K., Kramer, C., Thompson, T. A. \& Regan, M. W.} 2013, \apj, {\bf 779}, 42

\bibitem[\protect\citeauthoryear{Schlesinger \etal}{2012}]{Schl11}
Schlesinger, K. J.,\skip{ Johnson, J. A., Rockosi, C. M., Lee, Y. S., Morrison, H. L., Sch\"onrich, R., Allende Prieto, C., Beers, T. C., Yanny, B., Harding, P., Schneider, D/ P., Chiappini, C., da Costa, L. N., Maia, M. A. G., Minchev, I., Rocha-Pinto, H. \& Santiago, B. X.} 2012, \apj, {\bf 761}, 160

\bibitem[\protect\citeauthoryear{Sch\"onrich \& Binney}{2009a}]{SB09a}
Sch\"onrich, R. \& Binney, J. 2009a, \mnras, {\bf 396}, 203

\bibitem[\protect\citeauthoryear{Sch\"onrich \& Binney}{2009b}]{SB09b}
Sch\"onrich, R. \& Binney, J. 2009b, \mnras, {\bf 399}, 1145

\bibitem[\protect\citeauthoryear{Sch\"onrich \& Binney}{2012}]{SB12}
Sch\"onrich, R. \& Binney, J. 2012, \mnras, {\bf 419}, 1546

\bibitem[\protect\citeauthoryear{Sch\"onrich \etal}{2010}]{SBD10}
Sch\"onrich, R., Binney, J. \& Dehnen, W. 2010, \mnras, {\bf 403}, 829

\bibitem[\protect\citeauthoryear{Schwarz}{1981}]{Schw81}
Schwarz, M. P. 1981, \apj, {\bf 247}, 77

\bibitem[\protect\citeauthoryear{Schweizer}{1976}]{Schw76}
Schweizer, F. 1976, \apjs, {\bf 31}, 313

\bibitem[\protect\citeauthoryear{Seabroke \& Gilmore}{2007}]{SG07}
Seabroke, G. M. \& Gilmore, G. 2007, \mnras, {\bf 380}, 1348

\bibitem[\protect\citeauthoryear{Sellwood}{1980}]{Sell80}
Sellwood, J. A. 1980, \aap, {\bf 89}, 296

\bibitem[\protect\citeauthoryear{Sellwood}{1981}]{Sell81}
Sellwood, J. A. 1981, \aap, {\bf 99}, 362

\bibitem[\protect\citeauthoryear{Sellwood}{1983}]{Sell83}
Sellwood, J. A. 1983, \jcop, {\bf 50}, 337

\bibitem[\protect\citeauthoryear{Sellwood}{1985}]{Sell85}
Sellwood, J. A. 1985, \mnras, {\bf 217}, 127

\bibitem[\protect\citeauthoryear{Sellwood}{1989a}]{Sell89a}
Sellwood, J. A. 1989a, in {\it Dynamics of Astrophysical Discs}, ed.\ J. A. Sellwood (Cambridge: Cambridge University Press) p.~155

\bibitem[\protect\citeauthoryear{Sellwood}{1989b}]{Sell89b}
Sellwood, J. A. 1989b, \mnras, {\bf 238}, 115

\bibitem[\protect\citeauthoryear{Sellwood}{1996}]{Sell96}
Sellwood, J. A. 1996, in IAU Symp.\ {\bf 169}, {\it Unsolved Problems of the Milky Way\/}, ed.\ L. Blitz \& P. Teuben (Dordrecht: Kluwer) p.~31

\bibitem[\protect\citeauthoryear{Sellwood}{2000}]{Sell00}
Sellwood, J. A. 2000, \apss, {\bf 272}, 31 (astro-ph/9909093)

\bibitem[\protect\citeauthoryear{Sellwood}{2003}]{Sell03}
Sellwood, J. A. 2003, \apj, {\bf 587}, 638

\bibitem[\protect\citeauthoryear{Sellwood}{2006}]{Sell06}
Sellwood, J. A. 2006, \apj, {\bf 637}, 567

\bibitem[\protect\citeauthoryear{Sellwood}{2008a}]{Sell08a}
Sellwood, J. A. 2008a, \apj, {\bf 679}, 379

\bibitem[\protect\citeauthoryear{Sellwood}{2008b}]{Sell08b}
Sellwood, J. A. 2008b, in {\it Formation and Evolution of Galaxy Disks}, eds.\ J. G. Funes SJ \&  E. M. Corsini (San Francisco: ASP) {\bf 396}, p.~341 (arXiv:0803.1574)

\bibitem[\protect\citeauthoryear{Sellwood}{2009}]{Sell09}
Sellwood, J. A. 2009, in IAU Symp.\ {\bf 254}, {\it The Galaxy Disk in Cosmological Context}, ed.\ J. Andersen, J. Bland-Hawthorn \& B. Nordstr\"om, (Cambridge: Cambridge University Press) p.~73 (arXiv:0807.1973)

\bibitem[\protect\citeauthoryear{Sellwood}{2010}]{Sell10}
Sellwood, J. A. 2010, \mnras, {\bf 409}, 145

\bibitem[\protect\citeauthoryear{Sellwood}{2011}]{Sell11}
Sellwood, J. A. 2011, \mnras, {\bf 410}, 1637

\bibitem[\protect\citeauthoryear{Sellwood}{2012}]{Sell12}
Sellwood, J. A. 2012, \apj, {\bf 751}, 44

\bibitem[\protect\citeauthoryear{Sellwood}{2013a}]{Sell13a}
Sellwood, J. A. 2013a, in {\it Planets Stars and Stellar Systems}, v.{\bf 5}, eds.\ T. Oswalt \& G. Gilmore (Heidelberg: Springer) p.~923 (arXiv:1006.4855)

\bibitem[\protect\citeauthoryear{Sellwood}{2013b}]{Sell13b}
Sellwood, J. A. 2013b, \apjl, {\bf 769}, L24

\bibitem[\protect\citeauthoryear{Sellwood \& Balbus}{1999}]{SB99}
Sellwood, J. A. \& Balbus, S. A. 1999, \apj, {\bf 511}, 660

\bibitem[\protect\citeauthoryear{Sellwood \& Binney}{2002}]{SB02}
Sellwood, J. A. \& Binney, J. J. 2002, \mnras, {\bf 336}, 785

\bibitem[\protect\citeauthoryear{Sellwood \& Carlberg}{1984}]{SC84}
Sellwood, J. A. \& Carlberg, R. G. 1984, \apj, {\bf 282}, 61

\bibitem[\protect\citeauthoryear{Sellwood \& Debattista}{2006}]{SD06}
Sellwood, J. A. \& Debattista, V. P. 2006, \apj, {\bf 639}, 868

\bibitem[\protect\citeauthoryear{Sellwood \& Debattista}{2009}]{SD09}
Sellwood, J. A. \& Debattista, V. P. 2009, \mnras, {\bf 398}, 1279

\bibitem[\protect\citeauthoryear{Sellwood \& Evans}{2001}]{SE01}
Sellwood, J. A. \& Evans, N. W. 2001, \apj, {\bf 546}, 176

\bibitem[\protect\citeauthoryear{Sellwood \& James}{1979}]{SJ79}
Sellwood, J. A. \& James, R. A. 1979, \mnras, {\bf 187}, 483

\bibitem[\protect\citeauthoryear{Sellwood \& Kahn}{1991}]{SK91}
Sellwood, J. A. \& Kahn, F. D. 1991, \mnras, {\bf 250}, 278

\bibitem[\protect\citeauthoryear{Sellwood \& Merritt}{1994}]{SM94}
Sellwood, J. A. \& Merritt, D. 1994, \apj, {\bf 425}, 530

\bibitem[\protect\citeauthoryear{Sellwood \& Moore}{1999}]{SM99}
Sellwood, J. A. \& Moore, E. M. 1999, \apj, {\bf 510}, 125

\bibitem[\protect\citeauthoryear{Sellwood \etal}{1998}]{SNT98}
Sellwood, J. A., Nelson, R. D. \& Tremaine, S. 1998, \apj, {\bf 506}, 590

\bibitem[\protect\citeauthoryear{Sellwood \& Sparke}{1988}]{SS88}
Sellwood, J. A. \& Sparke, L. S. 1988, \mnras, {\bf 231}, 25P

\bibitem[\protect\citeauthoryear{Sellwood \& Wilkinson}{1993}]{SW93}
Sellwood, J. A. \& Wilkinson, A. 1993, \rpp, {\bf 56}, 173

\bibitem[\protect\citeauthoryear{Shaw \etal}{1993}]{Shaw93}
Shaw, M. A.,{ Combes, F., Axon, D. J. \& Wright, G. S.} 1993, \aap, {\bf 273}, 31

\bibitem[\protect\citeauthoryear{Shen \etal}{2010}]{Shen10}
Shen, J.,{ Rich, R. M., Kormendy, J., Howard, C. D., De Propris, R. \& Kunder, A.} 2010, \apjl, {\bf 720}, L72

\bibitem[\protect\citeauthoryear{Shen \& Debattista}{2009}]{ShD09}
Shen, J. \& Debattista, V. P. 2009, \apj, {\bf 690}, 758

\bibitem[\protect\citeauthoryear{Shen \& Sellwood}{2004}]{Shen04}
Shen, J. \& Sellwood, J. A. 2004, \apj, {\bf 604}, 614

\bibitem[\protect\citeauthoryear{Sheth \etal}{2012}]{Shet12}
Sheth, K.,{ Melbourne, J., Elmegreen, D. M., Elmegreen, B. G., Athanassoula, E., Abraham, R. G. \& Weiner, B. J.} 2012, \apj, {\bf 758}, 136

\bibitem[\protect\citeauthoryear{Sheth \etal}{2005}]{Shet05}
Sheth, K.,{ Vogel, S. N., Regan, M. W., Thornley, M. D. \& Teuben, P. J.} 2005, \apj, {\bf 632}, 217

\bibitem[\protect\citeauthoryear{Shetty \etal}{2007}]{Shet07}
Shetty, R.,{ Vogel, S. N., Ostriker, E. C. \& Teuben, P. J.} 2007, \apj, {\bf 665}, 1138

\bibitem[\protect\citeauthoryear{Shiidsuke \& Ida}{1999}]{SI99}
Shiidsuke, K. \& Ida, S. 1999, \mnras, {\bf 307}, 737

\bibitem[\protect\citeauthoryear{Shlosman \etal}{1989}]{SFB89}
Shlosman, I., Frank, J. \& Begelman, M. C. 1989, \nat, {\bf 338}, 45

\bibitem[\protect\citeauthoryear{Silk \& Mamon}{2012}]{SM12}
Silk, J. \&  Mamon, G. A. 2012, \raa, {\bf 12}, 917

\bibitem[\protect\citeauthoryear{Simkin \etal}{1980}]{Simk80}
Simkin, S. M., Su, H. J. \& Schwarz, M. P. 1980, \apj, {\bf 237}, 404

\bibitem[\protect\citeauthoryear{Skibba \etal}{2012}]{Skib12}
Skibba, R. A.,\skip{Masters, K. L., Nichol, R. C., Zehavi, I., Hoyle, B., Edmondson, E. M., Bamford, S. P., Cardamone, C. N., Keel, W. C., Lintott, C. \& Schawinski, K.} 2012, \mnras, {\bf 423}, 1485

\bibitem[\protect\citeauthoryear{Skokos \etal}{2002}]{Skok02}
Skokos, Ch., Patsis, P. A. \& Athanassoula, E. 2002, \mnras, {\bf 333}, 847

\bibitem[\protect\citeauthoryear{Smith \etal}{2012}]{Smit12}
Smith, M. C., Whiteoak, S. H. \& Evans, N. W. 2012, \apj, {\bf 746}, 181
	
\bibitem[\protect\citeauthoryear{Soderblom}{2010}]{Sode10}
Soderblom, D. R. 2010, \araa, {\bf 48}, 581

\bibitem[\protect\citeauthoryear{Sofue \& Rubin}{2001}]{SR01}
Sofue, Y. \& Rubin, V. 2001, \araa, {\bf 39}, 137

\bibitem[\protect\citeauthoryear{Solway \etal}{2012}]{SSS12}
Solway, M., Sellwood, J. A. \& Sch\"onrich, R. 2012, \mnras, {\bf 422}, 1363

\bibitem[\protect\citeauthoryear{Soubiran \etal}{2008}]{Soub08}
Soubiran, C.,{ Bienaym\'e, O., Mishenina, T. V. \& Kovtyukh, V. V.} 2008, \aap, {\bf 480}, 91

\bibitem[\protect\citeauthoryear{Sparke \& Sellwood}{1987}]{SS87}
Sparke, L. S. \& Sellwood, J. A. 1987, \mnras, {\bf 225}, 653

\bibitem[\protect\citeauthoryear{Sparke \etal}{2008}]{Spar08}
Sparke, L. S.,{ van Moorsel, G., Erwin, P. \& Wehner, E. M. H.} 2008, \aj, {\bf 135}, 99

\bibitem[\protect\citeauthoryear{Spitzer \& Schwarzschild}{1953}]{SS53}
Spitzer, L. \& Schwarzschild, M. 1953, \apj, {\bf 118}, 106

\bibitem[\protect\citeauthoryear{Springel}{2010a}]{Spri10a}
Springel, V. 2010a, \araa, {\bf 48}, 391

\bibitem[\protect\citeauthoryear{Springel}{2010b}]{Spri10b}
Springel, V. 2010b, \mnras, {\bf 401}, 791

\bibitem[\protect\citeauthoryear{Springel \etal}{2006}]{Spri06}
Springel, V., Frenk, C. S. \& White, S. D. M. 2006, \nat, {\bf 440}, 1137

\bibitem[\protect\citeauthoryear{Sridhar \& Touma}{1996}]{ST96}
Sridhar, S. \& Touma, J. 1996, \mnras, {\bf 279}, 1263

\bibitem[\protect\citeauthoryear{Steinmetz \etal}{2006}]{Stei06}
Steinmetz, M., 2006, \aj, {\bf 132}, 1645 

\bibitem[\protect\citeauthoryear{Stone \etal}{1998}]{Ston98}
Stone, J. M., Ostriker, E. C. \& Gammie, C. F. 1998, \apjl, {\bf 508}, L99

\bibitem[\protect\citeauthoryear{Struck}{2010}]{Stru10}
Struck, C. 2010, \mnras, {\bf 403}, 1516

\bibitem[\protect\citeauthoryear{Strutskie}{2006}]{Stru06}
Skrutskie, M. F.,\skip{ Cutri, R. M., Stiening, R., Weinberg, M. D., Schneider, S., Carpenter, J. M., Beichman, C., Capps, R., Chester, T., Elias, J., Huchra, J., Liebert, J., Lonsdale, C., Monet, D. G., Price, S., Seitzer, P., Jarrett, T., Kirkpatrick, J. D., Gizis, J. E., Howard, E., Evans, T., Fowler, J., Fullmer, L., Hurt, R., Light, R., Kopan, E. L., Marsh, K. A., McCallon, H. L., Tam, R., Van Dyk, S., Wheelock, S.} 2006, \aj, {\bf 131}, 1163

\bibitem[\protect\citeauthoryear{Tagger \etal}{1987}]{Tagg87}
Tagger, M., Sygnet, J. F., Athanassoula, E. \& Pellat, R. 1987, \apjl, {\bf 318}, L43

\bibitem[\protect\citeauthoryear{Tamburro \etal}{2009}]{Tamb09}
Tamburro, D.,{ Rix, H.-W., Leroy, A. K., Mac Low, M.-M., Walter, F., Kennicutt, R. C., Brinks, E. \& de Blok, W. J. G.} 2009, \aj, {\bf 137}, 4424

\bibitem[\protect\citeauthoryear{Tonini \etal}{2006}]{Toni06}
Tonini, C., Lapi, A. \& Salucci, P. 2006, \apj, {\bf 649}, 591

\bibitem[\protect\citeauthoryear{Toomre}{1964}]{Toom64}
Toomre, A. 1964, \apj, {\bf 139}, 1217

\bibitem[\protect\citeauthoryear{Toomre}{1966}]{Toom66}
Toomre, A. 1966, in {\it Geophysical Fluid Dynamics}, notes on the 1966 Summer Study Program at the Woods Hole Oceanographic Institution, ref. no. 66-46

\bibitem[\protect\citeauthoryear{Toomre}{1969}]{Toom69}
Toomre, A. 1969, \apj, {\bf 158}, 899

\bibitem[\protect\citeauthoryear{Toomre}{1981}]{Toom81}
Toomre, A. 1981, In ''The Structure and Evolution of Normal Galaxies'', Eds.~S. M. Fall \& D. Lynden-Bell (Cambridge, Cambridge Univ. Press) p.~111

\bibitem[\protect\citeauthoryear{Toomre}{1983}]{Toom83}
Toomre, A. 1983, in IAU Symposium {\bf 100}, {\it Internal Kinematics and Dynamics of Galaxies}, ed.\ E. Athanassoula (Dordrecht: Reidel) p~177

\bibitem[\protect\citeauthoryear{Toomre}{1990}]{Toom90}
Toomre, A. 1990, in {\it Dynamics \& Interactions of Galaxies}, ed.\ R. Wielen (Berlin, Heidelberg: Springer-Verlag), p.~292

\bibitem[\protect\citeauthoryear{Toomre \& Kalnajs}{1991}]{TK91}
Toomre, A. \& Kalnajs, A. J. 1991, in {\it Dynamics of Disc Galaxies}, ed.\ B. Sundelius (Gothenburg: G\"oteborgs University) p.~341

\bibitem[\protect\citeauthoryear{T\'oth \& Ostriker}{1992}]{TO92}
T\'oth, G. \& Ostriker, J. P. 1992, \apj, {\bf 389}, 5

\bibitem[\protect\citeauthoryear{Tremaine \& Weinberg}{1984a}]{TW84a}
Tremaine, S. \& Weinberg, M. D. 1984a, \apjl, {\bf 282}, L5

\bibitem[\protect\citeauthoryear{Tremaine \& Weinberg}{1984b}]{TW84b}
Tremaine, S. \& Weinberg, M. D. 1984b, \mnras, {\bf 209}, 729

\bibitem[\protect\citeauthoryear{Trujillo \etal}{2009}]{Truj09}
Trujillo, I.,{ Martinez-Valpuesta, I., Mart\'\i nez-Delgado, D. Pe\`narrubia, J., Gabany, R. J. \& Pohlen, M.} 2009, \apj, {\bf 704}, 618

\bibitem[\protect\citeauthoryear{Valenzuela \& Klypin}{2003}]{VK03}
Valenzuela, O. \& Klypin, A. 2003, \mnras, {\bf 345}, 406

\bibitem[\protect\citeauthoryear{van Albada \& Roberts}{1981}]{vAR81}
van Albada, G. D. \& Roberts, W. W. 1981, \apj, {\bf 246}, 740

\bibitem[\protect\citeauthoryear{van der Kruit \& Freeman}{2011}]{vdKF11}
van der Kruit, P. C. \& Freeman, K. C 2011, \araa, {\bf 49}, 301

\bibitem[\protect\citeauthoryear{van der Kruit \& Searle}{1981}]{vdKS81}
van der Kruit, P. C. \& Searle, L. 1981, \aap, {\bf 95}, 105

\bibitem[\protect\citeauthoryear{van de Ven \& Fathi}{2010}]{vF10}
van de Ven, G. \& Fathi, K. 2010, \apj, {\bf 723}, 767

\bibitem[\protect\citeauthoryear{van Leeuwen}{2007}]{vLee07}
van Leeuwen, F. 2007, \aap, {\bf 474}, 653

\bibitem[\protect\citeauthoryear{V\'asquez \etal}{2013}]{Vasq13}
V\'asquez, S.,\skip{ Zoccali, M., Hill, V., Renzini, A., Gonz\'alez, O. A., Gardner, E., Debattista, V. P., Robin, A. C., Rejkuba, M., Baffico, M., Monelli, M., Motta, V.\&  Minniti, D.} 2013, \aap, {\bf 555}, A91

\bibitem[\protect\citeauthoryear{Velazquez \& White}{1999}]{VW99}
Velazquez, H. \& White, S. D. M. 1999, \mnras, {\bf 304}, 254

\bibitem[\protect\citeauthoryear{Villalobos \& Helmi}{2008}]{Vill08}
Villalobos, \'A. \& Helmi A. 2008, \mnras, {\bf 391}, 1806

\bibitem[\protect\citeauthoryear{Villalobos \etal}{2010}]{Vill10}
Villalobos, \'A, Kazantzidis, S. \& Helmi, A. 2010, \apj, {\bf 718}, 314

\bibitem[\protect\citeauthoryear{Villa-Vargas \etal}{2009}]{VSH09}
Villa-Vargas, J., Shlosman, I. \& Helller, C. 2009, \apj, {\bf 707}, 218

\bibitem[\protect\citeauthoryear{Villa-Vargas \etal}{2010}]{VSH10}
Villa-Vargas, J., Shlosman, I. \& Helller, C. 2010, \apj, {\bf 719}, 1470

\bibitem[\protect\citeauthoryear{Villumsen}{1985}]{Vill85}
Villumsen, J. V. 1985, \apj, {\bf 290}, 75

\bibitem[\protect\citeauthoryear{Visser}{1978}]{Viss78}
Visser, H. C. D. 1978, \PhD, University of Groningen

\bibitem[\protect\citeauthoryear{Wada}{2004}]{Wada04}
Wada, K. 2004, In ``Coevolution of Black Holes and Galaxies''.  Ed.\ L. C. Ho (Cambridge, Cambridge University Press) p.~186 (astro-ph/0308134)

\bibitem[\protect\citeauthoryear{Wada \etal}{2011}]{Wada11}
Wada, K., Baba, J. \& Saitoh, T. R. 2011, \apj, {\bf 735}, 1

\bibitem[\protect\citeauthoryear{Wada \& Koda}{2001}]{WK01}
Wada, K. \& Koda, J. 2001, \pasj, {\bf 53}, 1163

\bibitem[\protect\citeauthoryear{Walker \etal}{1996}]{Walk96}
Walker, I. R., Mihos, J. C. \& Hernquist, L. 1996, \apj, {\bf 460}, 121

\bibitem[\protect\citeauthoryear{Wegg \& Gerhard}{2013}]{WG13}
Wegg, C. \& Gerhard, O. 2013, \mnras, to appear (arXiv1308.0593)

\bibitem[\protect\citeauthoryear{Weiland \etal}{1994}]{Weil94}
Weiland, J. L.,\skip{ Arendt, R. G., Berriman, G. B., Dwek, E., Freudenreich, H. T., Hauser, M. G., Kelsall, T., Lisse, C. M., Mitra, M., Moseley, S. H., Odegard, N. P., Silverberg, R. F., Sodroski, T. J., Spiesman, W. J. \& Stemwedel, S. W.} 1994, \apjl, {\bf 425}, L81

\bibitem[\protect\citeauthoryear{Weinberg}{1985}]{Wein85}
Weinberg, M. D. 1985, \mnras, {\bf 213}, 451

\bibitem[\protect\citeauthoryear{Weinberg}{1991}]{Wein91}
Weinberg, M. D. 1991, \apj, {\bf 373}, 391

\bibitem[\protect\citeauthoryear{Weinberg}{1998}]{Wein98}
Weinberg, M. D. 1998, \mnras, {\bf 297}, 101

\bibitem[\protect\citeauthoryear{Weinberg \& Katz}{2002}]{WK02}
Weinberg, M. D. \& Katz, N. 2002, \apj, {\bf 580}, 627

\bibitem[\protect\citeauthoryear{Weinberg \& Katz}{2007}]{WK07}
Weinberg, M. D. \& Katz, N. 2007, \mnras, {\bf 375}, 425

\bibitem[\protect\citeauthoryear{Weiner \etal}{2001}]{Wein01}
Weiner, B. J.,{ Williams, T. B., van Gorkom, J. H. \& Sellwood, J. A.} 2001, \apj, {\bf 546}, 916

\bibitem[\protect\citeauthoryear{Wielen}{1977}]{Wiel77}
Wielen, R. 1977, \aap, {\bf 60}, 263

\bibitem[\protect\citeauthoryear{Wilson \etal}{2011}]{Wils11}
Wilson, M. L.,\skip{ Helmi, A., Morrison, H. L., Breddels, M. A., Bienaym\'e, O., Binney, J., Bland-Hawthorn, J., Campbell, R., Freeman, K. C., Fulbright, J. P., Gibson, B. K., Gilmore, G., Grebel, E. K., Munari, U., Navarro, J. F., Parker, Q. A., Reid, W., Seabroke, G., Siebert, A., Siviero, A., Steinmetz, M., Williams, M. E. K., Wyse, R. F. G. \& Zwitter, T.} 2011, \mnras, {\bf 413}, 2235

\bibitem[\protect\citeauthoryear{Wyse}{2009}]{Wyse09}
Wyse, R. F. G. 2009, In ``The Galaxy Disk in Cosmological Context'', IAU Symposium {\bf 254}. Eds.~J. Andersen, J. Bland-Hawthorn \& B. Nordstr\"om (Cambridge, Cambridge University Press) p.~179 (arXiv:0809.4516)

\bibitem[\protect\citeauthoryear{Yoachim \& Dalcanton}{2006}]{YD06}
Yoachim, P. \& Dalcanton, J. J. 2006, \aj, {\bf 131}, 226

\bibitem[\protect\citeauthoryear{Yoachim \etal}{2012}]{Yoac12}
Yoachim, P., Ro\v skar, R. \& Debattista, V. P.	2012, \apj, {\bf 752}, 97

\bibitem[\protect\citeauthoryear{Yong \etal}{2012}]{Yong12}
Yong, D., Carney, B. W. \& Friel, E. D.	2012, \aj, {\bf 144}, 95

\bibitem[\protect\citeauthoryear{York \etal}{2000}]{York00}
York, D. G., 2000, \aj, {\bf 120}, 1579

\bibitem[\protect\citeauthoryear{Yu \etal}{2012}]{Yu12}
Yu, J-C.,{ Sellwood, J. A., Pryor, C. Hou, J-L. \& Li, C.}  2012, \apj, {\bf 754}, 124

\bibitem[\protect\citeauthoryear{Zang}{1976}]{Zang76}
Zang, T. A. 1976, \PhD, MIT

\bibitem[\protect\citeauthoryear{Z\'anmar S\'anchez \etal}{2008}]{ZSS08}
Z\'anmar S\'anchez, R.,{ Sellwood, J. A., Weiner B. J. \& Williams, T. B.} 2008, \apj, {\bf 674}, 797

\bibitem[\protect\citeauthoryear{Zibetti \etal}{2009}]{ZCR09}
Zibetti, S., Charlot, S. \& Rix, H.-W. 2009, \mnras, {\bf 400}, 1181

\end{thebibliography}
\end{document}